\documentclass[cernpreprint,texlive=2009,txfonts,UKenglish]{latex/atlasdoc}
\usepackage{latex/atlaspackage}
\usepackage{latex/atlasbiblatex}
\usepackage{latex/atlascontribute}
\usepackage{multirow}
\usepackage{latex/atlasphysics}
\graphicspath{{./}}
\usepackage{HIGG-2013-17-defs}

\def \cth {\cos\theta^{*}}
\def \cthone {\cos\theta_{1}}
\def \cthtwo {\cos\theta_{2}}
\def \mone {m_{12}}
\def \mtwo {m_{34}}
\def \mfl {m_{4\ell}}
\def \ptfl {p_{{\rm T},4\ell}}
\def \etafl {\eta_{4\ell}}
\def \tev{\ifmmode {\mathrm{\ Te\kern -0.1em V}}\else
                    \textrm{Te\kern -0.1em V}\fi}%

\hypersetup{pdftitle={ATLAS draft},pdfauthor={The ATLAS Collaboration}}

\AtlasTitle{Study of the spin and parity of the Higgs boson in diboson decays with the ATLAS detector}

\author{The ATLAS Collaboration}

 \PreprintIdNumber{ CERN-PH-EP-2015-114}

\AtlasAbstract{%
This note presents studies of the spin, parity and Lagrangian tensor structure of the Higgs boson  
in the \hZZ , \hWW\ and \hgg\ decay processes at the LHC.
The investigations are based on  $25\;{\rm fb}^{-1}$ of $pp$ collision data collected by the ATLAS experiment at
$\sqrt{s}=7$~\TeV\ and $\sqrt{s}=8$~\TeV. The Standard Model (SM) Higgs boson hypothesis, corresponding to the quantum numbers $J^{P}=0^{+}$, 
is tested against several alternative spin and parity models.
These include two non-SM spin-0 and spin-2 models with universal and non-universal couplings to 
fermions and vector bosons. 
The results presented here exclude all of the alternative models in
favour of the SM Higgs boson hypothesis at  more 
than $99\%$~confidence level. 
The tensor structure of the $HVV$ interaction in the spin-0 hypothesis is also investigated using the \hZZ\ and \hWW\ decays. 
The observed distributions of the variables sensitive to the ratios of the non-SM tensor couplings to the SM ones, \KtildeH\ and \KtildeA, 
are compatible with the SM prediction.
Assuming that the values of the \KtildeH\ and \KtildeA\ couplings are the same for the $HWW$ and $HZZ$ processes,
values of the non-SM couplings outside the intervals $-0.73< \KtildeH < 0.63$  and $-2.18< \KtildeA < 0.83$ are excluded at 95\% ~confidence level.
}

\hypersetup{colorlinks,breaklinks,pdftitle={ATLAS Draft Cover},pdfauthor={ATLAS Collaboration}}
\hypersetup{linkcolor=blue,citecolor=blue,filecolor=black,urlcolor=blue}

\usepackage{latex/atlascontribute}

\graphicspath{{logos/}}

\AtlasTitle{Study of the spin and parity of the Higgs boson in diboson decays with the ATLAS detector}

\AtlasRefCode{HIGG-2013-17}

\AtlasJournal{EPJC}

\AtlasAbstract{ 
Studies of the spin, parity and tensor couplings of the Higgs boson 
in the \hZZ , \hWW\ and \hgg\ decay processes at the LHC are presented. 
The investigations are based on  $25\;{\rm fb}^{-1}$ of $pp$ collision data collected by the ATLAS experiment at
$\sqrt{s}=7$~\TeV\ and $\sqrt{s}=8$~\TeV. The Standard Model (SM) Higgs boson hypothesis, corresponding to the quantum numbers $J^{P}=0^{+}$, 
is tested against several alternative spin scenarios,
including non-SM spin-0 and spin-2 models with universal and non-universal couplings to 
fermions and vector bosons. 
All tested alternative models are excluded in favour of the SM Higgs boson hypothesis at more 
than $99.9\%$~confidence level. 
Using the \hZZ\ and \hWW\ decays, 
the tensor structure of the interaction between the spin-0 boson and the SM vector bosons is also investigated.
The observed distributions of variables sensitive to the non-SM tensor couplings 
are compatible with the SM predictions and constraints on the non-SM couplings are derived. 
}
\begin{document}

\section{Introduction}
\label{sec:intro}
The discovery of a Higgs boson by the ATLAS~\cite{HiggsObservationATLAS} and CMS~\cite{HiggsObservationCMS} experiments at the 
Large Hadron Collider (LHC) at CERN marked the beginning of a new era of experimental
studies of the properties of this new particle.
In the Standard Model (SM), the Higgs boson is a CP-even scalar particle, $J^{CP}=0^{++}$.~\footnote{In the following, 
for brevity, only the $J^P$ label is used to indicate the spin and CP quantum numbers.}
Theories of physics beyond the SM (BSM) often require an extended Higgs sector featuring several neutral Higgs bosons. 
Such cases may include CP-mixing in the Higgs boson interactions, which could result in observable differences in the kinematics
of final-state particles produced in their decays.
A review of the phenomenology in the determination of Higgs boson spin and CP properties 
can be found in Ref.~\cite{YR3} and references therein.

Previous determinations of the Higgs boson spin and CP quantum numbers by the ATLAS and CMS Collaborations are reported in 
Refs.~\cite{HiggsSpin2013} and~\cite{CMS_Spin}.
Results on the same subject have also been published by the D0 and CDF Collaborations in Ref.~\cite{Aaltonen:2015mka}. 
All these studies indicate the compatibility of the spin and CP properties of the observed Higgs boson with the SM predictions.
The ATLAS measurement excluded several alternative spin and parity
hypotheses in favour of the quantum numbers predicted by the SM. In addition to the exclusion of several non-SM
spin hypotheses, the CMS measurement probed   
the tensor structure of the Higgs boson decay to SM vector bosons in the spin-0 scenario.
This paper complements the previous ATLAS study of the Higgs boson spin and parity.
The new study takes advantage of improvements to the analysis strategy and to the modelling used to describe 
alternative spin hypotheses, and includes studies on CP-mixing for the spin-0 scenario. 
The improved theoretical framework is based on the Higgs boson characterisation model described in Refs.~\cite{YR3,HC}. 
 
The study of the spin and parity properties of the Higgs boson presented in this paper is based on the \hgg , \hZZ\ and \hWW\ 
decay channels and their combination. The \hWW\ analysis is described in detail in a separate publication~\cite{spincp_ww_paper}.  
These analyses are based on 4.5~fb$^{-1}$ and 20.3~fb$^{-1}$ of $pp$ collision data collected by the ATLAS 
experiment at centre-of-mass energies of 7~\TeV\ and 8~\TeV , respectively. 
For the \hWW\ studies only the data collected at a centre-of-mass energy of 8 \TeV\ are used. 
The SM hypothesis $J^P=0^+$ is compared to alternative spin-0 models: 
a pseudoscalar boson $J^P=0^-$ and a BSM scalar boson $J^P=0^+_h$ \cite{JHU1,JHU2}, 
which describes the interaction of the Higgs boson with the SM vector bosons with higher-dimension operators 
discussed in Section~\ref{sec:spin0}. 
Graviton-like tensor models with $J^P=2^+$ with universal and non-universal couplings \cite{HC,YR3} are also considered. 
In these tests of fixed spin and parity hypotheses it is assumed that the
resonance decay involves only one CP eigenstate.

In addition to the fixed spin and parity hypothesis tests,  
the possible presence of BSM terms in the Lagrangian describing the $HVV$ 
vertex\footnote{In this paper the symbol $V$ is used to describe a massive SM vector boson, 
namely either a $W$ or a $Z$ boson.} of the spin-0 resonance is also investigated.  
The $HVV$ interaction is described in terms of an effective Lagrangian that contains the SM interaction 
and BSM CP-odd and CP-even terms~\cite{YR3,HC}.
The relative fractions of the CP-odd and CP-even BSM contributions to 
the observed Higgs boson decays are constrained, and limits on the corresponding BSM tensor couplings are derived.

This paper is organised as follows. In Section~\ref{sec:detector} 
the ATLAS detector is described. 
In Section~\ref{sec:theory} the theoretical framework used to derive the spin and parity models, as well as the
parameterisation used to describe the $HVV$ coupling tensor structure, are discussed. 
In Section~\ref{sec:MCsamples}, 
the choice of Monte Carlo generators for the simulation of signal and backgrounds is described. 
The analyses of fixed spin and parity hypotheses for the three decay channels and their combination are presented 
in Section~\ref{sec:fixed_hypo}.
Individual and combined studies of the tensor structure of the $HVV$ interaction are presented in Section~\ref{sec:tensor}. 
Concluding remarks are given in Section~\ref{sec:conclusion}.

\section{The ATLAS detector}
\label{sec:detector}
The ATLAS detector is described in detail in Ref.~\cite{atlas-det}. ATLAS is a multi-purpose detector with 
a forward-backward symmetric cylindrical geometry. 
It uses a right-handed coordinate system with its origin at the nominal interaction point (IP) in the centre of the 
detector and the $z$-axis along the beam pipe.  The $x$-axis points from the IP to the centre of the LHC ring, and the $y$-axis 
points upward. Cylindrical coordinates $(r,\phi)$ are used in the transverse plane, $\phi$ being the azimuthal angle around the beam pipe. 
The pseudorapidity is defined as $\eta = -\ln \tan (\theta /2)$, where $\theta$ is the polar angle.

At small radii from the beamline, the inner detector (ID), immersed in a $2$~T magnetic field produced by a thin superconducting solenoid
located in front of the calorimeter, is made up of fine-granularity pixel and microstrip detectors. 
These silicon-based detectors cover the range $|\eta|<2.5$. A gas-filled straw-tube transition-radiation tracker (TRT) 
complements the silicon tracker at larger radii and also provides electron identification based on transition radiation. 
The electromagnetic (EM) calorimeter is a lead/liquid-argon sampling calorimeter with an accordion geometry. The EM calorimeter 
is divided into a barrel section covering $|\eta| <1.475 $ and two end-cap sections covering 
$1.375<|\eta| <3.2 $. For $|\eta| <2.5 $ it is divided into three layers in depth, which are finely segmented in $\eta$ and $\phi$. 
An additional thin presampler layer, covering $|\eta| <1.8 $, is used to correct for 
fluctuations in energy losses of particles before they reach the calorimeter.
Hadronic calorimetry in the region $|\eta| <1.7 $ uses steel absorbers and scintillator tiles as the active medium. 
Liquid argon with copper absorbers is used in the hadronic end-cap calorimeters, which cover the region $1.5<|\eta| <3.2 $.  A forward 
calorimeter using copper or tungsten absorbers with liquid argon completes the calorimeter coverage up to $|\eta| =4.9 $. The muon spectrometer (MS) 
measures the deflection of muon trajectories with $|\eta| <2.7 $, using three stations of precision drift tubes, with cathode strip chambers in the innermost 
layer for $|\eta| > 2.0 $. The deflection is provided by a toroidal magnetic field with an integral of approximately $3$~Tm and $6$~Tm in the central and 
end-cap regions of the ATLAS detector, respectively. The muon spectrometer is also instrumented with dedicated trigger chambers, 
the resistive-plate chambers in the barrel and thin-gap chambers in the end-cap,  covering $|\eta| <2.4 $.

\section{Theoretical models}
\label{sec:theory}
In this section, the theoretical framework for the measurements of the spin and parity of the resonance is discussed. 
An effective field theory (EFT) approach is adopted to describe the interaction 
between the resonance and the SM vector bosons, following the Higgs boson characterisation model 
described in Refs.~\cite{YR3,HC}. 
Three possible BSM scenarios for the spin and parity of the boson are considered: 
\begin{itemize}
\item{the observed resonance is a spin-2 particle,} 
\item{the observed resonance is a pure BSM spin-0 CP-even or CP-odd Higgs boson,} 
\item{the observed resonance is a mixture of the SM spin-0 state and a BSM spin-0 CP-even or CP-odd state.}
\end{itemize}
The third case would imply CP-violation in the Higgs sector. 
In the case of CP mixing, the Higgs boson would be a mass eigenstate, but not a CP eigenstate. 
In all cases, only one resonance with a mass of about 125~\GeV\ is considered. 
It is also assumed that the total width of the resonance is small compared to the typical experimental resolution of the 
ATLAS detector (of the order of 1--2~\GeV\ in the four-lepton and $\gamma \gamma$ final states, as documented in Ref.~\cite{MassPaper}).
Interference effects between the BSM signals and SM backgrounds are neglected.

The EFT approach, used by the Higgs boson characterisation model, is only valid up to a certain energy  scale, $\Lambda$. 
The models described in Ref.~\cite{HC} assume that the resonance structure corresponds to one new boson ($X(J^P)$ with $J^{P} = 0^{\pm}$ or $2^+$), 
assuming that any other BSM particle only exists at an energy scale larger than $\Lambda$.
The $\Lambda$ scale is set to 1~\TeV\ to account 
for the experimental results obtained at the LHC and previous collider experiments, which do not show any 
evidence of new physics at lower energy scales. 

The case where the observed resonance has $J^{P} = 1^{\pm}$ is not studied in this paper.
The \hgg\ decay is forbidden by the Landau--Yang theorem~\cite{landau,yang1950} for a spin-1 particle. 
Moreover, the spin-1 hypothesis was already studied in 
the previous ATLAS publication~\cite{HiggsSpin2013} in the \hZZ\ and \hWW\ decays and excluded at a more than 99\% confidence level.

\subsection{The spin-0 hypothesis}\label{sec:spin0}
In the spin-0 hypothesis, models with fixed spin and parity, and models with mixed SM spin-0 and 
BSM spin-0 CP-even and CP-odd contributions are considered.
In Ref.~\cite{HC}, the spin-0 particle interaction with pairs of $W$ or $Z$ bosons is 
given through the following interaction Lagrangian:
\begin{eqnarray}
\mathcal L_0^{V} = &\left\{   \cos(\alpha) \kappa _{\rm SM}\left[\frac{1}{2}g_{HZZ} Z_{\mu} Z^ {\mu} + g_{HWW} W^{+}_{\mu} W^ {- \mu}\right]\right. \nonumber \\
\label{eq:spin0_l}
&\left.  -\frac{1}{4}\frac{1}{\Lambda}\left[ \cos(\alpha)  \kappa _{HZZ}  Z_{\mu\nu}Z^{\mu\nu} + \sin(\alpha)  \kappa _{AZZ}  Z_{\mu\nu} \tilde{Z}^{\mu\nu}\right]  \right.\\
& \left. -\frac{1}{2} \frac{1}{\Lambda} \left[  \cos(\alpha)  \kappa _{HWW} W^+_{\mu\nu}W^{-\mu\nu}+  \sin(\alpha)  \kappa _{AWW} W^+_{\mu\nu} \tilde{W}^{ - \mu\nu}\right] \right\} X_0. \nonumber
\end{eqnarray}  
Here $V^{\mu}$ represents the vector-boson field ${(V=Z,W^{\pm}})$, the $V^{\mu\nu}$ are the reduced field tensors 
and the dual tensor is defined as $\tilde{V}^{\mu\nu}= \frac{1}{2}\varepsilon^{\mu\nu\rho\sigma}V_{\rho\sigma}$. The 
symbol $\Lambda$ denotes the EFT energy scale. The symbols $\kappa_{\rm SM}$, $\kappa_{HVV}$ and $\kappa_{AVV}$ 
denote the coupling constants corresponding to the interaction of 
the SM, BSM CP-even or BSM CP-odd spin-0 
particle, represented by the $X_0$ field,  with  $ZZ$ or $WW$ pairs. 
To ensure that the Lagrangian terms are Hermitian, these couplings are assumed to be real.
The mixing angle $\alpha$ allows for production of CP-mixed states and implies CP-violation 
for $\alpha \neq 0$ and $\alpha \neq \pi$, provided the corresponding 
coupling constants are non-vanishing. 
The SM couplings, $g_{HVV}$, are proportional to the square of the vector boson masses: $g_{HVV} \propto  m^{2}_{V}$.  
Other higher-order operators described in Ref.~\cite{HC}, namely the derivative operators, 
are not included in Eq.~(\ref{eq:spin0_l}) and have been neglected in this analysis since they induce modifications 
of the discriminant variables well below the sensitivity achievable with the available data sample.

As already mentioned, for the spin-0 studies the SM Higgs boson hypothesis is compared to two alternatives: 
the CP-odd $J^P=0^-$ and the BSM CP-even $J^P=0^+_h$ hypotheses.  
All three models are obtained by selecting the corresponding parts of the Lagrangian described 
in Eq.~(\ref{eq:spin0_l}) while setting all other contributions
to zero. The values of the couplings corresponding to the different spin-0 models are listed in Table~\ref{tab:spin0_couplings}. 
\begin{table}[htbp]
  \begin{center}
    \begin{tabular}{lccccc}
     \hline\hline
     $J^P$ & Model & \multicolumn{4}{c}{Values of tensor couplings}\\
    &&$\kappa _{\rm SM}$& $\kappa _{HVV}$&$\kappa _{AVV}$&$\alpha$\\
      \hline
       $0^+$ & SM Higgs boson &  $1$ & $ 0$& $ 0$& $0$ \\
       $0^+_h$ & BSM spin-0 CP-even &   $0$& $1$& $ 0$& $0$\\
        $0^-$  & BSM spin-0 CP-odd  &    $0$& $ 0$& $1$& $ \pi /2$\\
      \hline\hline
    \end{tabular}
  \end{center}
  \caption{ Parameters of the benchmark scenarios for spin-0 boson tensor couplings used in tests (see Eq.~(\ref{eq:spin0_l})) 
of the fixed spin and parity models.  }
  \label{tab:spin0_couplings}
\end{table} 

The investigation of the tensor structure of the $HVV$ interaction is based on the assumption that the observed particle has spin zero.
Following the parameterisation defined in Eq.~(\ref{eq:spin0_l}), scenarios are considered where 
only one CP-odd or one CP-even BSM contribution at a time is present 
in addition to the SM contribution.  
To quantify the presence of  BSM contributions in $H \to ZZ^{*}$ and $H \to WW^{*}$ decays, the ratios of couplings
\KtildeA\ and \KtildeH\ are measured. 
Here $\tilde{\kappa}_{ AVV}$ and $\tilde{\kappa}_{HVV}$
are defined as follows:
\begin{equation}
 \tilde{\kappa}_{ AVV}=\frac{1}{4} \frac{ {\rm v} }{\Lambda} \kappa_{AVV}\;\; {\rm and}\;\; \tilde{\kappa}_{HVV}=\frac{1}{4} \frac{  {\rm v} }{\Lambda} \kappa_{HVV}, 
\end{equation}
where ${\rm v}$ is the vacuum expectation value~\cite{PDG-2014}  of the SM Higgs field. 

The mixing parameters \KtildeA\ and \KtildeH\ correspond to the ratios of tensor couplings $g_4/g_1$ and $g_2/g_1$ proposed in the anomalous coupling 
approach described in 
Refs.~\cite{JHU1,JHU2}. To compare the results obtained in this analysis to other existing studies, the final results are also expressed in 
terms of the effective cross-section fractions $(f_{g2},\phi_{g2})$ and $(f_{g4},\phi_{g4})$ proposed in Refs.~\cite{YR3} and~\cite{JHU1,JHU2}.  Further details of 
these conversions are given in Appendix~A.

The BSM terms described in Eq.~(\ref{eq:spin0_l}) are also expected to change the relative contributions of the vector-boson fusion (VBF) and vector-boson associated production ($VH$) processes with respect to the gluon-fusion (ggF) production process, which is predicted to be the main production mode for the 
SM Higgs boson at the LHC.
For large values of the BSM couplings, at the LHC energies, the VBF production mode can have a cross section that is comparable to the 
ggF process~\cite{MG5}.  This study uses only kinematic properties of particles from $H \to V V^{*}$ decays to derive information on the CP 
nature of the Higgs boson. 
The use of the signal rate information for different production modes, in the context of the EFT analysis, 
may increase the sensitivity to the BSM couplings at the cost of a loss in generality. 
For example the ratio of the VBF and $VH$ production modes with respect to the ggF one 
can be changed by a large amount for non-vanishing values of the BSM couplings. 
In the studies presented in this paper the predictions of the signal rates are not used to constrain the BSM couplings.  
 
As described in Section~\ref{sec:tensor_ww}, only events with no reconstructed jets (the 0-jet category) are used in the \hWW\ analysis 
for the studies of the tensor structure; hence this analysis has little sensitivity to the VBF production mode.
The \hZZ\ analysis also has little sensitivity to this production mode since it is mainly based on variables related to the four-lepton kinematics.
The Boosted Decision Tree (BDT) algorithm~\cite{Hocker:2007ht} used to discriminate signals from the $ZZ^{*}$ background, 
described in Sections~\ref{sec:fixed_zz} and  \ref{sec:tensor_zz}, 
includes the transverse momentum of the four-lepton system and is trained on simulated samples of ggF-produced signals.
An enhancement of the VBF production mode would improve the separation between background and signal since 
it predicts larger values of the transverse momentum spectrum for events produced via VBF than via ggF~\cite{YR3}.

\subsection{The spin-2 hypothesis}\label{sec:spin2}
In the Higgs boson characterisation model~\cite{HC}, the description of the  interaction of a spin-2 particle with fermions and vector bosons  
is described by the following Lagrangian:
\begin{equation}
 \mathcal L_2 = - \frac{1}{\Lambda}\left[ \sum_V \kappa_V {\cal T}^V_{\mu\nu}  X^{\mu\nu} +   \sum_f \kappa_f {\cal T}^f_{\mu\nu} X^{\mu\nu}  \right].
  \label{eq:spin2lagrangian}
\end{equation}
The spin-2 tensor field $X^{\mu\nu}$ is chosen to interact with the energy-momentum tensors, 
${\cal T}^V_{\mu\nu}$ and ${\cal T}^f_{\mu\nu}$, 
of any vector boson $V$ and fermion $f$, as inspired by gravitation theories. 
The strength of each interaction is determined by the couplings $\kappa_V$ and $\kappa_f$. In the simplest
formulation, all couplings are equal. 
This scenario is referred to as universal couplings (UC), while scenarios with different values of 
the couplings are referred to as non-universal couplings (non-UC). 
In the UC scenario, the production of a spin-2 particle in $pp$ collisions is expected to be dominated by QCD processes, 
with negligible contributions from electroweak (EW) processes (i.e.~from processes involving EW boson propagators). 
Simulation studies based on \mgaMC~\cite{MG5} , which implements the Lagrangian described in Eq.~(\ref{eq:spin2lagrangian}), predict for 
the production cross section in the UC scenario $\sigma_{\mathrm{EW}}/\sigma_{\mathrm{QCD}}\simeq 3 \times 10^{-4}$. 
These studies also show that EW production of the spin-2 resonance would occur mainly in association with a 
massive EW boson ($WX$, $ZX$). 
Present observations do not show a dominant $VH$ production mechanism, hence suggesting that $\sigma_{\mathrm{EW}}$ is significantly smaller than $\sigma_{\mathrm{QCD}}$. 
This paper considers only QCD production for all the spin-2 benchmark scenarios.
  
The UC models predict a branching ratio of about $5\%$ to photon pairs and negligible branching ratios to massive EW gauge boson pairs, 
$WW^*$ and $ZZ^*$. This prediction is disfavoured by the experimental measurements~\cite{Aad:2014eva,ATLAS:2014aga,ATLAScouplingsHgg}
and therefore the equality between all couplings $\kappa$ cannot hold. 
In the benchmark scenarios studied in this paper, each of the couplings $\kappa_W$, $\kappa_Z$, and $\kappa_\gamma$
is assumed to be independent of all the other couplings.
In the following, the UC scenario only refers to $\kappa_{q} = \kappa_{g}$, 
without implying the equality for the other $\kappa$ values.

The simplest QCD production processes, $gg\to X$ and $q\bar{q} \to X$ (where $q$ refers to light quarks), 
yield different polarisations for the spin-2 particle $X$, 
and hence different angular distributions of its decay products. 
These mechanisms are considered in the model of a graviton-like tensor with minimal couplings proposed in Refs.~\cite{JHU1,JHU2}, which
has been studied experimentally in Ref.~\cite{HiggsSpin2013}. 
The EFT Lagrangian, however, also allows for more complex processes with emission of one or more additional partons.
For instance, processes with one-parton emission, like $qg\to qX$ and $~\bar{q}g\to \bar{q}X$, can produce a spin-2 state 
through either a $qqX$ or a $ggX$ vertex. When two partons are emitted, as in $gg\to q\bar{q}X$ or $q\bar{q}\to q\bar{q}X$, the spin-2 production 
may occur through $qqX$ or $ggX$ vertices, respectively, such that the polarisation of $X$ is not uniquely determined by the initial state. 
Moreover, the EFT also allows for four-leg vertices like $qqgX$. These additional diagrams effectively change the polarisation of the particle $X$, 
compared to what is assumed by the model in Refs.~\cite{JHU1,JHU2}. 
As a consequence, the angular distributions of the decay products become harder to separate 
from those expected for a scalar resonance.

The QCD production of a spin-2 particle is driven by the values of the couplings $\kappa_g,~\kappa_q$.
Presently, there are no experimental constraints on the ratio $\kappa_q / \kappa_g$  from observed decay modes, 
since the separation of jets initiated by gluons or by light quarks is experimentally 
difficult and has not yet been attempted in Higgs boson studies. 
The ratio $\kappa_q/\kappa_g$ can thus be regarded as a free parameter.
When $\kappa_q\neq\kappa_g$, the spin-2 model predicts an enhancement of the tail of the distribution of the transverse momentum, $\pT^X$, 
of the spin-2 particle. Such a high-$\pT^X$ tail is not present for the $\kappa_q = \kappa_g$ (UC) case.
As stated before, however, the EFTs are valid only up to some energy scale, $\Lambda$. At higher energies, new physics phenomena are expected 
to enter to regularise the anomalous ultra-violet behaviour. 

In the present analysis, a selection $\pT^X<300$~\GeV\ is applied when investigating non-UC scenarios, $\kappa_q\neq\kappa_g$.
In addition, for the non-UC scenarios, analyses using a tighter selection $\pT^X<125~\GeV$ are also performed. 
This is a conservative choice for the $\pT^{X}$ selection, as the EFT must describe the physics at least up to the mass of the observed resonance. 
It has been verified that the choice of the $\pT^{X}$ selection does not affect the results for the UC scenario.
Even assuming the $\pT^X<300$~\GeV\ selection, some choices of $\kappa_q/\kappa_g$ produce high-$\pT^X$ 
tails incompatible with the observed differential distribution reported in Refs.~\cite{Aad:2014tca,Aad:2014lwa}. 
For this reason the investigated range of the $\kappa_q/\kappa_g$ ratio is limited to between zero and two.
The spin-2 scenarios considered in this study are presented in Table~\ref{tab:QCDcouplings}. 
The $\kappa_{q} = \kappa_{g}$ model is referred to hereafter as the UC scenario.  
The $\kappa_q=0$ case implies a negligible coupling to light quarks,  
whereas the $\kappa_q=2\kappa_g$ case is an alternative scenario with an enhanced coupling to quarks.

\begin{table}[htbp!]
  \begin{center}
    \begin{tabular}{lccc}
    \hline\hline
      \multicolumn{2}{c}{Values of spin-2 quark and gluon couplings} & \multicolumn{2}{c}{$\pT^X$ selections (\GeV)} \\
      \hline
       $\kappa_q=\kappa_g$  & Universal couplings & -- & -- \\
       $\kappa_q=0$         & Low light-quark fraction &  $<300$ & $<125$ \\
       $\kappa_q=2\kappa_g$ & Low gluon fraction&  $<300$ & $<125$ \\
      \hline\hline
    \end{tabular}
  \end{center}
  \caption{
Choices of the couplings to quarks  $\kappa_{q}$ and to gluons $\kappa_{g}$ 
studied for the spin-2 benchmark scenarios. The values of the selection 
criteria applied to the transverse momentum $\pT^X$ of the spin-2 resonance are also shown.
For the UC scenario no $\pT^X$ selection is applied. }
  \label{tab:QCDcouplings}
\end{table}

\section{Data and simulated samples}
\label{sec:MCsamples}
The data presented in this paper were recorded by the ATLAS detector during the 2012 LHC run
with proton--proton collisions at a centre-of-mass energy of 8~\TeV,
and correspond to an integrated luminosity of 20.3~fb$^{-1}$. 
For the \hgg\ and \hZZ\ channels, the data collected in 2011 at 
a centre-of-mass energy of 7~\TeV\ corresponding to an integrated
luminosity of 4.5~fb$^{-1}$, are also used. 
Data quality requirements are applied to reject events recorded when 
the relevant detector components were not operating correctly.
More than 90\% of the recorded luminosity is used in these studies.
The trigger requirements used to collect the data analysed in this paper are the same as
those described in previous publications~\cite{Aad:2014eva,ATLAS:2014aga,ATLAScouplingsHgg}. 
They are only briefly recalled in the following sections.
 
The Monte Carlo (MC) samples for the backgrounds and for the SM Higgs boson signal are the same as those
used for the analyses described in Refs.~\cite{Aad:2014eva,ATLAS:2014aga,ATLAScouplingsHgg}, whereas new non-SM signal 
samples have been simulated. An overview of the signal samples is given in Section~\ref{sec:BMSsamples}.

The effects of the underlying event  and of additional minimum-bias
interactions occurring in the same or neighbouring bunch crossings,
referred to as pile-up in the following, are modelled
with \PYTHIA8~\cite{pythia8}. The ATLAS detector response is
simulated~\cite{atlassim} using either \GEANT4~\cite{GEANT4}
alone or combined with a parameterised \GEANT4-based calorimeter
simulation~\cite{AFII}.

\subsection{SM Higgs boson and BSM signal samples}
\label{sec:BMSsamples}

The SM Higgs boson ggF production for all analyses is modelled
using the \POWHEG-Box~\cite{powheg} generator at next-to-leading order (NLO), 
interfaced to \PYTHIA8 for parton showering and hadronisation and to simulate multi-parton
interactions. To improve the modelling of the SM
Higgs boson \pT, a reweighting procedure is applied. 
This procedure applies a weight depending on the \pT\ of the Higgs boson to each event. 
The weights are chosen in order to reproduce the prediction of the next-to-next-to-leading-order (NNLO) and
next-to-next-to-leading-logarithms (NNLL) dynamic-scale calculation
given by the \textsc{hres}2.1 program~\cite{deFlorian:2011xf,Grazzini:2013mca}.

For the \hgg\ analysis, the signal samples are generated at several
values of the Higgs boson mass $m_H$ around 125~\GeV. 
The samples are used to obtain a parameterisation of the signal yields and of the invariant mass 
distribution of the two-photon system as continuous functions of $m_H$ (both inclusively and for each category in the analysis, 
as described in Section~\ref{sec:fixed_gg}).
The spin-2 samples are generated using the \mgaMC ~\cite{MG5} program with LO accuracy
for zero, one, and two additional partons, and with subsequent matching of the matrix-element calculation 
with a model of the parton shower, underlying event and hadronisation, using \PYTHIA6~\cite{pythia}.

In the \hZZ\ analysis the signal samples representing the production
and decay of Higgs bosons with spin-0 and different parities are
generated as follows. The SM Higgs boson production via
gluon fusion at the mass $m_H=125.5$~\GeV\ is simulated using the
\POWHEG-Box\ generator.  
For the non-SM signals, the decays of the generated Higgs bosons are 
simulated, according to the Higgs boson parity assumptions, using the JHU~\cite{JHU1,JHU2} MC generator at leading order (LO). 
The spin-2 samples are generated using the \mgaMC\ MC generator, 
as for the \hgg\ analysis.

For the \hWW\ analysis, the SM Higgs boson signal is generated at $m_H
= 125$~\GeV\ using the \POWHEG-Box\ Monte Carlo generator. The spin-0 BSM signal samples are generated using \mgaMC.  
The signal samples representing the
production and decay of Higgs bosons with spin-$2$ are generated using
the \mgaMC\ MC generator, as for the \hgg\ analysis.

For studies of the tensor structure of the $HVV$ decay, all simulated signal
samples are obtained by using the matrix element (ME) reweighting
method applied, as explained in the following, to a sample generated with non-zero values of the BSM couplings. 
The reweighting procedure is validated against samples produced at different values of the couplings, to ensure
that the distributions of the CP-sensitive final-state observables and
of their correlations are reproduced correctly. For the \hZZ\
analysis, the MC production is only performed for one set of
tensor couplings: $g_1=1$, $g_2=1+i$, $g_4= 1+i$. All other configurations of couplings are obtained by reweighting this
 sample at generator level. The ratios of the
corresponding squares of ME values calculated at LO are
used as weights. To calculate these ME values, the
JHUGenME~\cite{JHU2} program is used.  In the \hWW\ analysis, only one
MC sample is generated, using \mgaMC\ with parameters $\kappa_{\rm SM} =1$, $\kappa_{AWW} = 2$, $\kappa_{HWW} = 2$, 
$\cos(\alpha) = 0.3$, and all other samples are obtained from it by reweighting the events on
the basis of the ME amplitudes.

In all the analyses presented in this paper, the mass of the Higgs boson is 
fixed to 125.4 \GeV ~\cite{MassPaper}.

\subsection{Background samples}
\label{sec:BKGsamples}

The MC simulated samples for the backgrounds, as well as for the determinations of the corresponding cross sections, 
are the same as those adopted in Refs.~\cite{Aad:2014eva,ATLAS:2014aga,ATLAScouplingsHgg}.
In the \hgg\ analysis, the background is dominated by prompt $\gamma \gamma$ events, with smaller contributions 
from $\gamma$--jet events. For the \hZZ\ analysis, the major background is the non-resonant $ZZ^{*}$ process, 
with minor contributions from the $\ttbar$ and $Z$+jets processes. 
For the \hWW\ analysis, the dominant backgrounds are non-resonant $W$ boson pair ($WW$) production, 
  $\ttbar$ and single-top-quark production, and the $\ZDY$ process 
followed by the decay to $\tau\tau$ final states.

\section{Tests of fixed spin and parity hypotheses}
\label{sec:fixed_hypo}
The \hgg\ and \hZZ\ analyses are improved with respect to the previous ATLAS publication of Ref.~\cite{HiggsSpin2013}. 
These analyses are described in some detail in the following subsections. The spin and parity analysis in the \hWW\ channel has 
also been improved, as discussed in detail in a separate publication~\cite{spincp_ww_paper}.
In the following, only a brief overview of this analysis is given. The expected and observed results of the individual channels
and of their combination are presented in Section~\ref{sec:fixed_comb}. 

\subsection{Statistical treatment}
\label{sec:fixed_stat}
The analyses rely on discriminant observables chosen to be sensitive to the spin and parity of the signal. 

A likelihood function, ${\cal L}({\rm data}~|~\spin, \mu, \vec{\theta})$, that depends on the spin-parity assumption of the signal 
is constructed  as a product of conditional probabilities over binned distributions of the discriminant observables in each channel:
\begin{equation}
\begin{split}
& \mathcal{L}({\rm data}~|~ \spin, \mu, \vec{\theta}) =  \displaystyle  \prod_{j}^{N_{\mathrm{chann.}}} \prod_{i}^{N_{\mathrm{bins}}} 
 P\big(N_{i,j} ~| ~\mu_{j} \cdot S^{(\spin)}_{i, j}(\vec{\theta}) ~+ B_{i,j}(\vec{\theta}) \big)  \cdot  \mathcal{A}_{j}(\vec{\theta}) \;,
\end{split}
\label{eq:likelihood}
\end{equation}

\noindent where $\mu_{j}$ represents the parameter associated with the signal rate normalised to the SM prediction in each channel $j$.\footnote{Here channel can be used to indicate different categories in the same final state when producing results for individual decay channels, or different final states when combining them.} 
The symbol $\vec{\theta}$ represents all nuisance parameters. 
The likelihood function is a product of Poisson distributions $P$ corresponding 
to the observation of $N_{i,j}$ events in each bin $i$ of the discriminant observables,
given the expectations for the signal, $S_{i,j}^{(\spin)}(\vec{\theta})$, and for the background, $B_{i,j}(\vec{\theta})$. 
Some of the nuisance parameters are constrained by auxiliary measurements. Corresponding constraints are
represented by the functions $\mathcal{A}_{j}(\vec{\theta})$.

While the couplings are predicted for the SM Higgs boson, they are not known 
\textit{a priori} for the alternative hypotheses, defined as \spinalt, as discussed in Section~\ref{sec:theory}. 
In order to be insensitive to assumptions on the couplings of the non-SM resonance (the alternative hypotheses) to SM particles, 
the numbers of signal events in each channel, for each different LHC centre-of-mass energy and for each tested hypothesis, are 
treated as independent parameters in the likelihood and fitted to the data when deriving 
results on the spin and parity hypotheses.

The test statistic $\tilde{q}$ used 
to distinguish between the two spin-parity hypotheses is based on a ratio of profiled likelihoods~\cite{asimov,asimovErratum}:
\begin{equation}
\tilde{q} = \log \frac{\mathcal{L}(\spinSM ,  \hat{\hat{\mu}}_{\spinSM}, \hat{\hat{\theta}}_{\spinSM}) } {\mathcal{L}(\spinalt ,  \hat{\hat{\mu}}_{\spinalt}, \hat{\hat{\theta}}_{\spinalt} )}\;,
\label{eq:lambda}
\end{equation}

\noindent where $\mathcal{L}(\spin,  \hat{\hat{\mu}}_{\spin}, \hat{\hat{\theta}}_{\spin})$ is the maximum-likelihood estimator, evaluated under 
either the SM $\spinSM = 0^{+}$ or the alternative \spinalt\ spin-parity hypothesis. 
The parameters $\hat{\hat{\mu}}_{\spin}$ and $\hat{\hat{\theta}}_{\spin}$ represent the values of the 
signal strength and nuisance parameters fitted to the data under each spin and parity hypothesis.
The distributions of the test statistic for both 
hypotheses are obtained using ensemble tests of MC pseudo-experiments.
For each hypothesis test, about $70\,000$ pseudo-experiments were generated. 
The generation of the pseudo-experiments uses the numbers of signal and background events in each channel obtained 
from maximum-likelihood fits to data. In the fits of each pseudo-experiment, these and all other nuisance parameters are profiled, 
i.e. fitted to the value that maximises the likelihood for each value of the parameter of interest. 
When generating the distributions of the test statistic for a given spin-parity hypothesis, the expectation values of 
the signal strengths are fixed to those obtained in the fit to the data under the same spin-parity assumption.
The distributions of $\tilde{q}$ are used to determine the corresponding $p$-values $p(\spinSM)=p^{\rm SM}$ and $p(\spinalt)=p^{\rm alt}$. 
For a tested hypothesis \spinalt ,   the observed (expected) $p$-values are obtained by integrating the corresponding 
distributions of the test statistic above the observed value of $\tilde{q}$ (above the median of the \spinSM\ $\tilde{q}$ distribution).
When the measured data are in agreement with the tested hypothesis, 
the observed value of $\tilde{q}$ is distributed such that all $p$-values are equally probable.

Very small values of the integral of the distribution of the test statistic for the 
\spinalt\ hypothesis, corresponding to large values of $\tilde{q}$, are interpreted as the data 
being in disagreement with the tested hypothesis in favour of the SM hypothesis. 

The exclusion of the alternative \spinalt\ hypothesis in favour of the SM \spinSM\ hypothesis 
is evaluated in terms of the modified confidence level $\CLs(\spinalt)$, defined as~\cite{Read:2002hq}:

\begin{equation}
\CLs(\spinalt) = \frac{p (\spinalt)}{ 1- p (\spinSM)}\;.
\label{eq:cls}
\end{equation}
 
\subsection{Spin analysis in the \hgg\ channel }
\label{sec:fixed_gg}
The analysis in the \hgg\ channel is sensitive to a possible spin-2 state. 
Since the spin-2 models investigated in the present paper are different from those assumed in Ref.~\cite{HiggsSpin2013}, 
the analysis has been redesigned, to improve its sensitivity to the new models.

The selection of $H\to\gamma\gamma$ candidate events is based on the procedure of other recent ATLAS
\hgg\ analyses (see for example Ref.~\cite{ATLAScouplingsHgg}). 
Events are selected if they satisfy a diphoton trigger criterion requiring loose photon identification,
with transverse momentum \pT\ thresholds of 35~\GeV\ and 25~\GeV\ for the photon with the highest ($\gamma_1$) and second-highest  ($\gamma_2$) \pT , respectively. 
During the offline selection two photons are further required to be in a fiducial pseudorapidity region,
defined by $|\eta^\gamma|<2.37$, where the barrel/end-cap transition region $1.37 < |\eta^\gamma| < 1.56$ is excluded. 
The transverse momentum of the photons must satisfy $\pT^{\gamma_1}>0.35\cdot\mgg$ and $\pT^{\gamma_2}>0.25\cdot\mgg$, and only events with a diphoton 
invariant mass $\mgg$ between 105~\GeV\ and 160~\GeV\ are retained.
For the events passing this selection, a further requirement is applied on the diphoton transverse momentum, $\pTgg<300~\GeV$, 
motivated by the assumed validity limit of the spin-2 EFT model, as explained in Section~\ref{sec:theory}.
After this selection, $17\,220$ events are left at a centre-of-mass energy $\sqrt{s}=7~\TeV$ and $94\,540$ events at $\sqrt{s}=8~\TeV$. 

Kinematic variables sensitive to the spin of the resonance are the diphoton transverse momentum \pTgg\ and the production angle of the two photons, 
measured in the Collins--Soper frame~\cite{CollinsSoper}:
\begin{equation}
  \costs\ = \frac{|\sinh \left( \Delta\eta^{\gamma\gamma} \right) |}{\sqrt{1+ \left( \pT^{\gamma\gamma}/\mgg \right)^2}}\frac{2\pT^{\gamma_1}\pT^{\gamma_2}}{\mgg^2}\; ,
  \label{eq:hgg_costs}
\end{equation}
where $\Delta\eta^{\gamma\gamma}$ is the separation in pseudorapidity of
the two photons. 

\begin{figure}[htbp]
  \centering
  \subfloat[~] { \includegraphics[width=0.48\columnwidth]{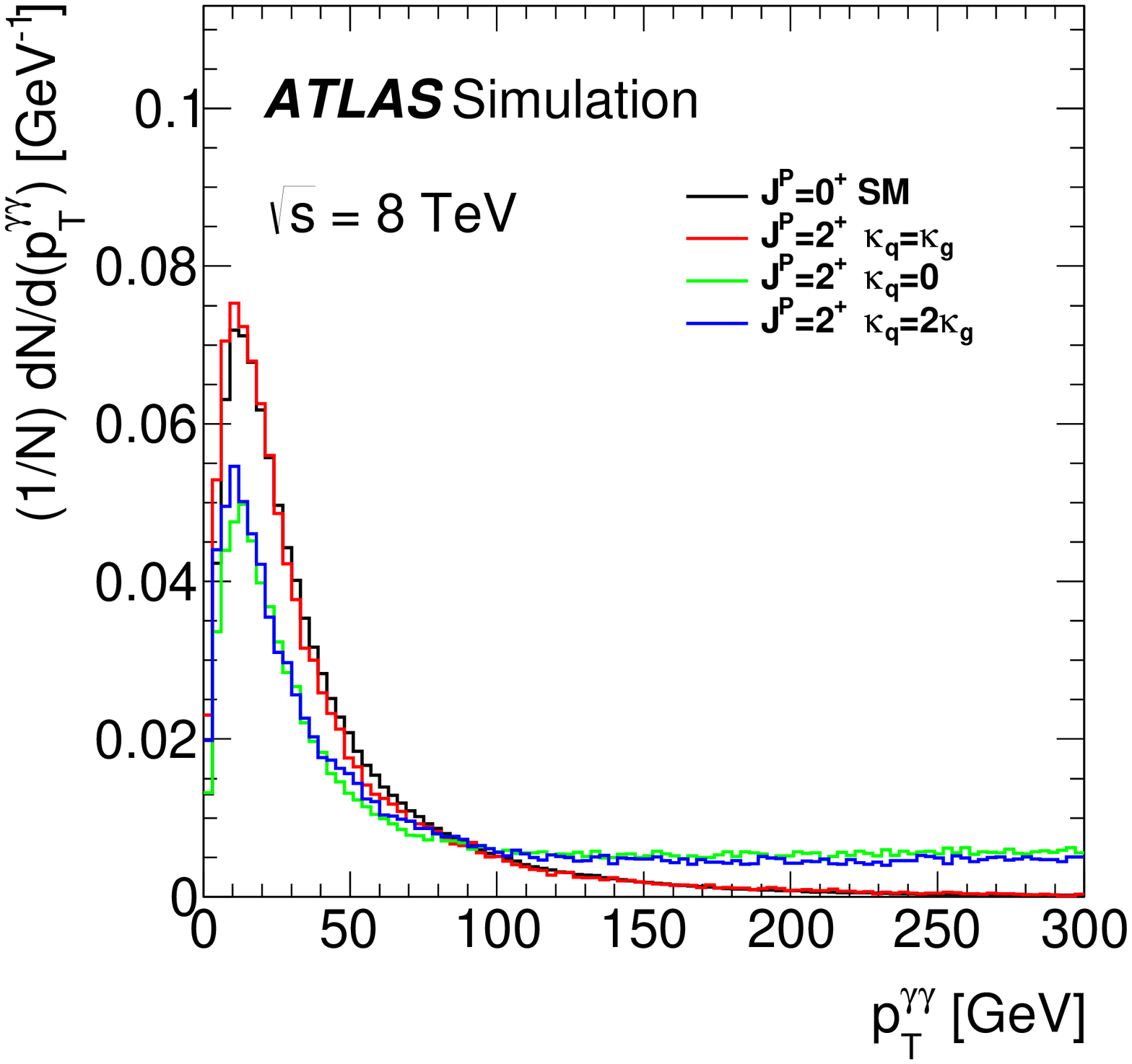} }
  \subfloat[~] { \includegraphics[width=0.48\columnwidth]{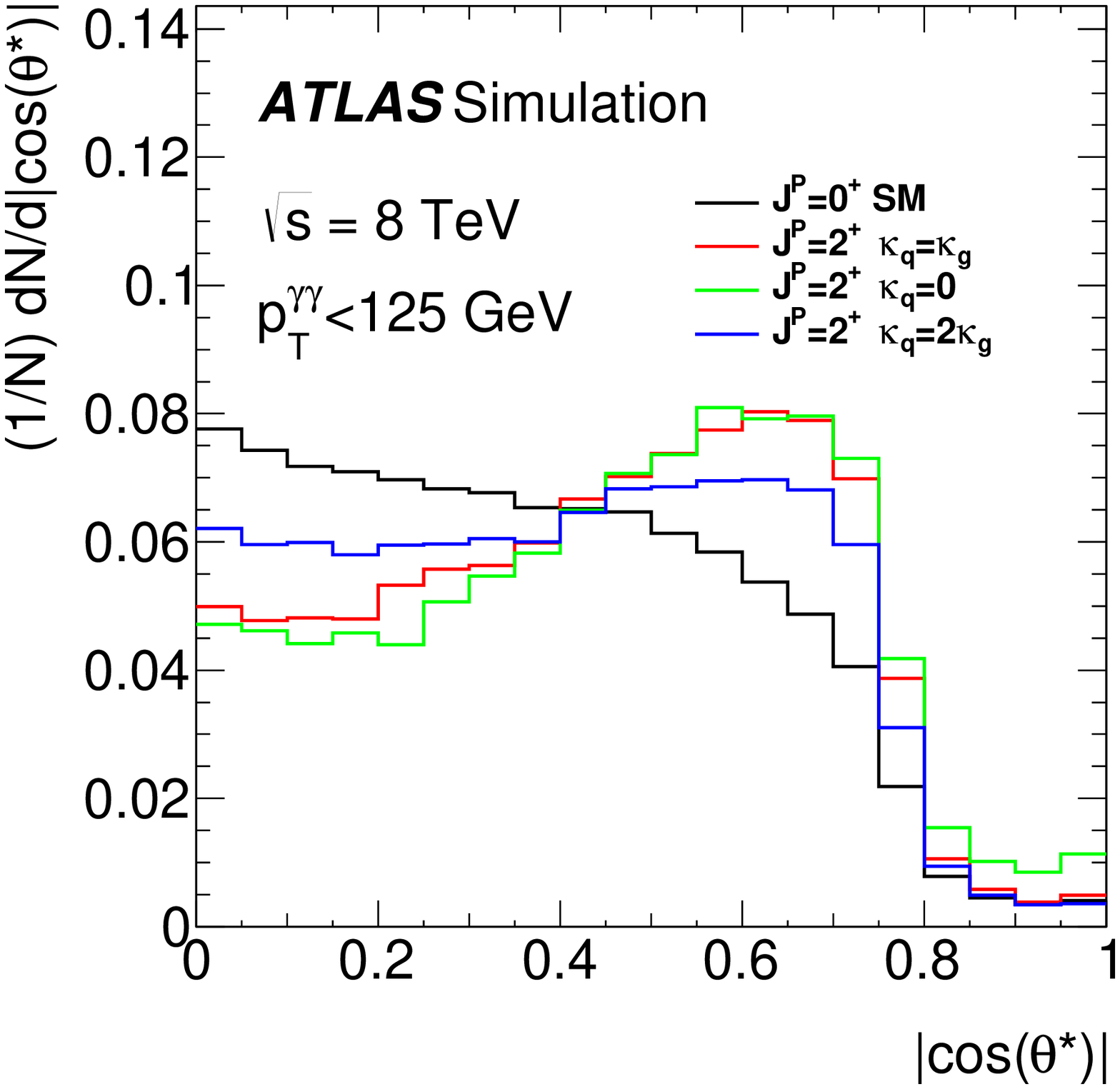} }
  \caption{
Expected distributions of kinematic variables sensitive  to the spin of the resonance considered in the \hgg\ analysis, 
(a) transverse momentum of the $\gamma \gamma$ system  \pTgg\  and (b) the production angle of the two photons in the Collins-Soper frame \costs ,
for a SM Higgs boson and for  spin-2 particles with three different choices of the QCD couplings. }
  \label{fig:PT_costheta}
\end{figure}
The predicted distributions of these variables, for events passing the selection, are shown in Figure~\ref{fig:PT_costheta}, for a SM Higgs boson 
and for a spin-2 particle with different QCD couplings. 
For the $\kappa_q\neq\kappa_g$ cases, the enhanced high-\pTgg\ tail offers the best discrimination, whereas for $\kappa_q=\kappa_g$ the most sensitive variable is \costs.

To exploit the signal distribution in both \pTgg\ and \costs, the selected events are divided into $11$ 
mutually exclusive categories: $10$ categories (labelled from C1 to C10)
collect events with $\pT^{\gamma\gamma}<125~\GeV$,  divided into 10 bins of equal size in \costs, 
while the $11^{\rm th}$ category (labelled C11) groups all events with $\pT^{\gamma\gamma}\geq125~\GeV$.
As described in Section~\ref{sec:theory}, for the non-UC spin-2 models the analysis is performed with two $\pTgg$ selections, 
namely $\pTgg<300~\GeV$ and $\pTgg<125~\GeV$: the 
latter case corresponds to not using the  $11^{\rm th}$ category.

The number of signal events above the continuum background can be estimated through a fit to the observed \mgg\ distribution in each category.
The \mgg\ distribution is modelled in each category as the sum of one-dimensional probability density 
functions (pdf) for signal and background distributions:
\begin{equation}
  f^{[c]}(\mgg|J) = \frac{n_B^{[c]}f_B^{[c]}(\mgg)+(n_J^{[c]}+n_{\rm bias}^{[c]})f_S^{[c]}(\mgg)}{n_B^{[c]}+n_J^{[c]}+n_{\rm bias}^{[c]}},
  \label{eq:mgg_pdf_cat}
\end{equation}
where $J$ is the spin hypothesis,
$n_B^{[c]}$ and $n_J^{[c]}$ are the background and the signal yield in category $c$, 
and $f_B^{[c]}(\mgg),~f_S^{[c]}(\mgg)$ are the \mgg\ pdfs for the background and the signal, respectively. 
The signal pdf $f_S^{[c]}(\mgg)$ is modelled as a weighted sum of a 
Crystal Ball function, describing the core and the lower mass tail, and of a Gaussian component 
that improves the description of the tail for higher mass values. 
For each category, $f_S^{[c]}(\mgg)$ is fitted to the simulated \mgg\ distribution of the SM Higgs boson and verified to be consistent also with the spin-2 models. 
The background pdf $f_B^{[c]}(\mgg)$ is empirically modelled as an exponential of a first- or second-degree polynomial. 
The choice of such a parameterisation can induce a bias (``spurious signal'') in the fitted signal yield, which is accounted for by the term $n_{\rm bias}^{[c]}$.
The size of the expected bias is determined as 
described in Refs.~\cite{Aad:2014lwa,ATLAScouplingsHgg},
and ranges between 0.6 and 4 events, depending on the category (with the signal ranging from 15 to more 
than 100 events). In the statistical analysis, $n_{\rm bias}^{[c]}$ is constrained for each category by 
multiplying the likelihood function by a Gaussian function centred at zero and with a width determined by the size of the expected bias.

\begin{figure}[htbp]
  \centering
  \subfloat[~] { \includegraphics[width=0.48\columnwidth]{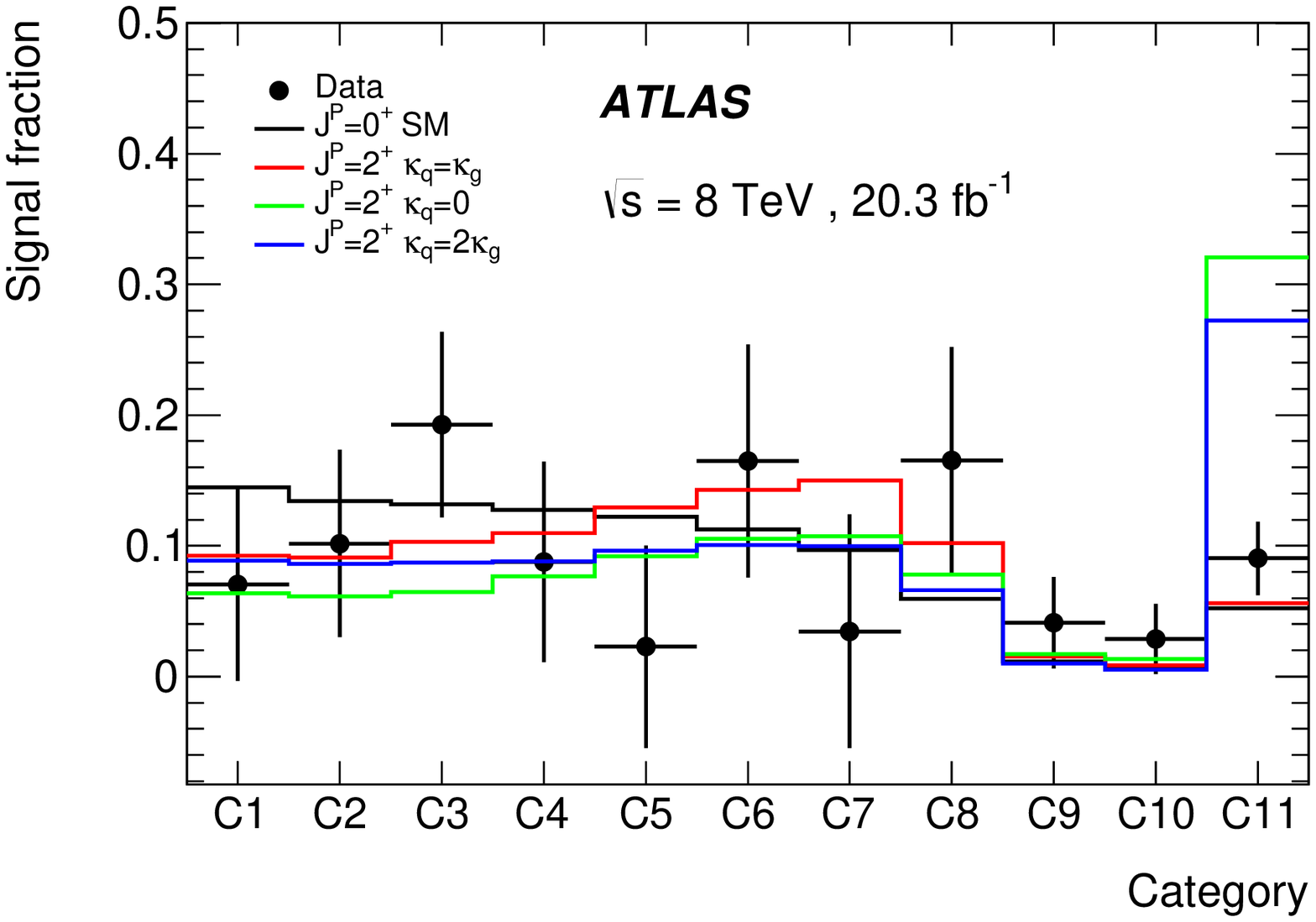} }
  \subfloat[~] { \includegraphics[width=0.48\columnwidth]{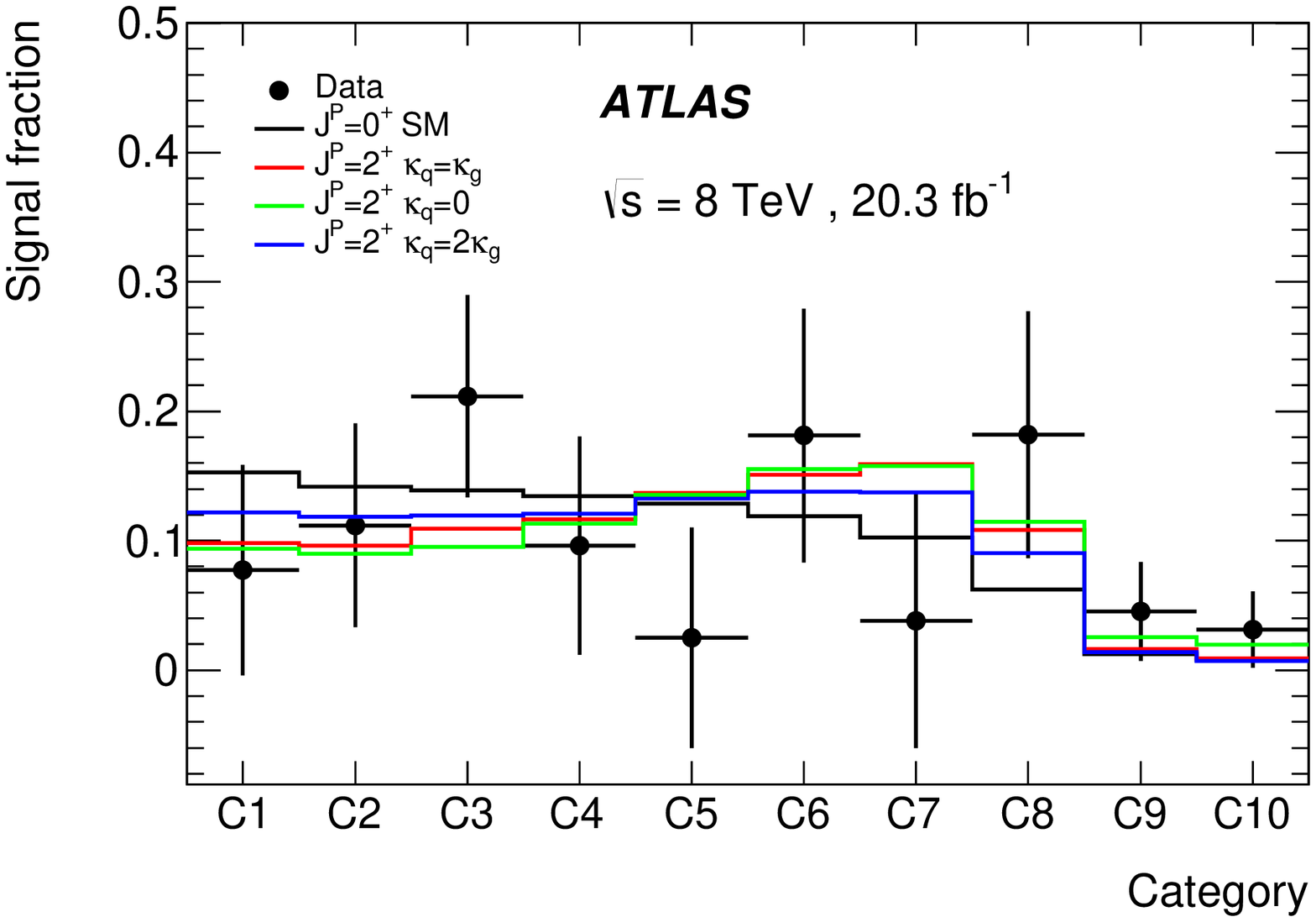} }
  \caption{
    Observed signal fraction per category for the \hgg\ analysis, and comparison to expected values for a SM Higgs boson and for a spin-2 particle with different choices of QCD couplings. 
  (a) the 11 categories described in the text are displayed, corresponding to the $\pTgg<300~\GeV$ selection;
  (b) the high-\pTgg\ category is discarded and the signal fractions are renormalised over the 10 remaining categories, corresponding 
to the $\pTgg<125~\GeV$ selection. 
  }
  \label{fig:fit_cat11_indep}
\end{figure}

Defining $n_S$ as the total signal yield (summed over all categories), 
the expected fraction of signal events belonging to each category, 
$\displaystyle\Phi_J^{[c]}\equiv\frac{n_J^{[c]}}{n_S}$, depends on the spin hypothesis $J$. The values of $\Phi_J^{[c]}$ extracted from the data can be compared to their 
expected values for each spin hypothesis, as shown in Figure~\ref{fig:fit_cat11_indep} for the data collected at $\sqrt{s}=8~\TeV$. 

For the non-UC scenario the  $11^{\rm th}$ (high-\pTgg) category provides strong discrimination power against the non-SM hypothesis, 
as visible in Figure~\ref{fig:fit_cat11_indep}(a). 

To discriminate between the SM spin-0 ($\spinSM = 0^{+}$) and alternative spin-2 hypotheses (\spinalt), 
two likelihood functions ${\cal L}_{\spinSM},~{\cal L}_{\spinalt}$ are built, following 
the general approach described in Eq.~(\ref{eq:likelihood}):
\begin{equation}
   -\ln {\cal L}_J = \sum_{c} \left\{
    \left( n_B^{[c]}+n_S\Phi_J^{[c]}+n_{\rm bias}^{[c]} \right) -
    \sum_{e\in[c]} \ln~\left[ n_B^{[c]}f_B^{[c]}(\mgg^{(e)})+(n_S\Phi_J^{[c]}+n_{\rm bias}^{[c]})f_S^{[c]}(\mgg^{(e)}) \right] 
  \right\}
  \label{eq:LH_comb_nosyst}
\end{equation}
where $\sum_{c}$ runs over all categories and $\sum_{e\in[c]}$ runs over all events in category $c$.
The total signal yield $n_S$ is a free parameter in the likelihood model.
The spin hypothesis being tested enters the likelihood function through the fractions of signal per category, $\Phi_J^{[c]}$. 

Several systematic uncertainties enter this model. They are implemented for each spin hypothesis as 
nuisance parameters, $\theta_J$, constrained by multiplicative Gaussian terms in the likelihood 
function (not included in Eq.~(\ref{eq:LH_comb_nosyst}) for simplicity). 

The signal fractions, $\Phi_J^{[c]}$, for the SM Higgs boson are affected by uncertainties on the $\pT$ spectrum of the resonance and on the size of 
the interference between the resonance and continuum production.
The former is computed as described in Ref.~\cite{ATLAScouplingsHgg}. 
The relative impact on the signal fractions is 
less than $\pm 1\%$ for categories 1 to 8 ($\pTgg<125~\GeV$ and $\costs<0.8$), and becomes as 
large as $\pm 13\%$ for categories 10 and 11.
The correction for the interference is evaluated according to Refs.~\cite{LDixon_interference_2003,LDixon_interference_2013}.
The systematic uncertainty is conservatively assumed to equal the correction itself, and its relative impact ranges between $\pm0.1\%$ and $\pm1.8\%$.

No systematic uncertainty is assigned to the simulated  $\pT^{X} $ distribution of the spin-2 models. 
The effect of the interference between the resonance and continuum production is essentially not known, 
as it depends on the width, $\Gamma_X$, of the resonance, which is unknown. 
The results presented here only hold under the assumption of a narrow width for the resonance, such that interference effects can be neglected.

Additional systematic uncertainties come from the calibration of the photon energy scale and energy resolution and affect 
the signal parameterisation $f_S^{[c]}$. 
These uncertainties are evaluated as described in Ref.~\cite{MassPaper}.

\subsection{Spin and parity analysis in the \hWW\ channel}
\label{sec:fixed_ww}
The analysis of the spin and parity in the \hWW\ channel is described in detail in a separate publication~\cite{spincp_ww_paper}.
In the following a brief summary is provided. The selection is restricted to events containing two charged leptons of different 
flavour (one electron and one muon).  The $e \nu \mu \nu$ channel is the most sensitive one ~\cite{ATLAS:2014aga}. 
The same-flavour channels  ($e \nu e \nu$ and $\mu \nu \mu \nu$) are not expected to add much in terms of sensitivity due to 
the presence of large backgrounds that cannot be removed without 
greatly reducing the acceptance of the alternative models considered in this analysis. 
The leading lepton is required to have $\pT>22$~\GeV\ and to match the object reconstructed by the 
trigger, while the sub-leading lepton needs to have $\pT>15$~\GeV. 
While the  spin-0 analyses select only events with no jets in the final state (no observed jets with $\pt >25$~\GeV\ 
within $|\eta| < 2.5$ or with $\pt >30$~\GeV\  within ${2.5 < |\eta| < 4.5}$), the spin-2 analysis enlarges the acceptance 
by allowing for zero or one jet (selected according to the above mentioned criteria).

The major sources of background after the dilepton selection are $Z/\gamma^{*}$+jets (Drell--Yan) events, diboson ($WW$, $WZ/ \gamma^*$, $ZZ/ \gamma^{*}$), 
top-quark ($t\bar{t}$ and single
top) production, and $W$ bosons produced in association with hadronic jets ($W$+jets), where a jet is misidentified as a lepton.
The contribution from misidentified leptons is significantly reduced by the requirement of two high-$\pT$ isolated leptons.
Drell--Yan events are suppressed through requirements on some of the dilepton variables\footnote{Throughout this section, the following 
variables are used: \ptll\ and \mll\  are the transverse momentum and the invariant mass of the two-lepton system, respectively, 
\dphill\ is the azimuthal angular difference between the two leptons, \mT\ is the transverse mass of  the reconstructed 
Higgs boson decay system, \dpt\ is the absolute value of the difference between the momenta of the two leptons 
and $ E_{\ell \ell \nu \nu} =p_{\text T}^{{\ell}_1} -0.5p_{\text T}^{{\ell}_2}+0.5 p_{\text T}^{miss}$, where $p_{\text T}^{miss}$ is the missing transverse momentum.} 
(\ptll\ $> 20$~\GeV, \dphill\ $< 2.8$), while a cut on \mll\ ($\mll < 80$~\GeV) targets the $WW$ background. For alternative spin models with non-universal couplings, as discussed in 
Section~\ref{sec:theory},
an additional upper bound is imposed on the Higgs boson $\pT$, reconstructed as the transverse component of 
the vector sum of the momenta of the two charged leptons and the missing transverse momentum.
Additionally, for events containing one jet, which include substantial top-quark and $W$+jets backgrounds, $b$-jet and \Ztautau\ vetoes are applied, 
together with transverse mass requirements: 
the larger of the transverse masses of the two $W$ bosons (each computed using the corresponding lepton and the missing transverse momentum) in the 
event is required to be larger than 50~\GeV, 
while the total transverse mass of the $WW$ system (defined with the two leptons and the missing transverse momentum) is required to be below 150~\GeV. 

Control regions (CRs) are defined for the $WW$, top-quark and Drell--Yan backgrounds, which are the most important ones after the topological selection 
described above. The CRs are used to normalise the background event yields with a fit to the rates observed in data. 
The simulation is then used to transfer these normalisations to the signal region (SR). 
The \Wjets\ background is estimated entirely from data, while non-$WW$ diboson backgrounds are estimated 
using MC simulation and cross-checked in a validation region.

After the signal region selection, 4730 and 1569 candidate events are found in data in the 0-jet and 1-jet categories, respectively. For the latter category, 
the number decreases to 1567  and 1511 events when applying a selection on the Higgs boson $\pT$ of less than $300$~\GeV\ and less than $125$~\GeV, respectively. 
In total 218 (77) events are expected from a SM Higgs boson signal in the 0-jet (1-jet) category, while about 4390 (1413) events are expected for the total background.
 
A BDT algorithm is used in both the fixed spin hypothesis tests and the tensor structure analyses. 
For spin-2 studies, the strategy follows the one adopted in Ref.~\cite{HiggsSpin2013}, 
with the main difference being that the 1-jet channel has been added. Two BDT discriminants
are trained to distinguish between the SM hypothesis and the background (BDT$_0$), and the alternative spin hypothesis and the background (BDT$_2$). 
Both BDTs employ the same variables, namely \mll, \ptll, \dphill\ and \mT, which provide the best discrimination between signal hypotheses and backgrounds, 
also in the presence of one jet in the final state. 
All background components are used in the trainings. In total, five BDT$_2$ trainings are performed for the alternative spin hypotheses 
(one for the spin-2 UC scenario and two for each of the two spin-2 non-UC hypotheses corresponding to the different $\pT^{X}$ selections), 
plus one training of BDT$_0$ for the SM Higgs boson hypothesis. 
 
For the spin-0 fixed hypothesis test and $HWW$ tensor structure studies, the first discriminant, BDT$_0$, is the same as the one used for 
the spin-2 analysis, trained to disentangle the SM hypothesis from the background.
A second BDT discriminant, \bdtcp, 
is obtained by training the SM signal versus the alternative signal sample (the pure CP-even or CP-odd BSM hypotheses), 
and then applied to all CP-mixing fractions. No background component is involved in this case. 
The variables used for the \bdtcp\ trainings are \mll, \dphill, \ptll\, and the missing transverse momentum for the CP-even 
analysis and \mll, \dphill, \Efun\, and \dpt\ for the CP-odd analysis.
The training strategy is different from the one used in the spin-2 analysis because,
while the spin-2 signal is very similar to the background, 
the spin-0 signals are all similar to each other, while being different from the main background components. 
Therefore, in the latter case, training the signal hypotheses against each other improves the sensitivity. 
The resulting BDT variable is afterwards used in binned likelihood fits to test the data for compatibility 
with the presence of a SM or BSM Higgs boson.

Several sources of systematic uncertainty are considered, both from experimental and theoretical sources, and are described in detail in Ref.~\cite{spincp_ww_paper}.
The correlations induced among the different background sources by the presence of other processes in the 
control regions are fully taken into account in the statistical procedure. 
The most important systematic uncertainties are found to be those related to the modelling of the $WW$ background, 
to the estimate of the $W$+jets background (originating from the data-driven method employed) and, for the spin-2 results in particular, 
to the $Z\rightarrow \tau\tau$ modelling.

\subsection{Spin and parity analysis in the \hZZ\ channel }
\label{sec:fixed_zz}
The reconstruction of physics objects and event selection used for
the \hZZ\ analysis is identical to the one presented in Ref.~\cite{MassPaper}.
The main improvement with respect to the previous ATLAS publication of Ref.~\cite{HiggsSpin2013} is the
introduction of a BDT discriminant designed to optimise the separation between the signal and the 
most relevant background process.

Events containing four reconstructed leptons (electrons or muons) in the final state are selected
using single-lepton and dilepton triggers. The selected events are classified according to
their final state: $4\mu,~2e2\mu,~2\mu2e$ and $4e$,
where for the decay modes $2e2\mu$ and 
$2\mu2e$ the first pair is defined to be the one with the dilepton mass closest to the 
$Z$ boson mass.
Each muon (electron) must satisfy $\pt>$~6~\GeV\ ($\pt>$~7~\GeV) and be measured in the 
pseudorapidity range $|\eta|<$~2.7 ($|\eta|<$~2.47). 
Higgs boson candidates are formed by selecting two same-flavour, opposite-charge lepton pairs 
in an event. The lepton with the highest $\pt$ in the quadruplet must have $\pt>20$~\GeV, and the leptons 
with the second- and third-highest $\pt$ must have $\pt>15$~\GeV\ and $\pt>$10~\GeV, respectively.
The lepton pair with the mass closest to the $Z$ boson mass 
is referred to as the leading lepton pair and its invariant mass as $m_{12}$. The 
requirement $50~\GeV < m_{12} < 106~\GeV$ is applied. The other lepton pair is 
chosen from the remaining leptons as the pair closest in mass to the $Z$ boson.
Its mass, denoted hereafter by $m_{34}$, must satisfy $12~\GeV < m_{34} < 115 ~\GeV$.
Further requirements are made on the impact parameters of the leptons relative to the interaction vertex and 
their isolation in both the tracker and calorimeter.

The main background process affecting the selection of \hZZ\ events is the non-resonant production
of $ZZ^*$ pairs. This background has the same final state as the signal events and 
hereafter is referred to as the irreducible background. It is estimated from 
simulation and normalised to the expected SM cross section calculated at NLO~\cite{Melia:2011tj,gg2ZZ}. 
The reducible sources of background come from $Z$+jets and $t\bar{t}$ processes, where 
additional leptons arise due to misidentified jets or heavy-flavour decays.  
The rate and composition of the reducible backgrounds are evaluated using data-driven techniques, separately
for the two final states with sub-leading muons $ \ell \ell +\mu\mu$ and those with 
sub-leading electrons $\ell \ell +ee$.
 
Only events with an invariant mass of the four-lepton system,
denoted by $m_{4\ell}$, satisfying the signal region definition 115~\GeV\ $<m_{4\ell}<$130~\GeV\ are selected. 
The expected signal and background yields in the signal region and the observed events in 
data are reported in Table \ref{tab::events_zz}. 
\begin{table}[ht]
\begin{center}
      \begin{tabular}{@{\hskip 0.5cm}l@{\hskip 0.5cm}c@{\hskip 0.5cm}c@{\hskip 0.5cm}c@{\hskip 0.5cm}c@{\hskip 0.5cm}c}
      \hline\hline
            & SM Signal          &$ZZ^*$          & $t\bar{t}, Z+{\rm jets}$ & Total expected & Observed\\
                 \hline
                 \multicolumn{5}{c}{$\sqrt{s} =7$~\TeV}\\
                 \hline
                $4\mu$     &1.02$\pm$0.10   &0.65$\pm$0.03   &0.14$\pm$0.06    &      1.81$\pm$0.12   &3\\
                $2\mu 2e$  &0.47$\pm$0.05   &0.29$\pm$0.02   &0.53$\pm$0.12     &      1.29$\pm$0.13   &1\\
                $2e 2\mu$  &0.64$\pm$0.06   &0.45$\pm$0.02   &0.13$\pm$0.05     &      1.22$\pm$0.08   &2\\
                $4e$       &0.45$\pm$0.04   &0.26$\pm$0.02   &0.59$\pm$0.12     &      1.30$\pm$0.13   &2\\
                Total      &2.58$\pm$0.25   &1.65$\pm$0.09   &1.39$\pm$0.26     &      5.62$\pm$0.37   &8\\
                \hline
                \multicolumn{5}{c}{$\sqrt{s} =8$~\TeV}\\
                \hline
                $4\mu$     &5.81$\pm$0.58    &3.36$\pm$0.17   &0.97$\pm$0.18    &    10.14$\pm$0.63     &13\\
                $2\mu 2e$  &3.00$\pm$0.30    &1.59$\pm$0.10   &0.52$\pm$0.12    &     5.11$\pm$0.34    &8\\
                $2e 2\mu$  &3.72$\pm$0.37    &2.33$\pm$0.11   &0.84$\pm$0.14    &     6.89$\pm$0.41    &9\\
                $4e$       &2.91$\pm$0.29    &1.44$\pm$0.09   &0.52$\pm$0.11    &     4.87$\pm$0.32    &7\\
                Total      &15.4 $\pm$1.5   &8.72$\pm$0.47   &2.85$\pm$0.39     &    27.0 $\pm$1.6    &37\\
                \hline\hline
                \end{tabular}
        \caption{\label{tab::events_zz} Expected signal, background and total yields, including their total uncertainties, and
        observed events in data, in the 115~\GeV\ $< m_{4\ell} <$ 130~\GeV\ signal region. The number of expected signal events is
        given for a SM Higgs boson mass of 125.5~\GeV .}
\end{center}
\end{table}

\begin{figure}[htbp]
 \centering
   \includegraphics[width=0.6\linewidth]{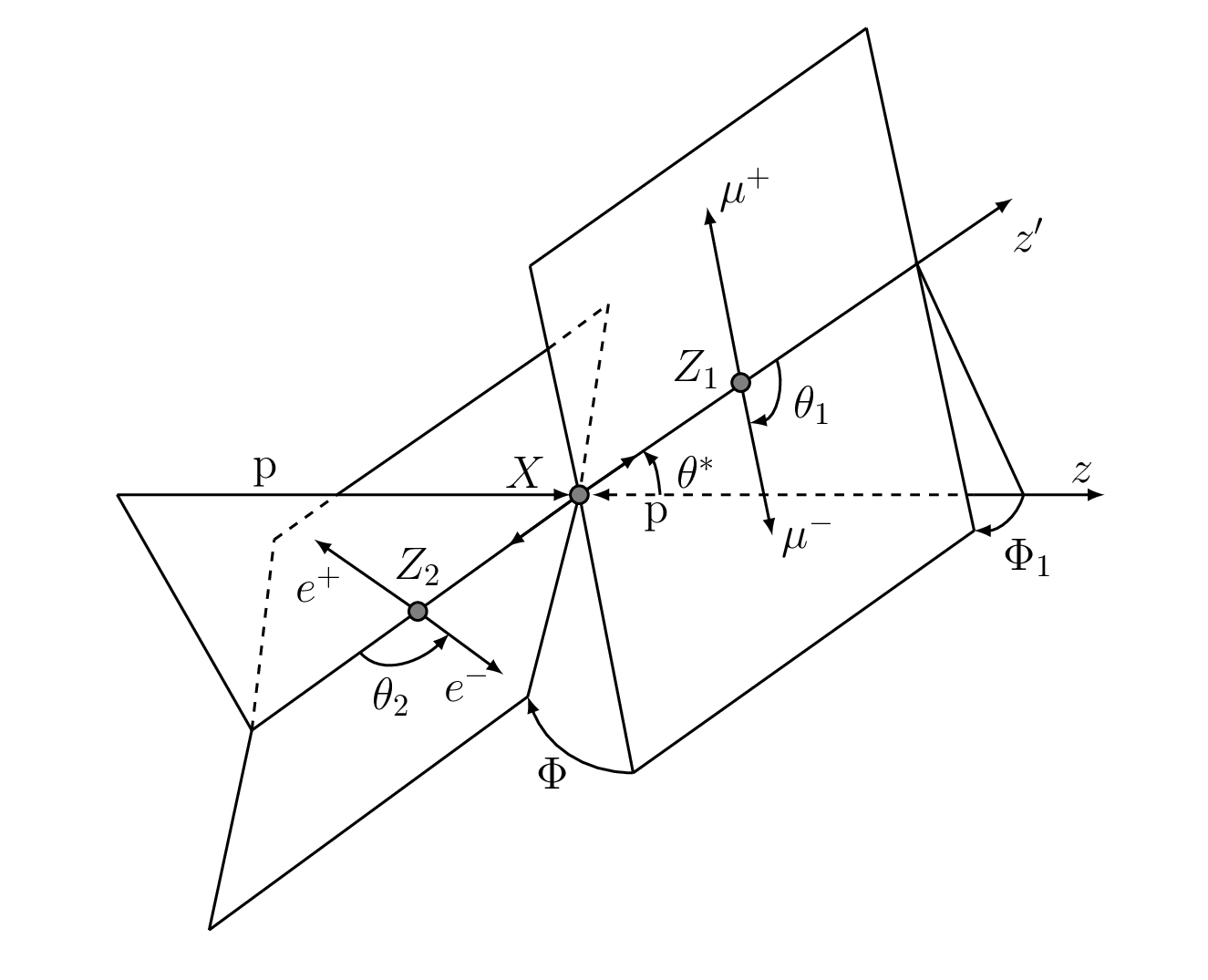}
   \caption{ \label{fig:angles}  Definitions of the angular observables
   sensitive to the spin and parity of the resonance in the \\  $X \to ZZ^{*}\to 4\ell$ decay. }
\end{figure}

The choice of production and decay angles used in this
analysis is presented in Figure~\ref{fig:angles}, where the following
definitions are used:
\begin{itemize}
\item{$\theta _1$ and $\theta _2$ are defined as the angles between
  final-state leptons with negative charge and the direction of flight of their
  respective $Z$ bosons, in the four-lepton rest frame;}
\item{$\Phi$ is the angle between the decay planes of two lepton pairs (matched to the two $Z$ boson decays) expressed in the four-lepton rest frame;}
\item{$\Phi _1$ is the angle between the decay plane of the
  leading lepton pair and a plane defined by the $Z_1$ momentum (the $Z$ boson associated with the leading lepton pair) 
  in the four-lepton rest frame and the positive direction of the
  collision axis;}
\item{$\theta ^{*}$ is the production angle of the $Z_1$ defined in the four-lepton rest frame.}
\end {itemize}

\begin{figure*}[htb]
\centering
\begin{minipage}[h]{0.31\linewidth}
\subfloat[~]{\includegraphics[width=1\linewidth]{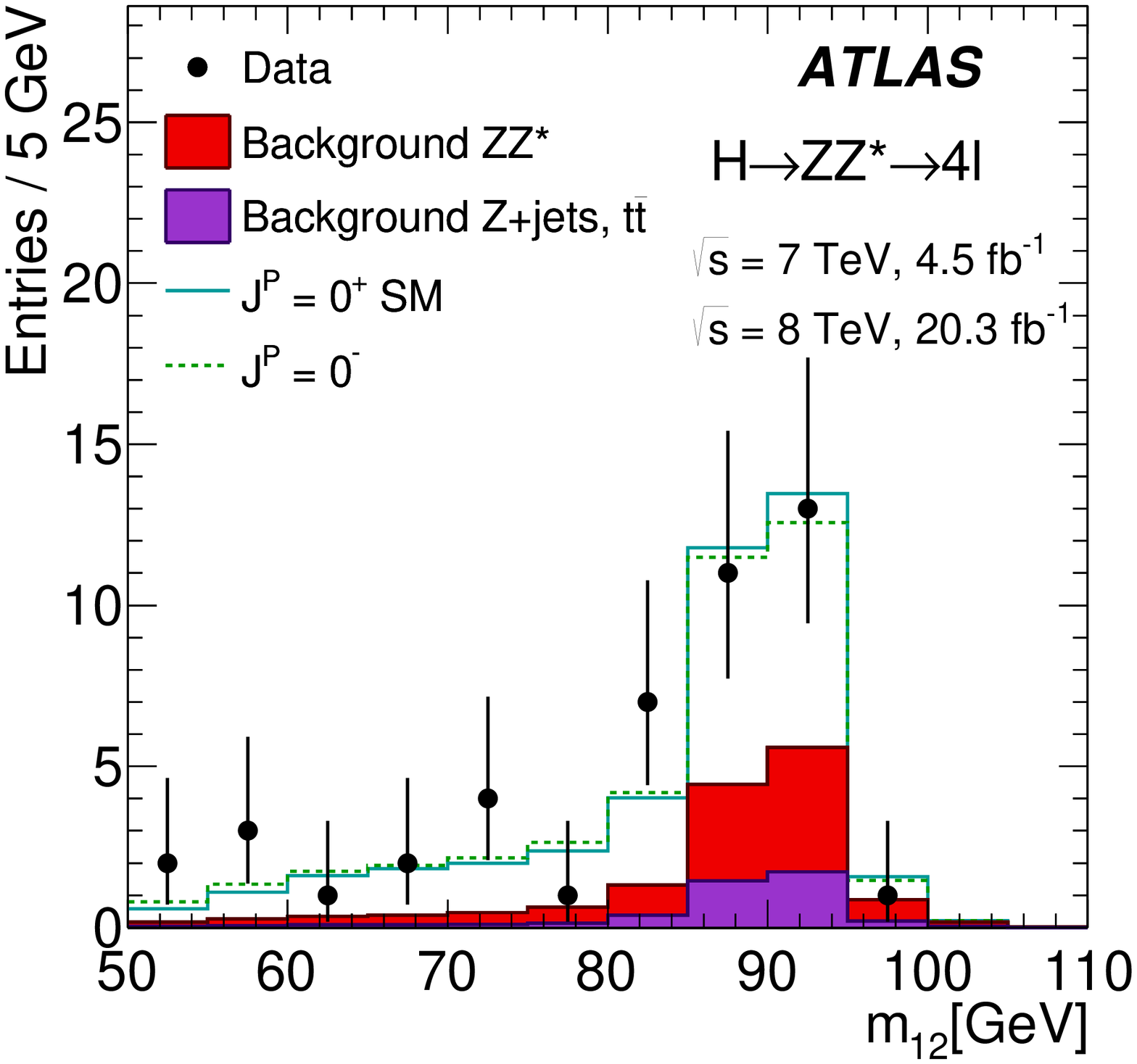}}
\end{minipage}
\hfill
\begin{minipage}[h]{0.31\linewidth}
\subfloat[~]{\includegraphics[width=1\linewidth]{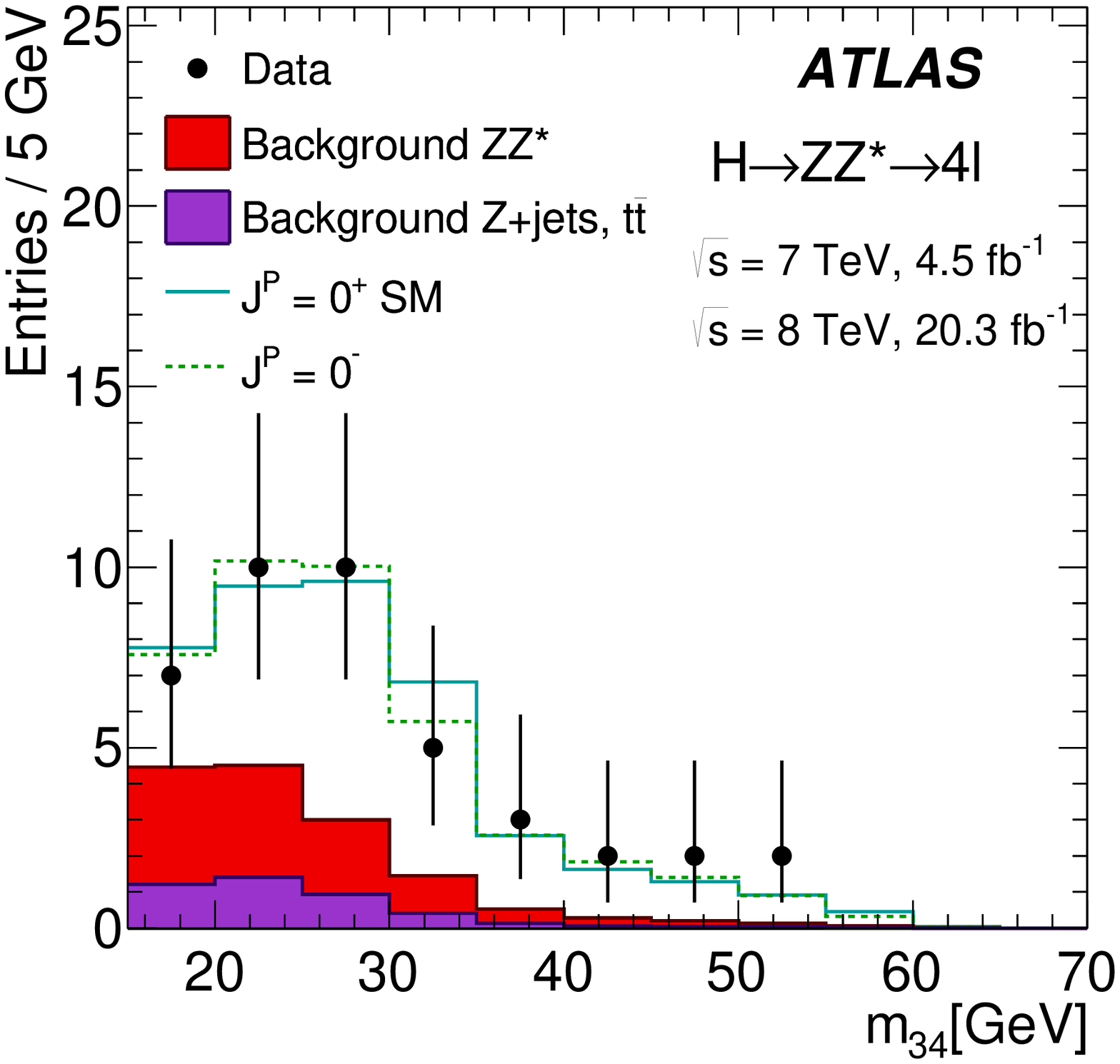}}
\end{minipage}
\hfill
\begin{minipage}[h]{0.31\linewidth}
\subfloat[~]{\includegraphics[width=1\linewidth]{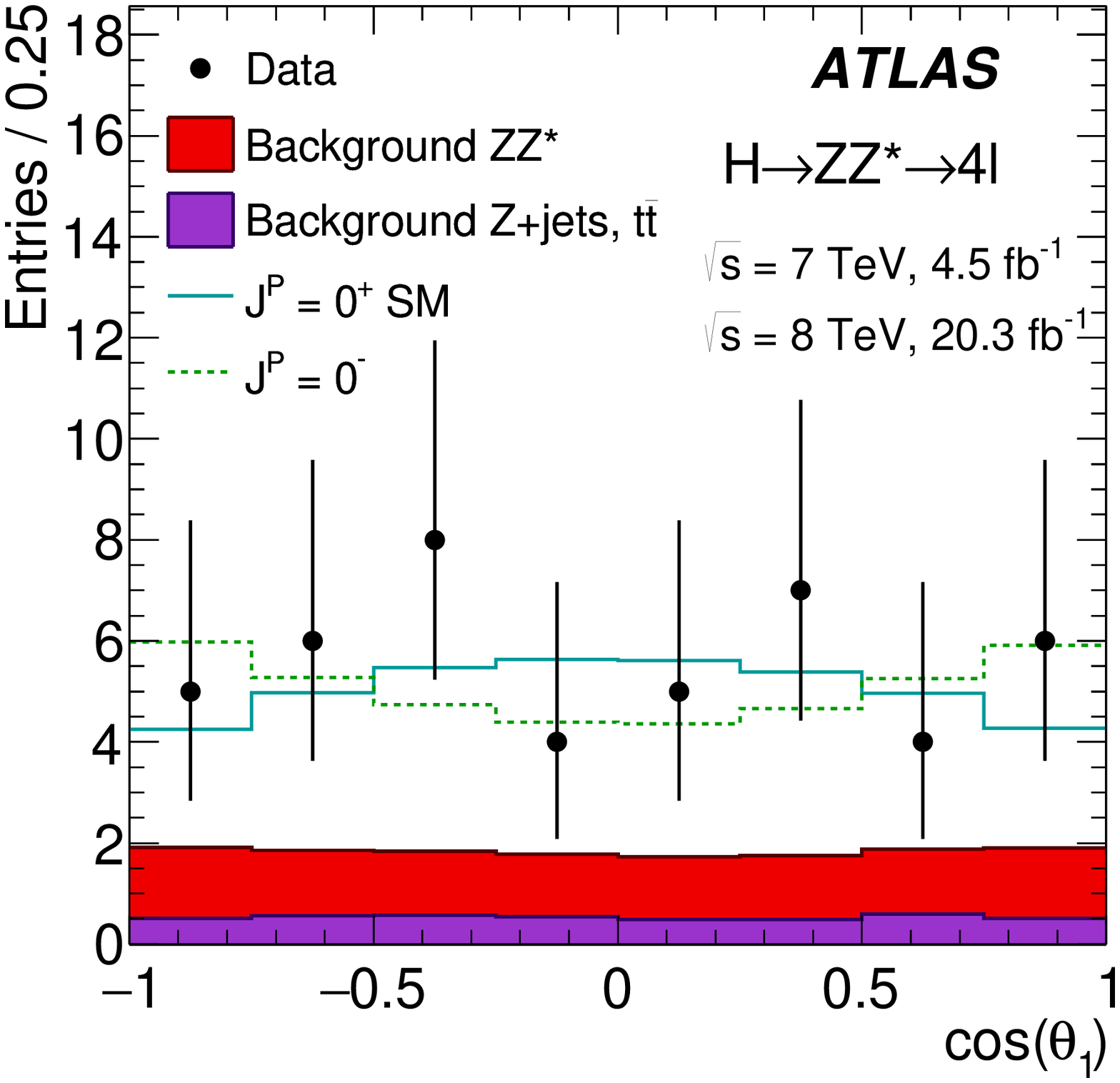}}
\end{minipage}
\vfill
\begin{minipage}[h]{0.31\linewidth}
\subfloat[~]{\includegraphics[width=1\linewidth]{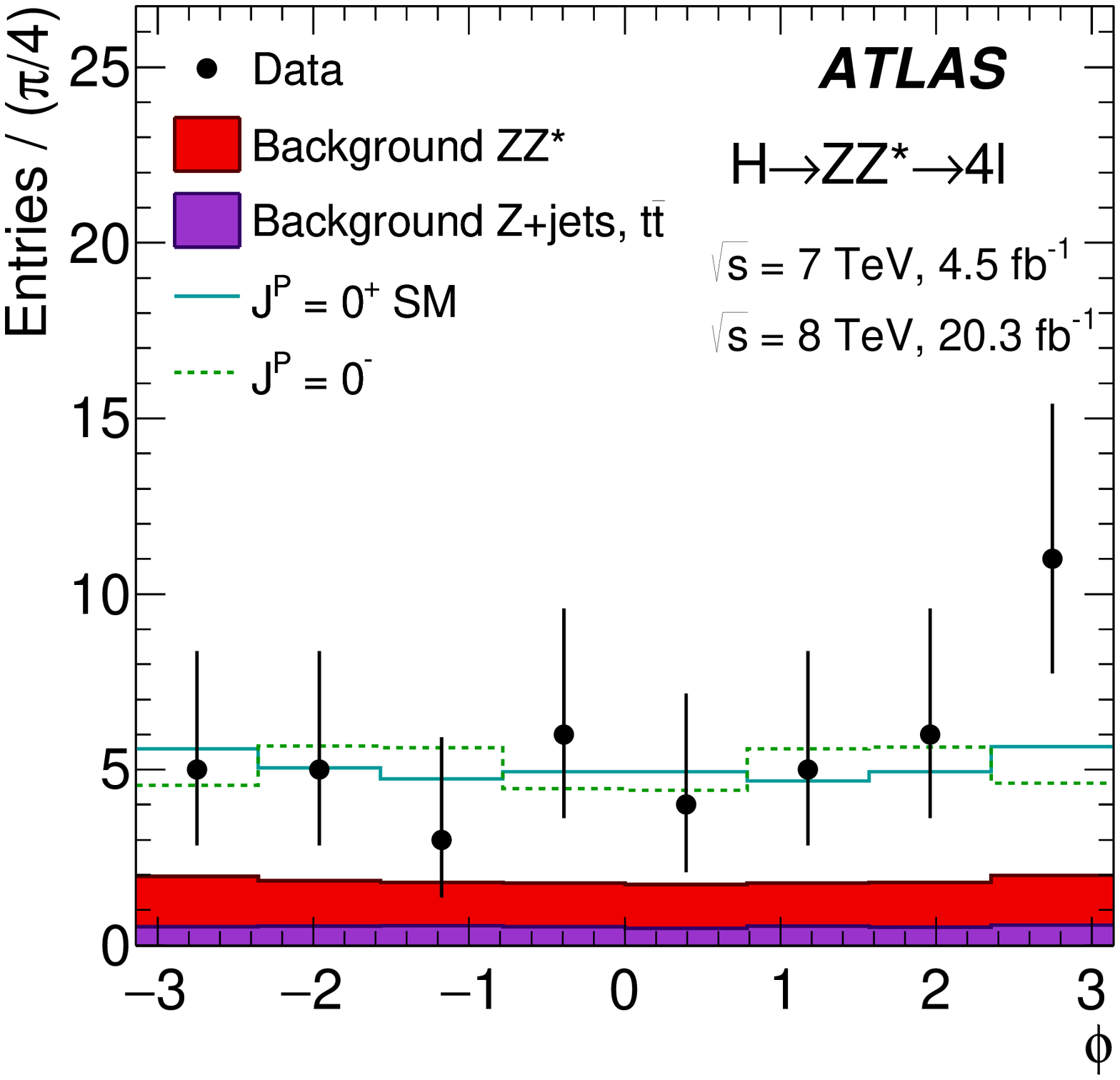}}
\end{minipage}
\hfill
\begin{minipage}[h]{0.31\linewidth}
\subfloat[~]{\includegraphics[width=1\linewidth]{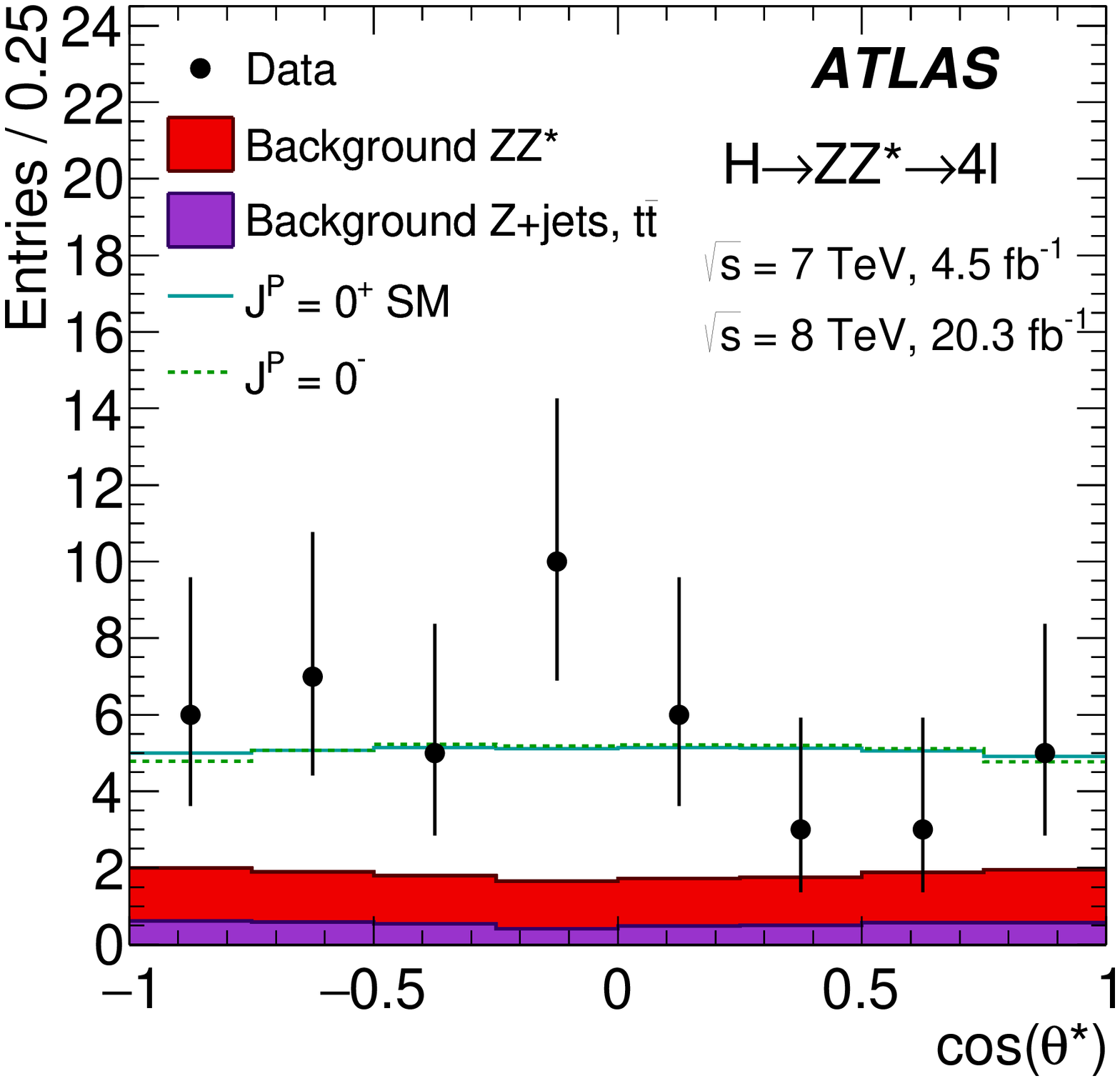}}
\end{minipage}
\hfill
\begin{minipage}[h]{0.31\linewidth}
\subfloat[~]{\includegraphics[width=1\linewidth]{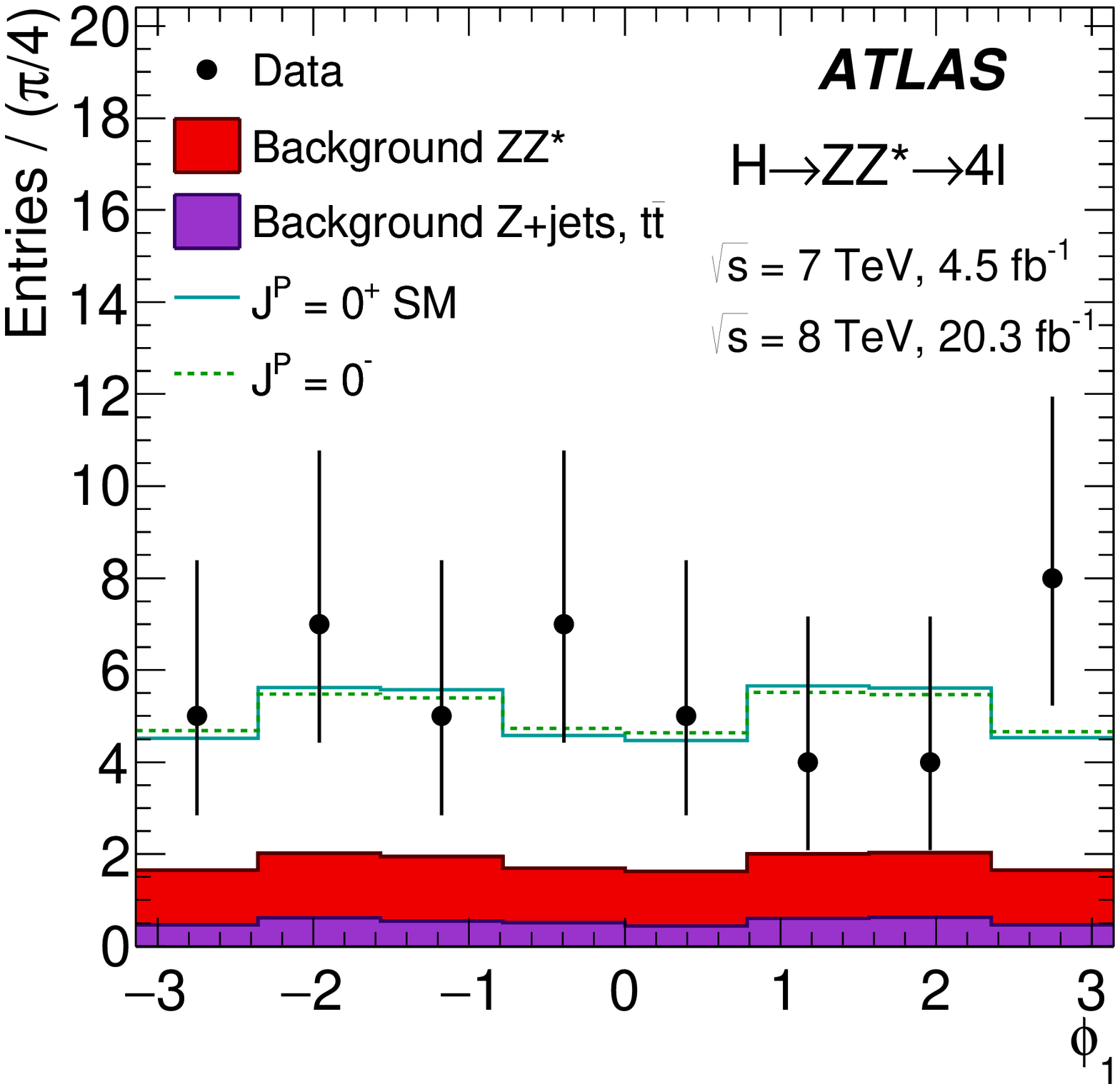}}
\end{minipage}
\caption{\label{fig:fs_1} Distributions of some of the final-state observables sensitive
to the spin and parity of the resonance in the $H\to ZZ^{*}\to 4\ell$ signal region $115~\GeV<m_{4\ell}< 130~\GeV$ for 
 data (points with errors),  backgrounds (filled histograms) and 
 predictions for two spin hypotheses (SM solid line and alternatives dashed lines).
 (a), (b) and (c):  invariant masses $m_{12}$ , $m_{34}$ and decay $\cos \theta _1$, respectively;  
 (d), (e) and (f): $\Phi$, $\cos \theta ^*$ \ and $\Phi _1$, respectively. }
\end{figure*}

The final-state observables sensitive to the spin and parity of a boson
decaying to $ZZ^{*}\to 4\ell$ are the two production angles $\theta ^{*}$
and $\Phi _1$ and the three decay angles $\Phi$, $\theta _1$ and
$\theta_2$. In the case of a spin-0 boson, the differential production cross section 
does not depend on the production variables $\cos(\theta ^{*})$ and $\Phi _1$.  
It should be noted that, as the Higgs boson mass is below
$2m_Z$, the shapes of the mass distributions of the intermediate $Z$
bosons, $m_{12}$ and $m_{34}$, are sensitive to the spin and parity of
the resonance.  In Figure~\ref{fig:fs_1} the distributions
of the final-state observables sensitive to the spin and parity of the 
decaying resonance are presented.  The distributions are shown for the
SM $J^P=0^+$ and $J^P=0^-$ simulated events, as well as for
$ZZ^{*}$ production and reducible backgrounds in the signal region
$115~\GeV<m_{4\ell}< 130~\GeV$. The events observed in data are
superimposed on each plot.

Two approaches were pursued to develop the discriminants used to
distinguish between different spin and parity hypotheses. The first
uses the theoretical differential decay rate for the
final-state observables sensitive to parity 
to construct a matrix-element-based likelihood
ratio analysis ($J^P$--MELA). 
The second approach is based on a BDT.

For the $J^P$--MELA approach~\cite{JHU1,YR3}, the probability of observing an event with given kinematics can be calculated. 
This probability is corrected for detector acceptance and analysis
selection, which are obtained from the simulated signal MC
samples. The full pdf also includes a term for incorrect pairing
of the leptons in the $4\mu$ and $4e$ channels. For a given pair of spin-parity 
hypotheses under test, the final discriminant is defined as the ratio of the pdf for a given hypothesis 
to the sum of the pdfs for both hypotheses.

For the BDT approach, a $J^P$ discriminant is formed for each pair of
spin-parity states to be tested, by training a BDT on the variables of
simulated signal events which fall in the signal mass window
$115~\GeV <m_{4\ell} <130~\GeV$. For the $0^+$ versus $0^-$
test, only the parity-sensitive observables $\Phi$, $\theta _1$,
$\theta _ 2$, $m_{12}$ and $m_{34}$ are used in the BDT training. For the
spin-2 test, the production angles $\theta ^*$ and $\Phi _1$ are also
included.

\begin{figure}[htb]
\centering
\center{\includegraphics[width=0.5\linewidth]{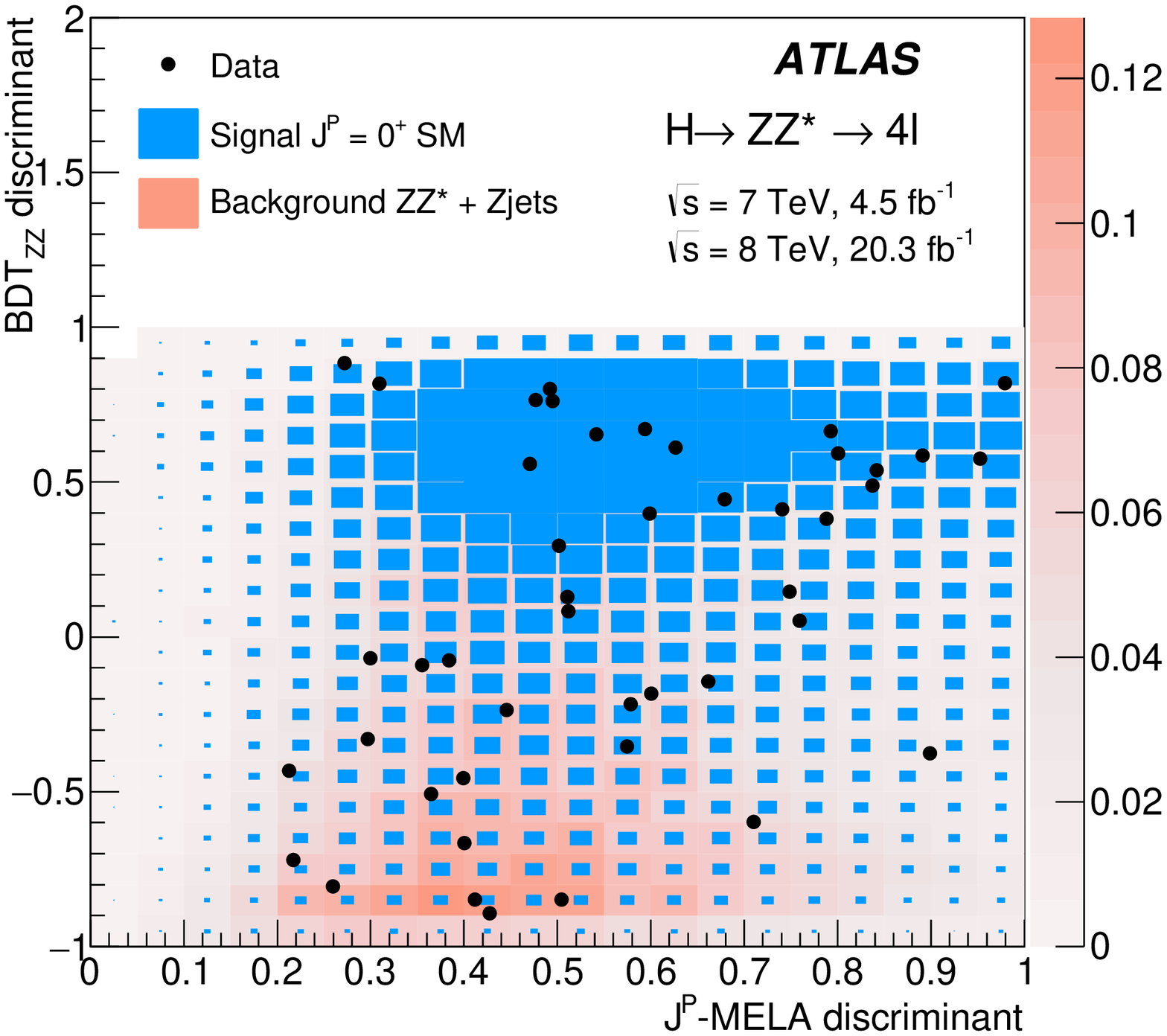}}
\caption{ \label{fig:melabdt2dshapes} 
The distributions of the discriminant BDT$_{ZZ}$ versus the $J^P$--MELA discriminant 
for the SM $J^P=0^+$ Higgs boson and for the backgrounds in the \hZZ\ signal region  $115~\GeV\ <m_{4\ell} <130~\GeV$.}
\end{figure}

\begin{figure*}[htb]
\centering
\begin{minipage}[h]{0.31\linewidth}
\subfloat[~]{\includegraphics[width=1\linewidth]{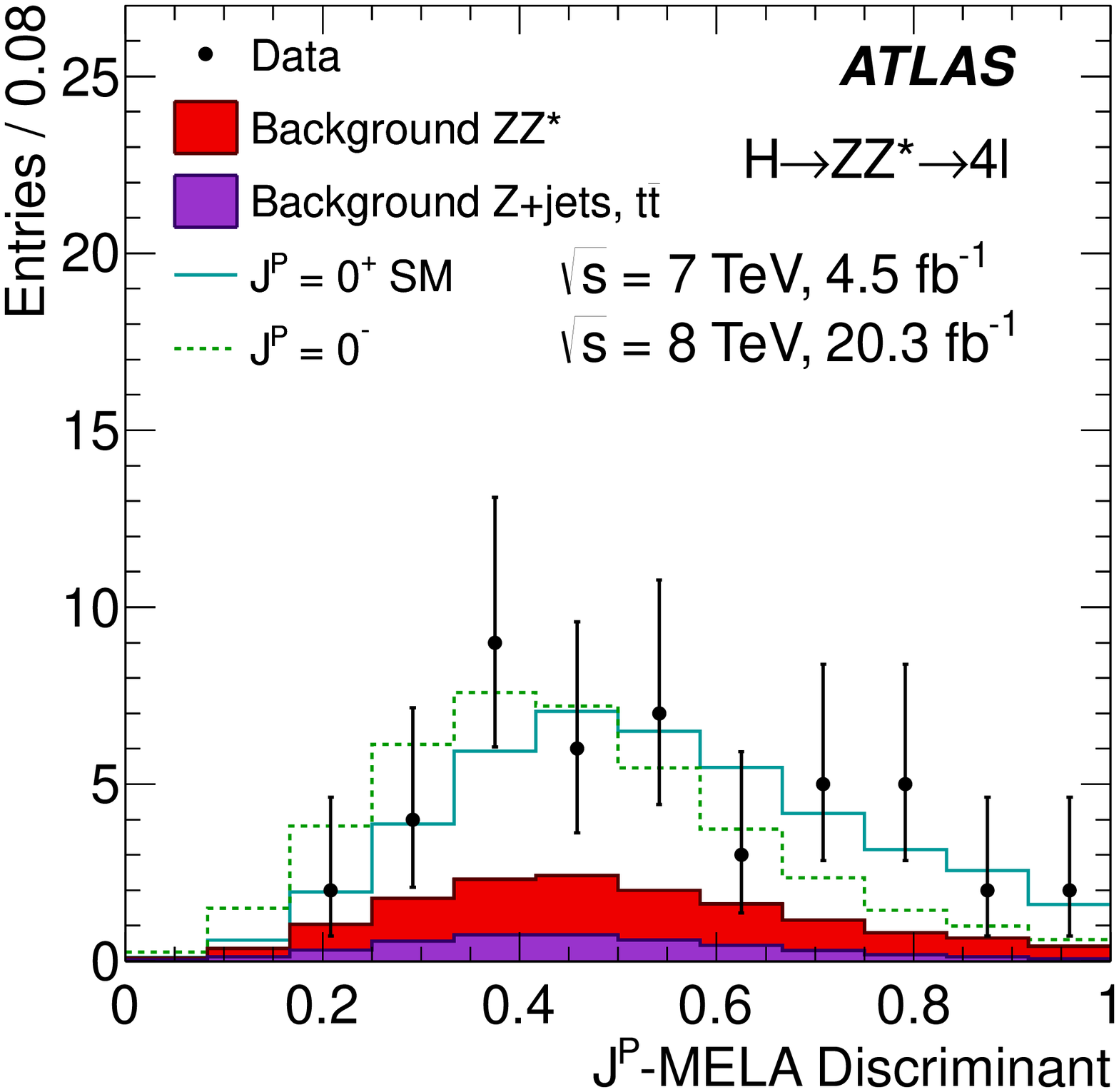}}
\end{minipage}
\hfill
\begin{minipage}[h]{0.31\linewidth}
\subfloat[~]{\includegraphics[width=1\linewidth]{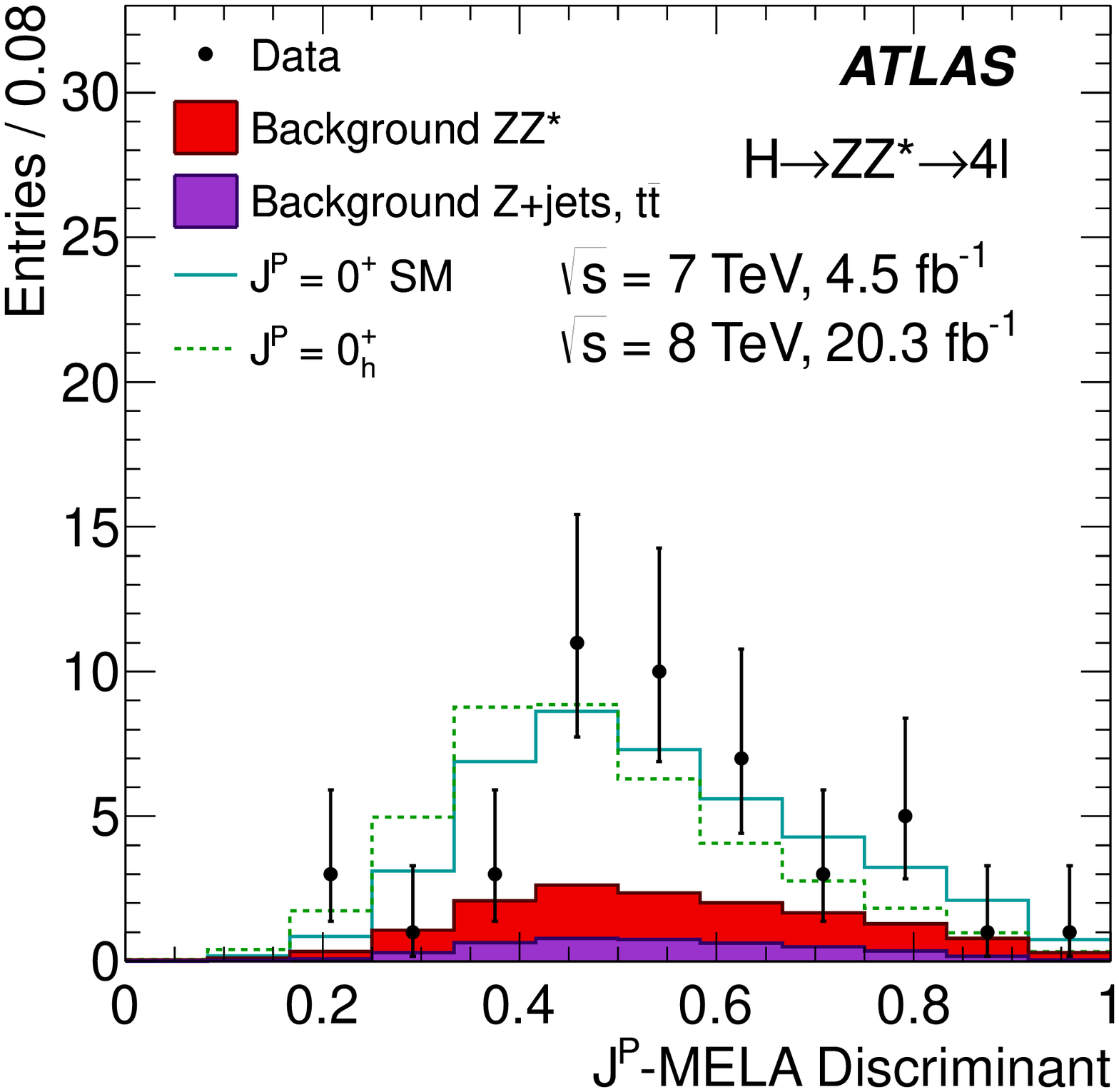}}
\end{minipage}
\hfill
\begin{minipage}[h]{0.31\linewidth}
\subfloat[~]{\includegraphics[width=1\linewidth]{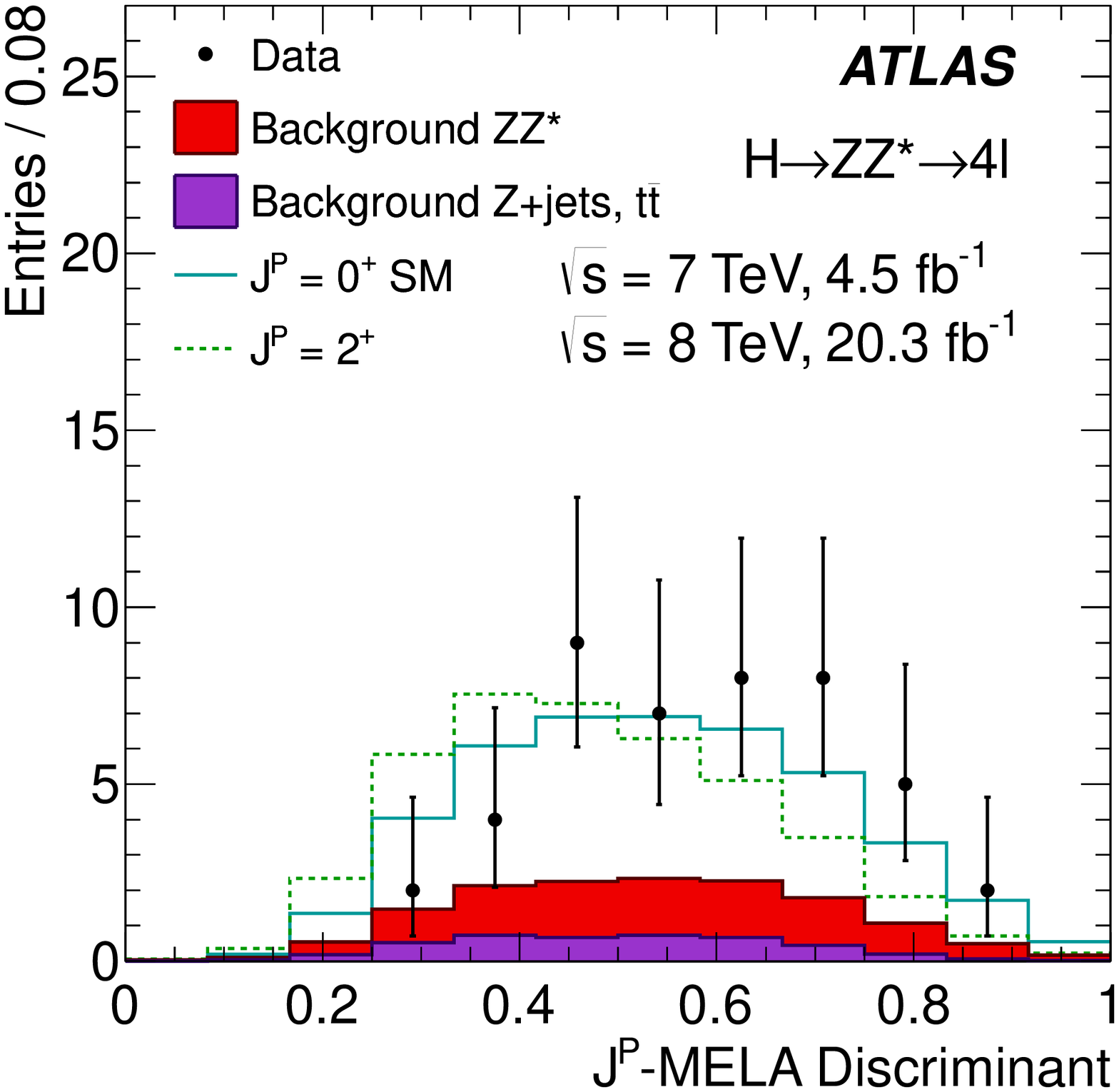}}
\end{minipage}
\vfill
\begin{minipage}[h]{0.31\linewidth}
\subfloat[~]{\includegraphics[width=1\linewidth]{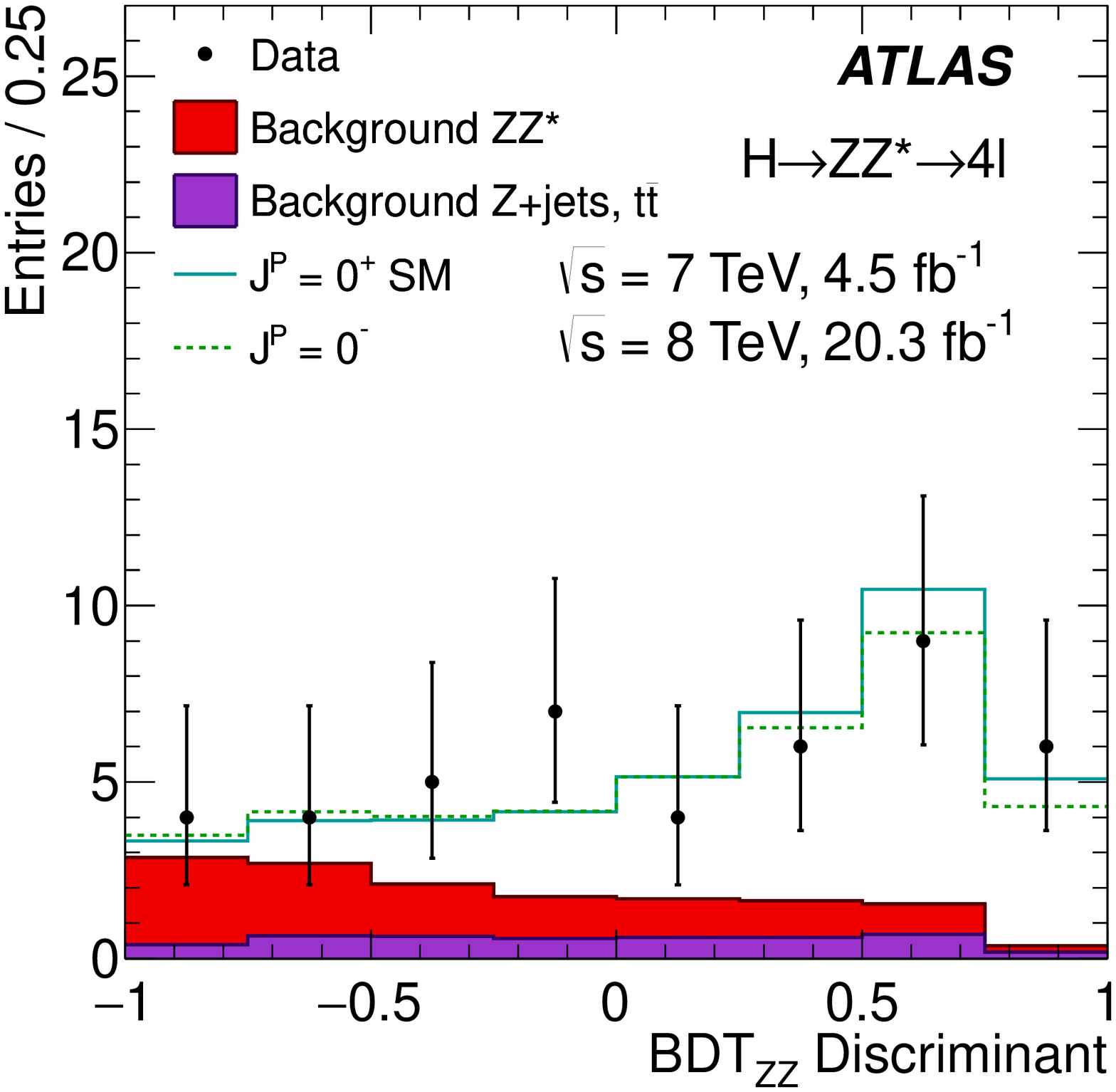}}
\end{minipage}
\hfill
\begin{minipage}[h]{0.31\linewidth}
\subfloat[~]{\includegraphics[width=1\linewidth]{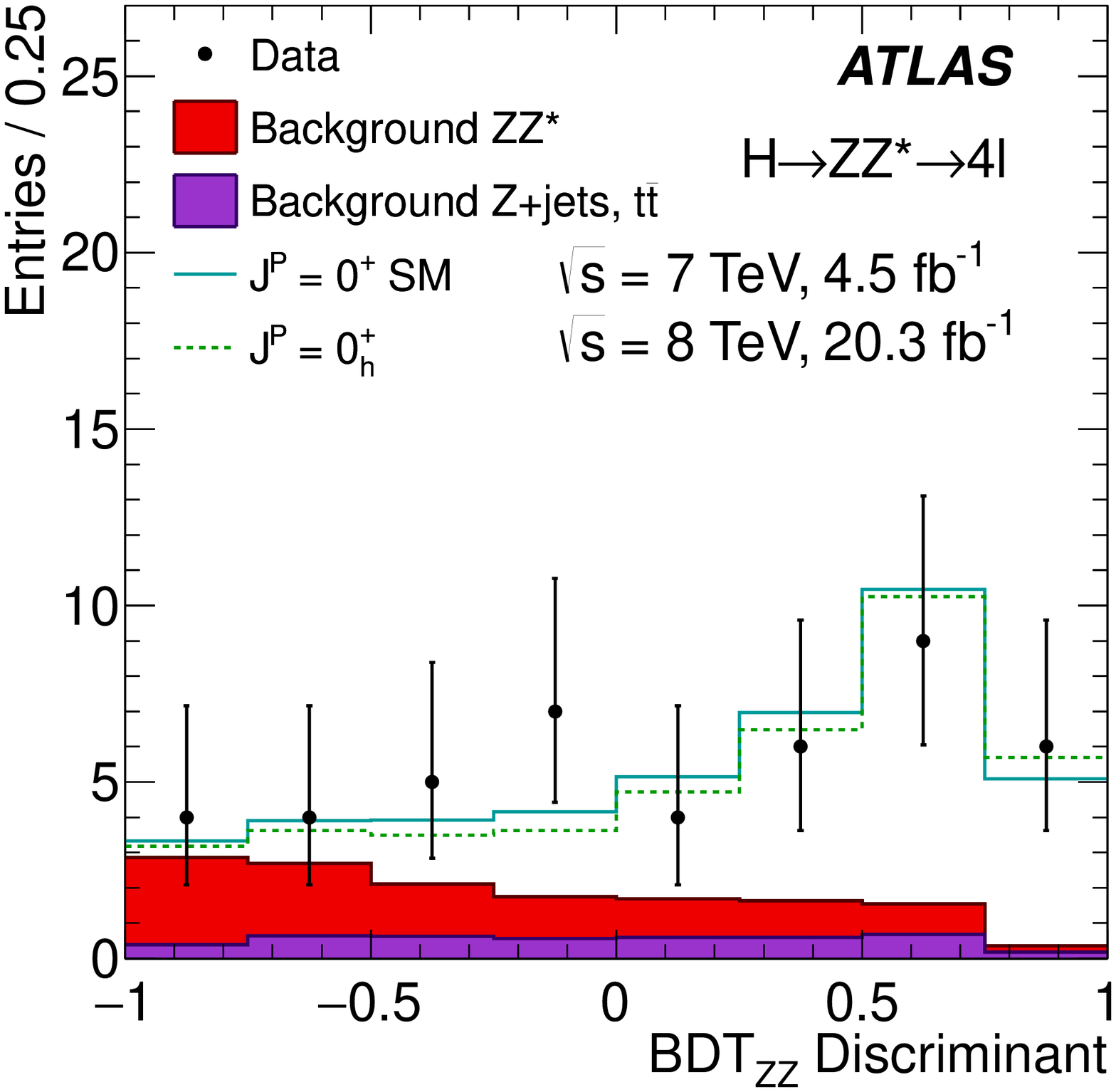}}
\end{minipage}
\hfill
\begin{minipage}[h]{0.31\linewidth}
\subfloat[~]{\includegraphics[width=1\linewidth]{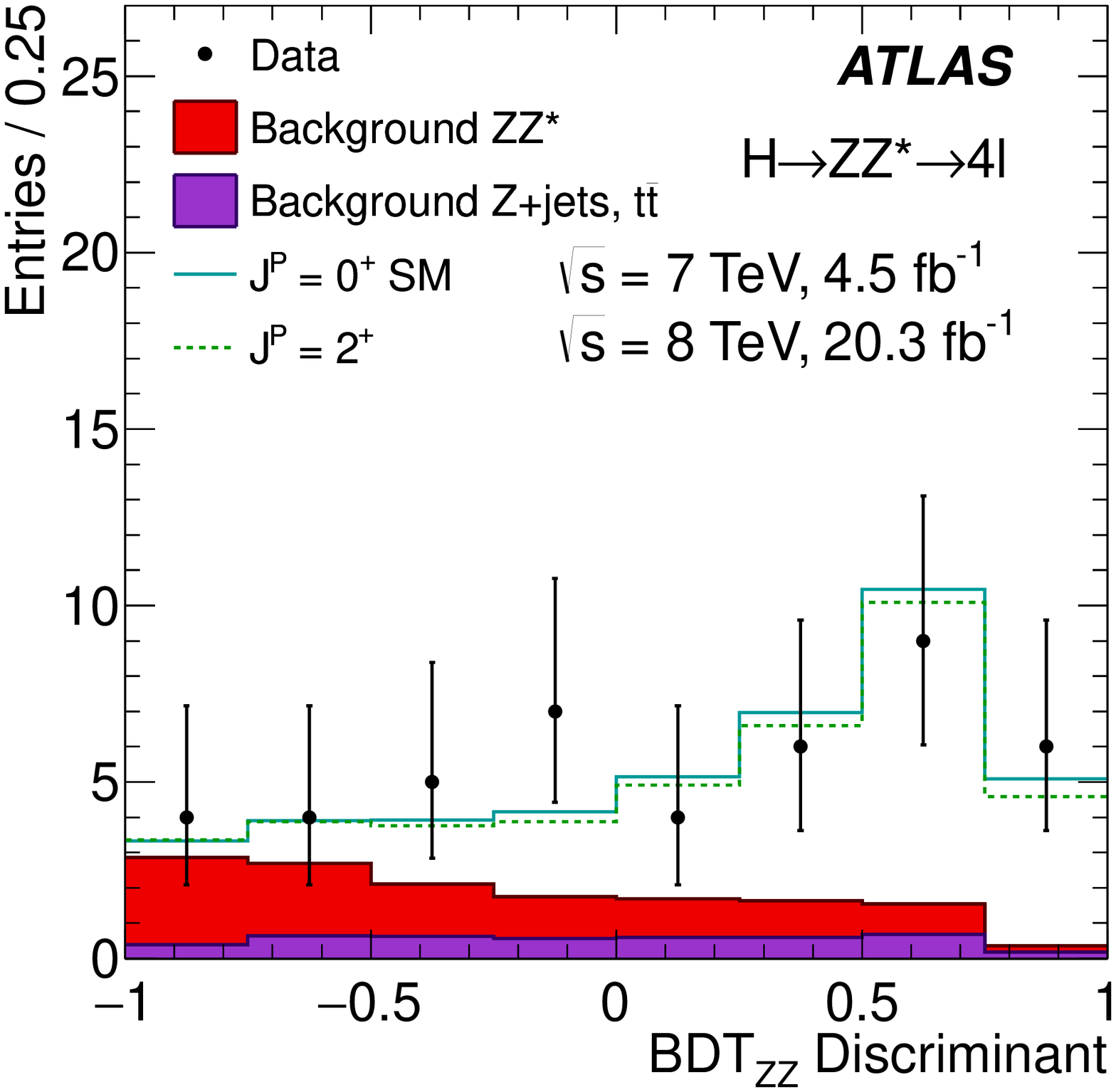}}
\end{minipage}
\caption{\label{fig:fs_2} 
Distributions of the $J^P$--MELA and of the ${\rm BDT}_{ZZ}$ discriminants in the 
$H\to ZZ^{*}\to 4\ell$ signal region $115~\GeV<m_{4\ell}< 130~\GeV$ for 
the data (points with errors),  the backgrounds (filled histograms),
and for predictions for several spin  and parity hypotheses.  
The SM hypothesis is shown by the solid line while the alternative hypotheses are shown by the dashed lines. 
The signal distributions are normalised to the signal strength fitted in data.
(a), (b) and (c): $J^P$--MELA discriminants for  $0^+$ SM vs $0^-$, $0^+$ SM vs $0^+_h$ and  $0^+$ SM vs $2^+$, respectively; 
(d), (e) and (f): ${\rm BDT}_{ZZ}$   discriminant for  $0^+$ SM vs $0^-$, $0^+$ SM vs $0^+_h$ and  $0^+$ SM vs $2^+$, respectively.}
\end{figure*}

Both analyses are  complemented with a BDT discriminant designed to separate the signal from the $ZZ^*$ background.
These discriminants are hereafter referred to as BDT$_{ZZ}$.
For the $J^P$--MELA analysis, the BDT$_{ZZ}$ discriminant is fully equivalent to the 
one described in Refs.~\cite{MassPaper,Aad:2014eva}. 
For the BDT analysis the discriminating variables used for the background BDT$_{ZZ}$ are the
invariant mass, pseudorapidity, and transverse momentum of the four-lepton system, and a matrix-element-based kinematic discriminant 
$K_D$ defined in Ref.~\cite{MG5}.
The results from both methods are obtained from likelihood fits to the two-dimensional distributions of the 
background BDTs and of the spin- and parity-sensitive discriminants. 
In this way, the small correlation between these variables are taken into account in the analyses.
The distribution of the background discriminant BDT$_{ZZ}$ versus the $J^P$--MELA discriminant
is presented in Figure~\ref{fig:melabdt2dshapes} for the SM $J^P=0^+$ signal, the backgrounds, and the data.
The projections of this distribution on the $J^P$--MELA and the BDT$_{ZZ}$ variables,
for different signal hypotheses, the backgrounds, and the data, are shown in Figure~\ref{fig:fs_2}.  
In this paper, only results based on the $J^P$--MELA approach are reported. 
The BDT approach was used as a cross-check and produced compatible results.

Two general types of systematic effects impact the analyses using fixed spin and
parity hypotheses: uncertainties on discriminant shapes due to experimental effects, and uncertainties on 
background normalisations from theory uncertainties and data-driven background estimates.
The systematic uncertainties on the shape are included in the analysis by creating discriminant shapes corresponding 
to variations of one standard deviation in the associated sources of systematic uncertainty. The systematic uncertainties 
on the normalisation are included as additional nuisance parameters
in the likelihood.

The list of sources of systematic uncertainty common to all ATLAS \hZZ\ analyses is presented in
Ref.~\cite{Aad:2014eva}.  The relative impact of these sources on the final
separation for all tested hypotheses is evaluated and sources affecting
the final separation (given in Section \ref{sec:fixed_comb}) 
by less than $\pm 0.5\%$ are neglected.

The main sources of systematic uncertainties are related to the experimental 
error on the Higgs boson mass, 
the modelling of the irreducible $ZZ^{*}$ 
background, the uncertainty on the integrated luminosity and the experimental uncertainties on 
the electron and muon reconstruction. 
The uncertainty on the Higgs boson mass affects the
final result since it impacts the shapes of the $m_{12}$, $m_{34}$, $\cos \theta _{1}$ and $\cos \theta _{2}$ variables. 
For the $J^P$--MELA method, the uncertainty on the estimate of the
fraction of $4\mu$ and $4e$ candidates with an incorrect pairing of
leptons is also considered. This uncertainty is derived by comparing
the corresponding prediction obtained from the \POWHEG\ and JHU
MC generators for the SM hypothesis. A
variation of $\pm 10\%$ of the incorrect pairing fraction
is applied to all spin and parity hypotheses.

\begin{table}[h!]
  \begin{center}        
 \begin{tabular}{lc}
 \hline\hline
 Source of the systematic uncertainty & Relative impact\\
 \hline
 Higgs boson mass experimental uncertainty               &  $\pm 2\%$\\
 $ZZ^{*}$  pdf                       &  $\pm 0.8\%$\\
 Muon momentum scale                &   $\pm 0.7\%$   \\
 $Zbb\to \ell \ell \mu \mu$  normalisation  &  $\pm 0.6\%$ \\
 $ZZ^{*}$ scale                       &  $\pm 0.6 \%$ \\
 Luminosity                           & $\pm 0.6\%$ \\
$e/\gamma$ resolution model (sampling term) &  $\pm 0.5\%$ \\
$e/\gamma$ resolution model (constant term) &  $\pm 0.5\%$ \\
$Z\to \ell \ell ee$ normalisation                   &  $\pm 0.5\%$ \\
 \hline
 Fraction of wrongly paired $4\ell$ candidates &  $\pm 0.4\%$\\
 \hline\hline
 \end{tabular}
 \caption{\label{tab:syst} Relative impact of the main systematic uncertainties on the expected separation (expressed in terms of numbers of standard deviations) 
between the SM $J^P=0^+$ and $J^P=0^-$ hypotheses for the \hZZ\ $J^P$--MELA analysis. 
   } 
\end{center}    
\end{table}
The influence of the main systematic uncertainties on the separation between the SM $J^P=0^+$
and $J^P=0^-$ hypotheses for the $J^P$--MELA
analysis is presented in Table~\ref{tab:syst}.
The total relative impact of all systematic uncertainties on the separation between the hypotheses (expressed in terms of numbers of standard deviations) 
is estimated to be about $\pm 3\%$.

\subsection{Individual and combined results}
\label{sec:fixed_comb}
\begin{figure}[htbp!]
  \centering    
  \subfloat[~]{\includegraphics[width=0.49\textwidth]{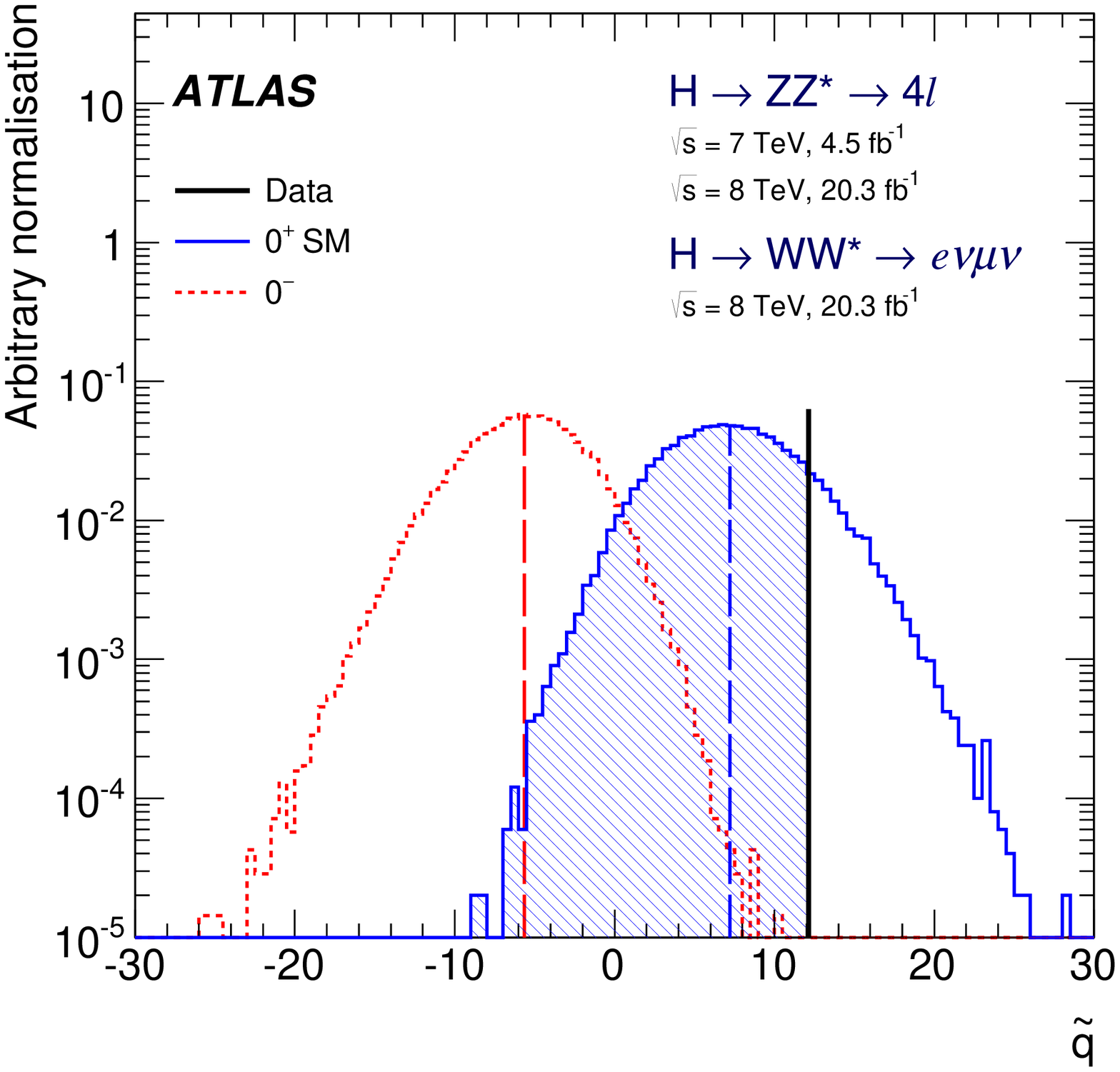}}
  \subfloat[~]{\includegraphics[width=0.49\textwidth]{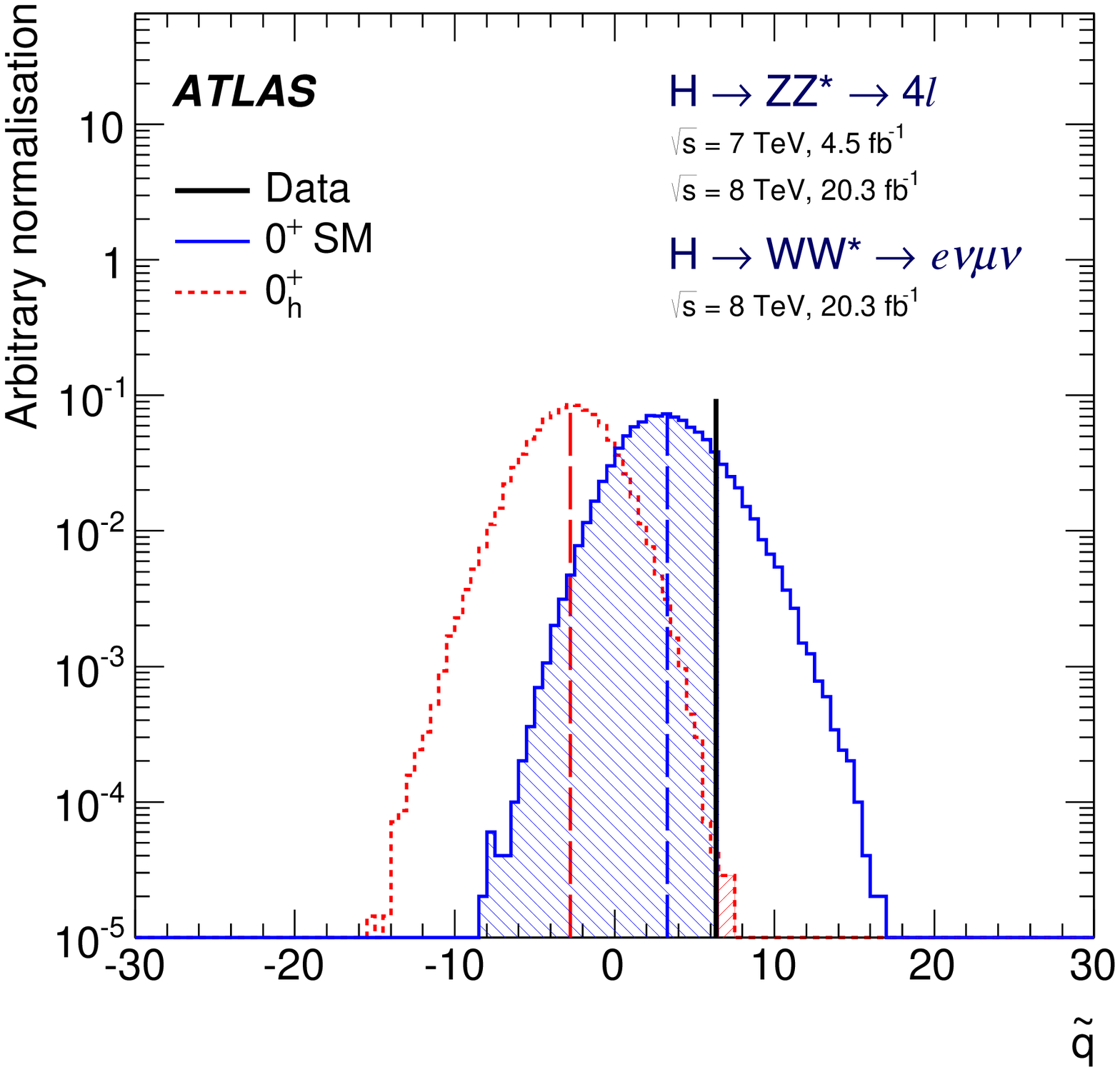}}\\
  \subfloat[~]{\includegraphics[width=0.49\textwidth]{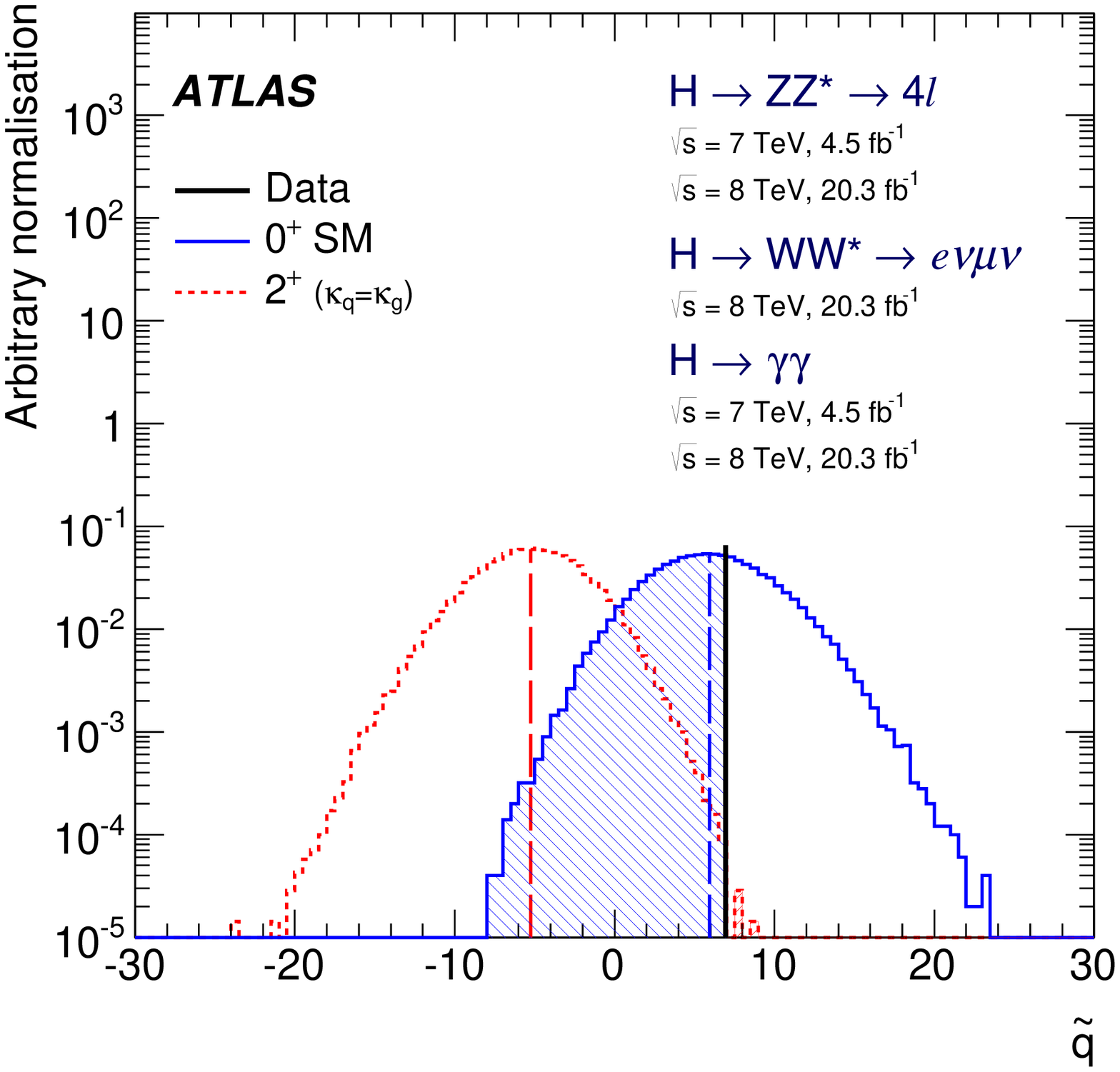}}
  \subfloat[~]{\includegraphics[width=0.49\textwidth]{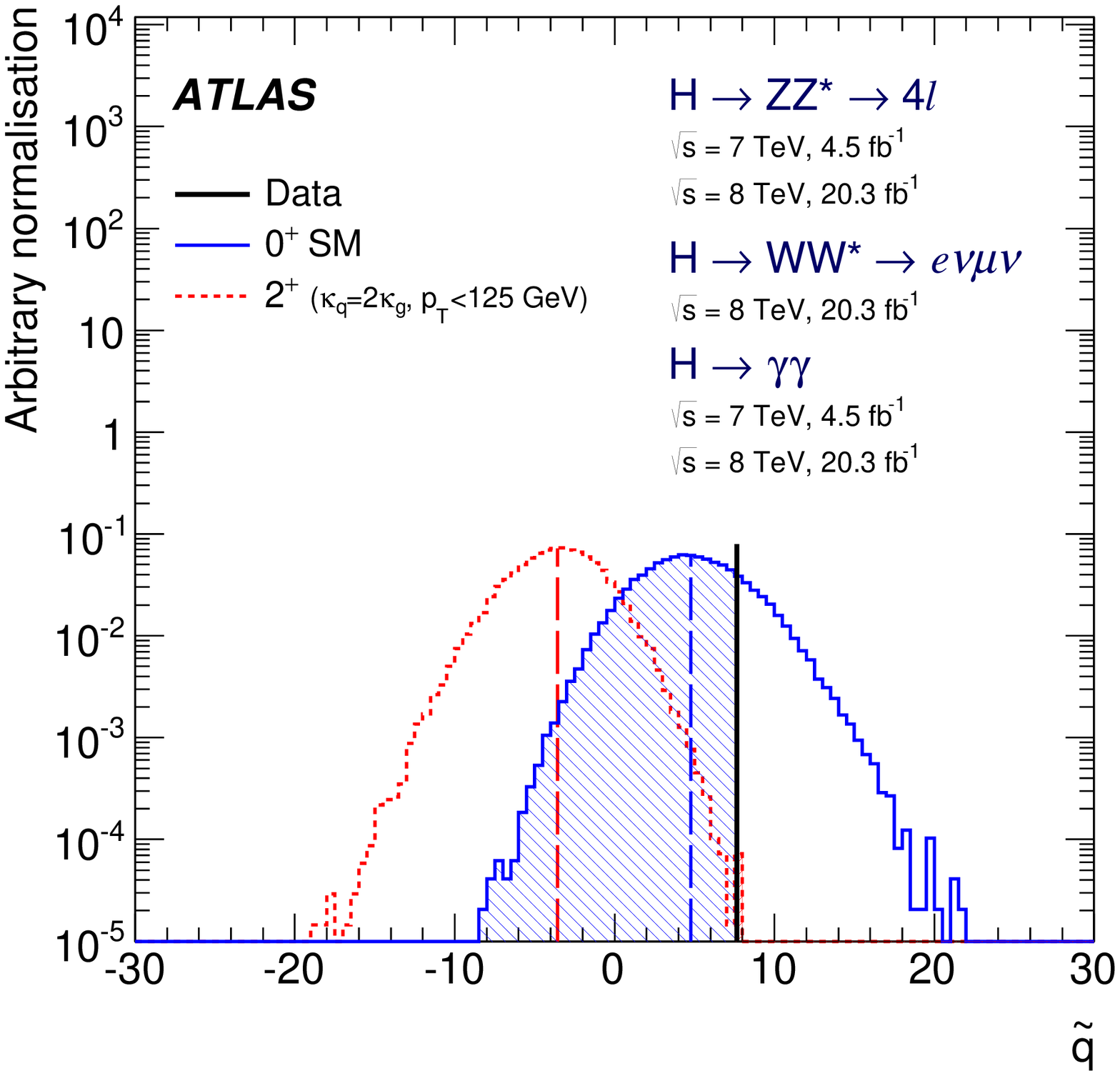}}

  \caption{Examples of distributions of the test statistic $\tilde{q}$ defined in Section~\ref{sec:fixed_stat}, 
    for the combination of decay channels.
    (a): $0^+$ versus  $0^-$; (b): $0^+$ versus  $0^+ _h$;  (c): $0^+$ versus the
    spin-2 model with universal couplings ($\kappa_{q} = \kappa_{g}$);  (d): $0^+$ versus the spin-2 model with
    $\kappa_{q} = 2 \kappa_{g}$ and the $\pt$ selection at $125$~GeV.
    The observed values are indicated by the vertical solid line and the expected medians by the dashed lines. 
    The shaded areas correspond to the integrals of the expected distributions used to compute the $p$-values 
    for the rejection of each hypothesis. 
    \label{fig:fixedhypo_teststat}}
\end{figure}

The distributions of discriminant variables in data agree with the SM predictions for all three channels, and 
exclusion ranges for alternative spin hypotheses are derived. 
Some examples of distributions of the test statistic $\tilde{q}$ (defined in Section~\ref{sec:fixed_stat}) 
used to derive the results are presented in Figure~\ref{fig:fixedhypo_teststat}.
In this figure, the observed value is indicated by the vertical  solid line and the expected medians by the dashed lines. 
The shaded areas correspond to the integrals of the expected distributions used to compute the $p$-values 
for the rejection of each hypothesis. 
The signal strengths per decay channel and per centre-of-mass energy are treated as independent parameters in each fit. 
Their values are compatible with the SM predictions.

The results obtained from the fit to the data, expressed in terms of $p$-values for different tested hypotheses and observed \CLs\ 
for the alternative hypotheses, are summarised in Tables~\ref{tab:channels_fixed_hypo_separations} and \ref{tab:fixed_hypo_separations}.
As shown in Table~\ref{tab:channels_fixed_hypo_separations}, the sensitivity to reject alternative 
hypotheses is driven by the \hZZ\ and the \hWW\ channels. The \hgg\ channel has sizeable sensitivity only to spin-2 models  
where the $\pT^{X}<125 \GeV$ selection is not applied. 
In all cases the data prefer the SM hypothesis to the alternative models,
with the exception of some of the spin-2 models for the \hgg\ channel. 
In this case both hypotheses have similar observed $p$-values, but neither of the two is below 10\% . 

As summarised in Table~\ref{tab:fixed_hypo_separations}, the $p$-values of the combined results for the three channels show good agreement between the 
data and the SM hypothesis for all performed tests. All tested alternative hypotheses are rejected at a more than 99.9\% confidence level (CL) in favour of the SM hypothesis.

\begin{table}[ht]
\centering
\begin{tabular} {|c|c|c|c|c|c|}
\hline
   &  \multicolumn{5}{c|}{$\hgg$} \\
Tested Hypothesis & $p^{\rm alt}_{{\rm exp},\mu=1}$ & $p^{\rm alt}_{{\rm exp},\mu=\hat{\mu}}$ & $p^{\rm SM}_{\rm obs}$
   &  $p^{\rm alt}_{\rm obs}$   & Obs. \CLs\ (\%) \\
\hline
 $2^+ (\kappa_q  = \kappa_g)$                & 0.13 & $7.5\cdot10^{-2}$ & 0.13 & 0.34 & 39 \\
 $2^+(\kappa _q =0;\; \pt<300 \GeV)$  & $4.3\cdot10^{-4}$ & $<3.1\cdot10^{-5}$  & 0.16 & 2.9$\cdot10^{-4}$  & 3.5$\cdot10^{-2}$ \\
 $2^+(\kappa _q =0;\; \pt<125 \GeV)$  & $9.4\cdot10^{-2}$ & 5.6$\cdot10^{-2}$ & 0.23 & 0.20 & 26 \\
 $2^+(\kappa _q =2 \kappa _g ;\; \pt<300\GeV)$  & $9.1\cdot10^{-4}$  & $<3.1\cdot10^{-5}$ & 0.16 & 8.6$\cdot10^{-4}$  & 0.10 \\
 $2^+(\kappa _q =2 \kappa _g ;\; \pt<125\GeV)$  & 0.27  & 0.24 & 0.20 & 0.54 & 68 \\

\hline
   &  \multicolumn{5}{c|}{$\hWW$} \\
Tested Hypothesis & $p^{\rm alt}_{{\rm exp},\mu=1}$ & $p^{\rm alt}_{{\rm exp},\mu=\hat{\mu}}$ & $p^{\rm SM}_{\rm obs}$
   &  $p^{\rm alt}_{\rm obs}$   & Obs. \CLs\ (\%) \\
\hline
 $0^+_h$              & 0.31 & 0.29  & 0.91 & 2.7$\cdot10^{-2}$ & 29 \\
 $0^-$                & 6.4$\cdot10^{-2}$ & 3.2$\cdot10^{-2}$   & 0.65 & 1.2$\cdot10^{-2}$ & 3.5  \\
 $2^+(\kappa_q  = \kappa_g)$                & 6.4$\cdot10^{-2}$ & 3.3$\cdot10^{-2}$   & 0.25 & 0.12 & 16 \\
 $2^+(\kappa _q =0;\; \pt<300\GeV)$  & 1.5$\cdot10^{-2}$ & 4.0$\cdot10^{-3}$     & 0.55 & 3.0$\cdot10^{-3}$ & 0.6  \\
 $2^+(\kappa _q =0;\; \pt<125\GeV)$  & 5.6$\cdot10^{-2}$ & 2.9$\cdot10^{-2}$   & 0.42 & 4.4$\cdot10^{-2}$ & 7.5  \\
 $2^+(\kappa _q =2 \kappa _g ;\; \pt<300\GeV)$  & 1.5$\cdot10^{-2}$ & 4.0$\cdot10^{-3}$     & 0.52 & 3.0$\cdot10^{-3}$ & 0.7  \\
 $2^+(\kappa _q =2 \kappa _g ;\; \pt<125\GeV)$  & 4.4$\cdot10^{-2}$ & 2.2$\cdot10^{-2}$   & 0.69 & 7.0$\cdot10^{-3}$ & 2.2  \\
\hline
   &  \multicolumn{5}{c|}{$\hZZ$} \\
Tested Hypothesis & $p^{\rm alt}_{{\rm exp},\mu=1}$ & $p^{\rm alt}_{{\rm exp},\mu=\hat{\mu}}$ & $p^{\rm SM}_{\rm obs}$
   &  $p^{\rm alt}_{\rm obs}$   & Obs. \CLs\ (\%)\\
\hline
 $0^+_h$              &  $3.2\cdot10^{-2}$ & $5.2\cdot10^{-3}$  & 0.80  &  $3.6\cdot10^{-4}$   &  0.18 \\
 $0^-$                &  $8.0\cdot10^{-3}$  & $3.6\cdot10^{-4}$  & 0.88  &  $1.2\cdot10^{-5}$   &  1.0$\cdot10^{-2}$ \\
 $2^+(\kappa_q  = \kappa_g)$                &  $3.3\cdot10^{-2}$ & $5.7\cdot10^{-4}$  & 0.91  &  $3.6\cdot10^{-5}$   &  4.0$\cdot10^{-2}$ \\
 $2^+(\kappa _q =0;\; \pt<300\GeV)$  &  $3.9\cdot10^{-2}$  & $9.0\cdot10^{-3}$  & 0.95  &  $2.7\cdot10^{-5}$  &  5.4$\cdot10^{-2}$ \\
 $2^+(\kappa _q =0;\; \pt<125\GeV)$   &  $4.6\cdot10^{-2}$  & $1.1\cdot10^{-2}$  & 0.93  &  $3.0\cdot10^{-5}$  &  4.3$\cdot10^{-2}$ \\
 $2^+(\kappa _q =2 \kappa _g ;\; \pt<300\GeV)$   &  $4.6\cdot10^{-2}$  & $1.1\cdot10^{-2}$ & 0.66  &  $3.3\cdot10^{-3}$  &  0.97 \\
 $2^+(\kappa _q =2 \kappa _g ;\; \pt<125\GeV)$ &  $5.0\cdot10^{-2}$  & $1.3\cdot10^{-2}$ & 0.88  &  $3.2\cdot10^{-4}$  &  0.27 \\
\hline
\end{tabular}
\caption{Expected and observed $p$-values for different spin-parity hypotheses, for each of the three 
channels $\hgg$, $\hZZ$, and $\hWW$. The observed \CLs\ for the alternative hypotheses are reported in the last column. 
The expected and observed $p$-values and the observed \CLs\ are defined in Section \ref{sec:fixed_comb} and the alternative hypotheses
are those described in Section \ref{sec:theory}. }
\label{tab:channels_fixed_hypo_separations}
\end{table}

\begin{table}[ht]
\centering
\begin{tabular} {lccccc}
\hline\hline
Tested Hypothesis & $p^{\rm alt}_{{\rm exp},\mu=1}$ &  $p^{\rm alt}_{{\rm exp},\mu=\hat{\mu}}$ & $p^{\rm SM}_{\rm obs}$ &  $p^{\rm alt}_{\rm obs}$   & Obs. \CLs\ (\%) \\
\hline
 $0^+_h$              &  $2.5\cdot10^{-2}$ &  $4.7\cdot10^{-3}$  & $0.85$ & $7.1\cdot10^{-5}$   &  $4.7\cdot10^{-2}$ \\              
 $0^-$                &  $1.8\cdot10^{-3}$ &  $1.3\cdot10^{-4}$  & $0.88$ & $<3.1\cdot10^{-5}$  &  $<2.6\cdot10^{-2}$ \\                         
 $2^+ (\kappa_q  = \kappa_g)$                &  $4.3\cdot10^{-3}$ & $2.9\cdot10^{-4}$  & $0.61$  &  $4.3\cdot10^{-5}$  &  $1.1\cdot10^{-2}$  \\         
 $2^+ (\kappa _q =0;\; \pt<300\GeV)$  &  $<3.1\cdot10^{-5}$  & $<3.1\cdot10^{-5}$ & $0.52$ & $<3.1\cdot10^{-5}$  &   $<6.5\cdot10^{-3}$ \\
 $2^+ (\kappa _q =0;\;\pt<125\GeV)$   &  $3.4\cdot10^{-3}$  & $3.9\cdot10^{-4}$  & $0.71$ & $4.3\cdot10^{-5}$  & $1.5\cdot10^{-2}$ \\
 $2^+ (\kappa _q =2 \kappa _g ;\; \pt<300\GeV)$  &  $<3.1\cdot10^{-5}$  & $<3.1\cdot10^{-5}$  & $0.28$ & $<3.1\cdot10^{-5}$  & $<4.3\cdot10^{-3}$ \\
 $2^+ (\kappa _q =2 \kappa _g;\; \pt<125\GeV)$  & $7.8\cdot10^{-3}$  & $1.2\cdot10^{-3}$  & $0.80$ & $7.3\cdot10^{-5}$  & $3.7\cdot10^{-2}$ \\

\hline\hline
\end{tabular}
\caption{Expected and observed $p$-values for different spin-parity hypotheses, for the combination of the three channels:
 $\hgg$ , $\hZZ$ and $\hWW$. The observed \CLs\ for the alternative hypothesis is reported in the last column. 
The expected and observed $p$-values and the observed \CLs\ are defined in Section~\ref{sec:fixed_comb}. 
The definitions of alternative hypotheses are given in Section~\ref{sec:theory}.}
\label{tab:fixed_hypo_separations}
\end{table}

\section{Study of CP-mixing and of the $HVV$ interaction tensor structure }
\label{sec:tensor}
Following the discussion in Section~\ref{sec:theory}, measurements of the $HVV$ 
interaction tensor couplings $\kappa _{\rm SM}$,  $\kappa _{AVV}$, $\kappa _{HVV}$ and of the mixing 
angle $\alpha$ are performed. The measurements consist of fitting the ratios of couplings $(\tilde{\kappa} _{AVV}/\kappa _{\rm SM}) \cdot \tan \alpha$ 
and $\tilde{\kappa} _{HVV}/\kappa _{\rm SM}$ to the discriminant observables for the \hWW\ and \hZZ\ processes and in their combination. 
In the fitting procedure only one ratio of couplings $(\tilde{\kappa} _{AVV}/\kappa _{\rm SM}) \cdot \tan \alpha$ 
or $\tilde{\kappa} _{HVV}/\kappa _{\rm SM}$ is considered at a time, while the other one is assumed to be absent.  

\subsection{Statistical treatment}
\label{sec:tensor_stat}
The measurement of the tensor structure of the $HVV$ interaction is based on a profiled likelihood~\cite{asimov,asimovErratum} 
that contains the discriminant observables sensitive to the EFT couplings. 
The signal rates in the different channels and for different centre-of-mass energies are treated as independent parameters. Therefore,
the global signal normalisation is not used to constrain the EFT couplings. 
The ratios of the BSM to SM couplings, \KtildeH\ and \KtildeA, are each separately fit  to the  discriminant observables in data.
The test statistic used to derive the confidence intervals on the parameters of interest is $q' = -2 \ln (\lambda)$, 
where $\lambda$ is the profiled likelihood~\cite{asimov,asimovErratum}. 
The results presented in the following rely on the asymptotic approximation~\cite{asimov,asimovErratum} for the test statistic. 
This approximation was cross-checked with Monte Carlo ensemble tests that confirm its validity in
the range of the parameters for which the 95\% CL limits are derived.

\subsection{Tensor structure analyses in the \hWW\ channel}
\label{sec:tensor_ww}
The \hWW\ analysis used to study the spin-0 tensor structure is already 
described in Section~\ref{sec:fixed_ww} and detailed in Ref.~\cite{spincp_ww_paper}. 
Only the 0-jet category is considered and the BDT$_0$ and \bdtcp\ are used as discriminant variables
in the likelihood defined to measure the spin-0 tensor structure couplings.
The only difference with respect to the spin hypothesis test is that, in this analysis, the BSM spin-0 
couplings are treated as continuous variables in the test statistic.

\subsection{Tensor structure analyses in the \hZZ\ channel }
\label{sec:tensor_zz}
To allow for a cross-check and validation of the obtained results, two different
fitting methods based on the analytical calculation of the leading-order matrix element of the \hZZ\
process are used. 

The method of the matrix-element-observable fit is based on modelling the distributions of the final-state 
observables in each bin of coupling ratios using Monte Carlo simulation.  
Using the Lagrangian defined in Eq.~(\ref{eq:spin0_l}), which is linear in the coupling constants $\kappa _{\rm SM}$, $\kappa _{HVV}$ and $\kappa _{AVV}$, the 
differential cross section at each point in the phase space can be expressed as a term corresponding to the SM amplitude, plus two additional terms, linear and
quadratic in the coupling constants. In this way it is possible to define two observables for each coupling, the so-called 
first- and second-order optimal observables, upon which the amplitude depends at each point of the phase space. 
For each event, they contain the full kinematic information about the couplings, which can thus be extracted from a fit to their shapes.
More details of the method can be found in Refs.~\cite{Atwood:1991ka, Davier:1992nw, Diehl:2002nj, OptObs}.

The observables sensitive to the presence and structure of  $\kappa _{\rm SM}$, $\kappa _{HVV}$ and $\kappa _{AVV}$   
considered in the current analysis are defined as follows:
\begin{eqnarray}
 O_1(\kappa_{HVV}) = & \frac{2\Re[{\rm ME}(\kappa _{\rm SM} \neq 0;\; \kappa _{HVV}, \kappa _{AVV}  = 0;\; \alpha=0 )^* \cdot {\rm ME}( \kappa _{HVV} \neq 0;\; \kappa _{\rm SM},  \kappa _{AVV}=0;\; \alpha=0)] }
  {|{\rm ME}(\kappa _{\rm SM} \neq 0;\; \kappa _{HVV}, \kappa _{AVV}  = 0;\; \alpha=0)|^2}, \nonumber \\
 O_2(\kappa _{HVV}) = & \frac{|{\rm ME}(\kappa _{HVV} \neq 0;\; \kappa _{\rm SM},  \kappa _{AVV}=0;\; \alpha=0 )|^2} {|{\rm ME}(\kappa _{\rm SM} \neq 0;\; \kappa _{HVV}, \kappa _{AVV}  = 0;\; \alpha=0)|^2}, \nonumber \\
 O_1(\kappa _{AVV},\alpha) =& \frac{2\Re[{\rm ME}(\kappa _{\rm SM} \neq 0;\; \kappa _{HVV}, \kappa _{AVV}  = 0;\; \alpha=0 )^* 
 \cdot {\rm ME}( \kappa _{AVV} \neq 0;\; \kappa _{\rm SM},  \kappa _{HVV}=0;\; \alpha=\pi/2)]  } 
 {|{\rm ME}(\kappa _{\rm SM} \neq 0;\; \kappa _{HVV}, \kappa _{AVV}  = 0;\; \alpha=0)|^2}, \nonumber \\
  O_2(\kappa _{AVV},\alpha) =&\frac{|{\rm ME}( \kappa _{AVV} \neq 0;\; \kappa _{\rm SM},  \kappa _{HVV}=0;\; \alpha=\pi/2)|^2} {|{\rm ME}(\kappa _{\rm SM} \neq 0;\; \kappa _{HVV}, \kappa _{AVV}  = 0;\; \alpha=0)|^2}. 
 \label{eq:opt}
\end{eqnarray}
Here ${\rm ME}(\kappa _{\rm SM},\kappa _{HVV},\kappa _{AVV},\alpha)$ denotes the leading-order matrix element of the \hZZ\ process.  
These definitions correspond to the first- and second-order optimal observables for a BSM amplitude with a three-component structure. 

The observables $O_{1,2}(\kappa _{HVV})$ and $O_{1,2}(\kappa _{AVV},\alpha)$ are used for the
$\tilde{\kappa} _{HVV}/\kappa _{\rm SM}$ and $(\tilde{\kappa} _{AVV}/\kappa _{\rm SM}) \cdot \tan \alpha$ individual fits respectively.  
In order to suppress the $ZZ^{*}$ background, a kinematic BDT discriminant similar to those described in
Section~\ref{sec:fixed_zz} is used as an additional observable in all fits.  The BDT training is performed independently for each final state
using observables with small sensitivity to parity: $\etafl$, $\ptfl$, $\mfl$, $\cos (\theta^*)$ and $\Phi _1$.  This BDT discriminant is 
denoted hereafter by ${\rm BDT(}ZZ\rm{)} $.

\begin{figure*}[htbp]
\centering
\begin{minipage}[h]{0.4\linewidth}
\subfloat[~]{\includegraphics[width=1\linewidth]{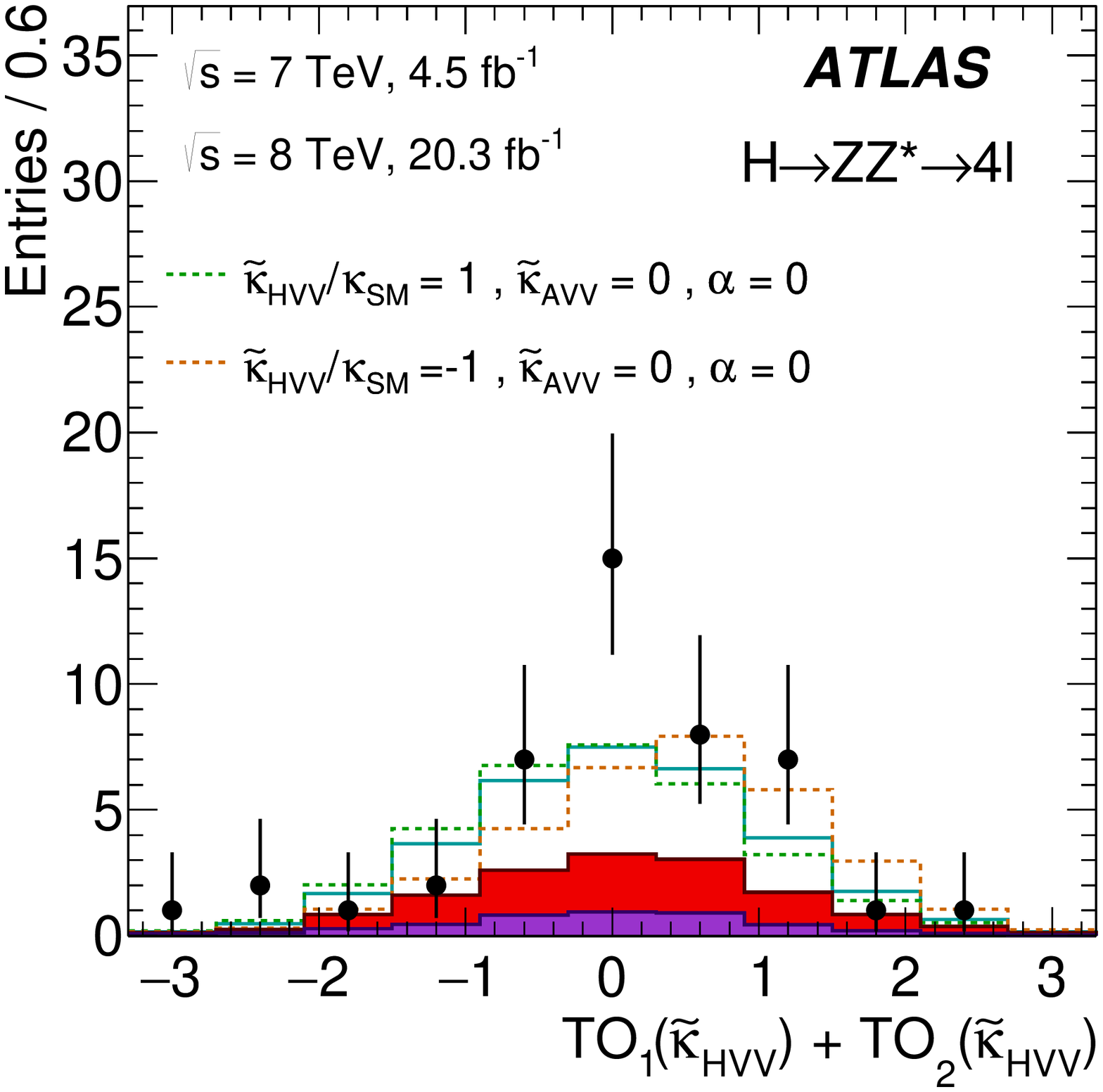}}  
\end{minipage}
\hfill
\begin{minipage}[h]{0.4\linewidth}
\subfloat[~]{\includegraphics[width=1\linewidth]{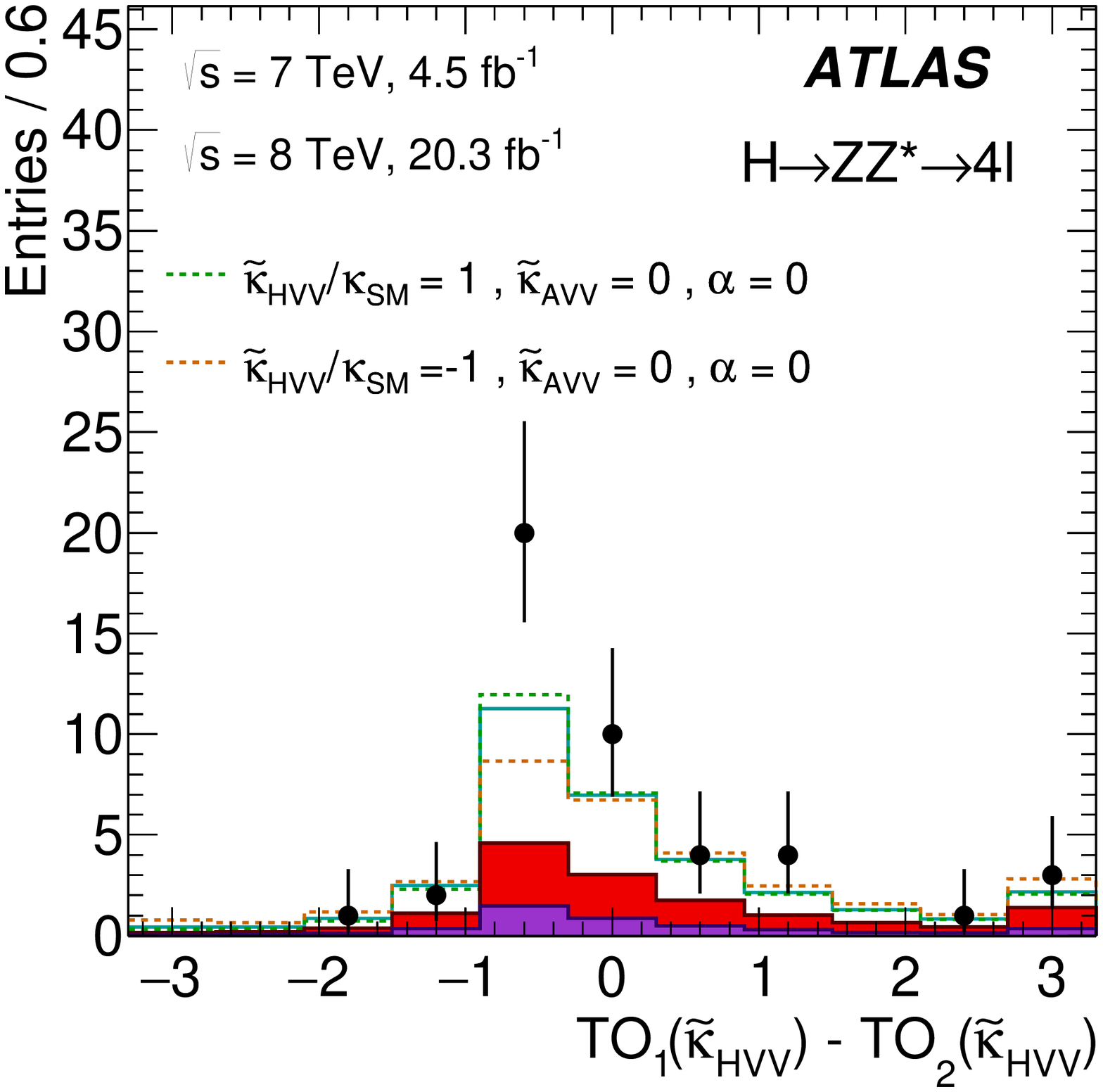}}  
\end{minipage}
\hfill
\begin{minipage}[h]{0.4\linewidth}
\subfloat[~]{\includegraphics[width=1\linewidth]{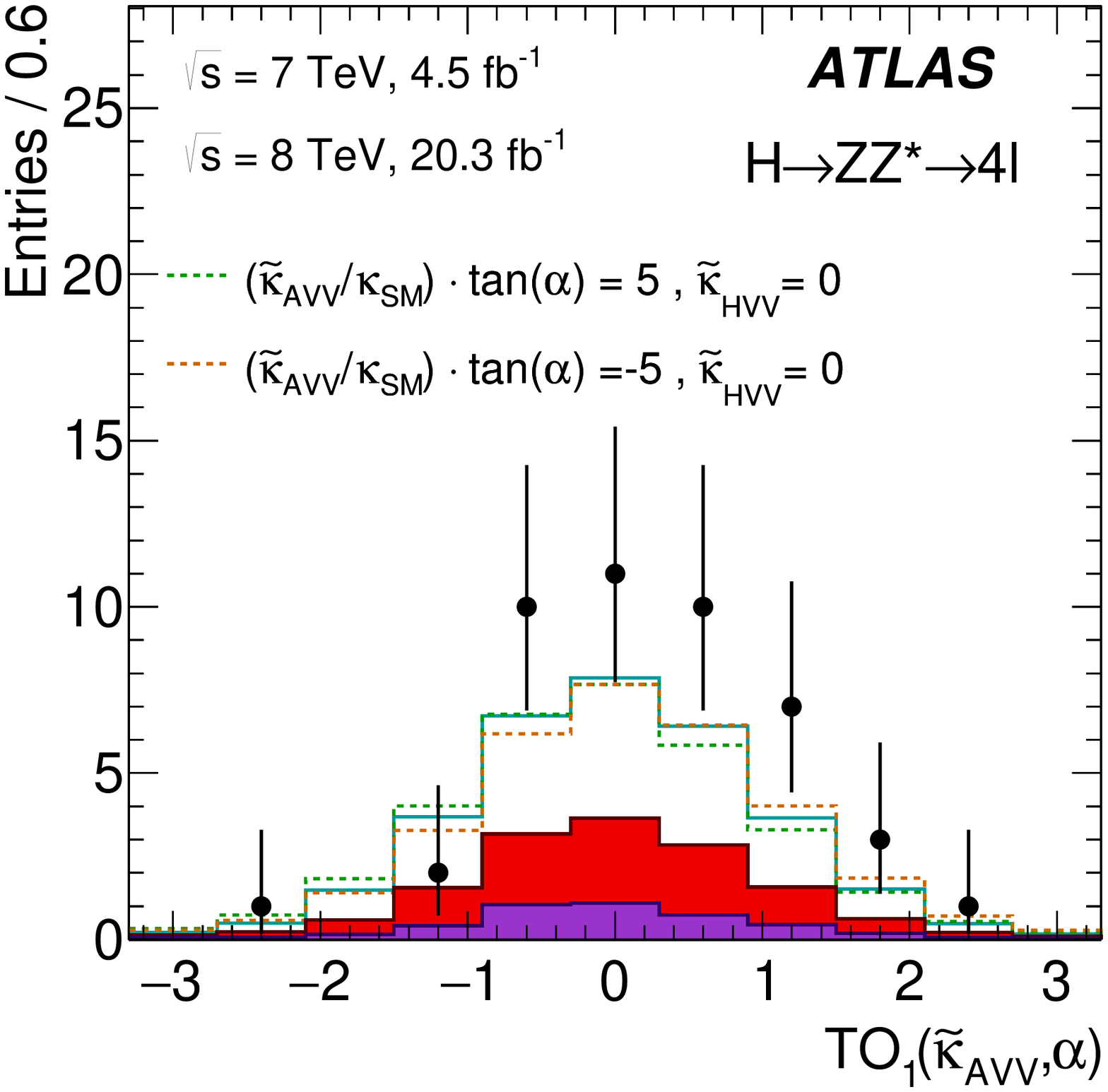}}  
\end{minipage}
\hfill
\begin{minipage}[h]{0.4\linewidth}
\subfloat[~]{\includegraphics[width=1\linewidth]{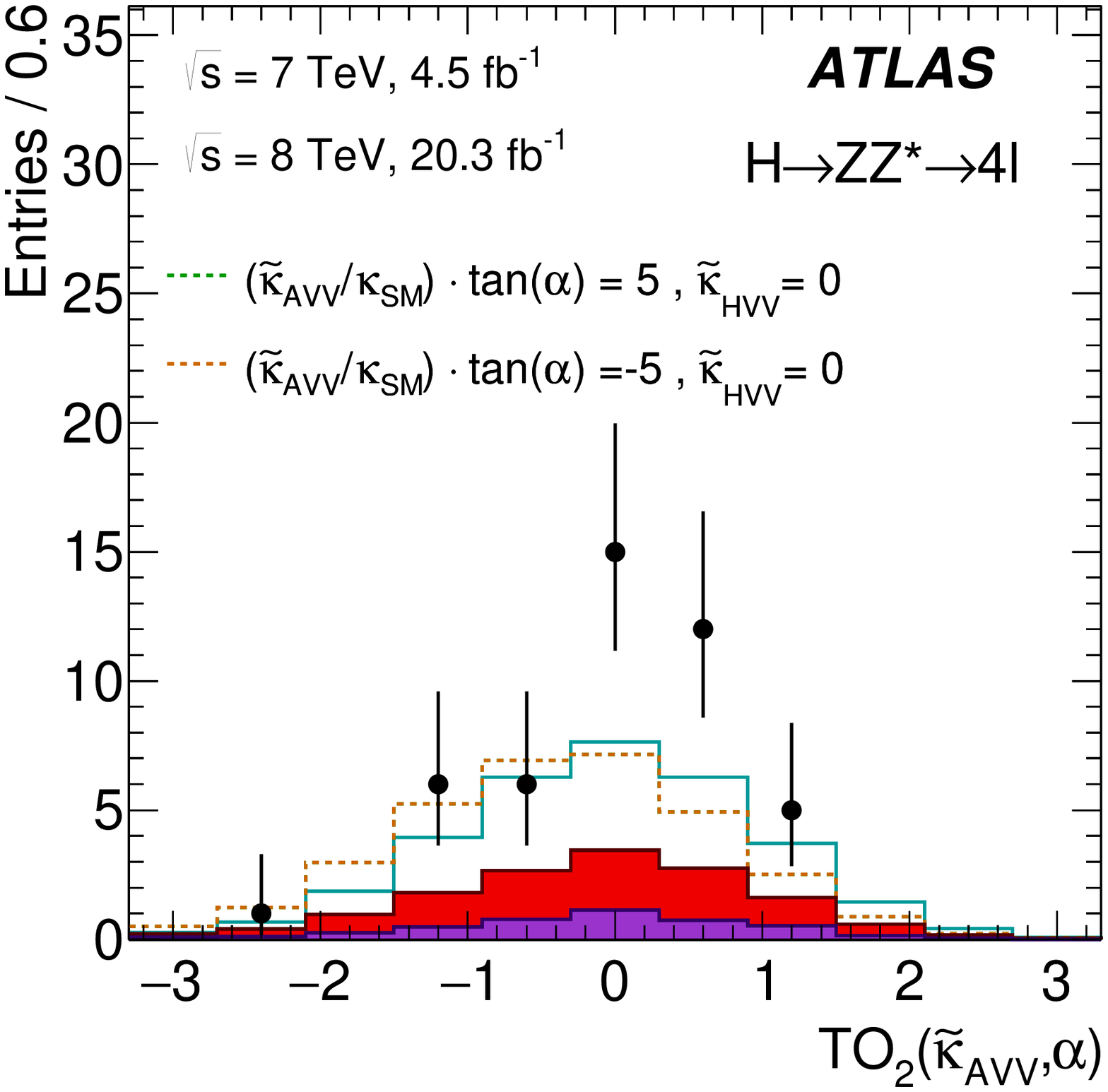}}  
\end{minipage}
\hfill
\begin{minipage}[h]{0.4\linewidth}
\subfloat[~]{\includegraphics[width=1\linewidth]{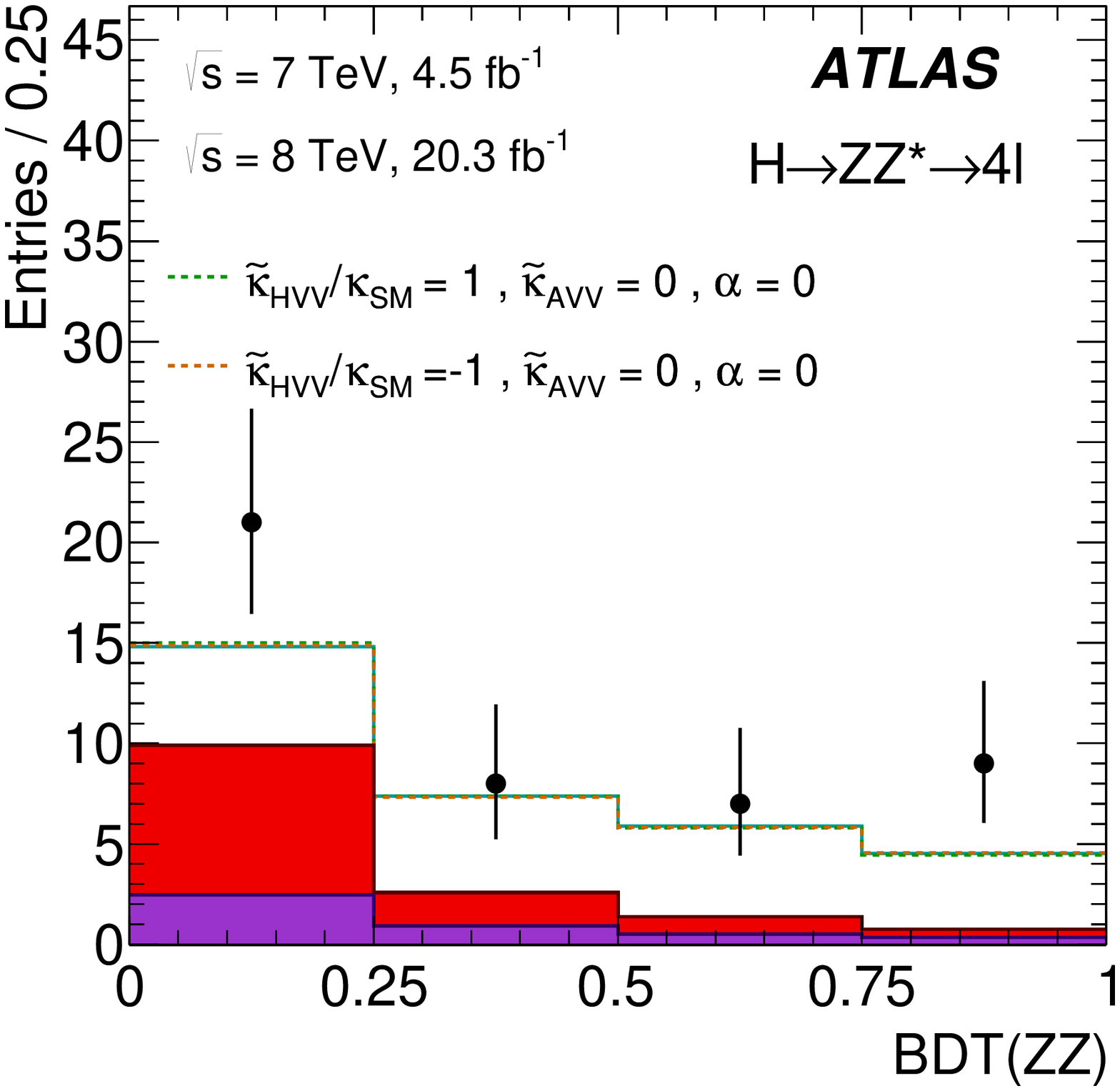}}  
\end{minipage}
\hfill
\begin{minipage}[h]{0.4\linewidth}
\center{\includegraphics[width=1\linewidth]{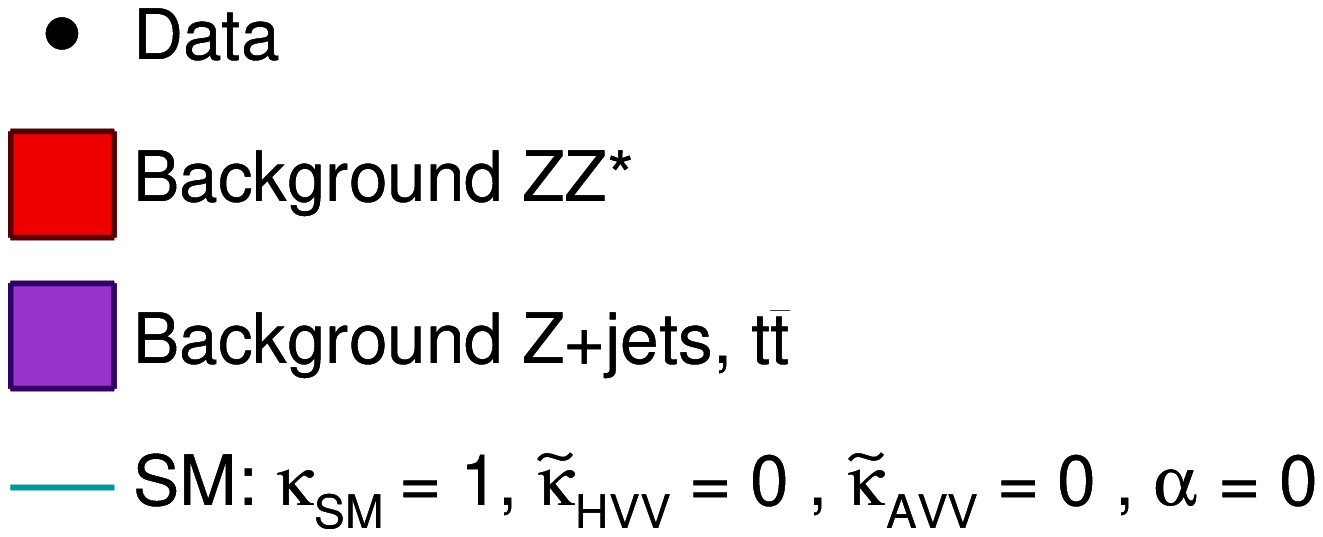}}  
\end{minipage}
\caption{\label{fig:oo_g4}
Distributions of the observables used in the matrix-element-observable fit. 
(a): 
$TO_1(\tilde{\kappa}_{HVV}) + TO_2(\tilde{\kappa}_{HVV})$, 
(b): $TO_1(\tilde{\kappa}_{HVV}) - TO_2(\tilde{\kappa}_{HVV})$,
for the Monte Carlo signal generated with 
$(\tilde{\kappa} _{HVV}/\kappa _{\rm SM} =0,\pm 1; \kappa _{AVV} =0)$. 
(c):  
$TO_{1}(\tilde{\kappa}_{AVV} , \alpha)$, (d): $TO_{2}(\tilde{\kappa}_{AVV} , \alpha)$  for the Monte Carlo signal generated  
with $((\tilde{\kappa}_{AVV}/\kappa _{\rm SM}) \cdot \tan \alpha =0,\pm 5;  \kappa _{HVV} =0)$. 
(e): ${\rm BDT(}ZZ\rm{)} $ for the Monte Carlo signal generated with 
$(\tilde{\kappa} _{HVV}/\kappa _{\rm SM} =0,\pm 1; \kappa _{AVV} =0)$.
The expected background contributions are shown as filled histograms on each plot.}
\end{figure*}

To simplify their use in the analysis, all observables defined in Eq.~(\ref{eq:opt})  undergo a pdf transformation
 such that each observable becomes  normally distributed in the Standard Model case. These transformed observables 
 are referred to hereafter as $TO_{1,2}(\kappa _{HVV})$  and $TO_{1,2}(\kappa _{AVV},\alpha)$ respectively.
The distributions of transformed observables
 for the Monte Carlo signal samples generated with 
$(\tilde{\kappa} _{HVV}/\kappa _{\rm SM} =0,\pm 1; \tilde{\kappa} _{AVV} =0)$ and $((\tilde{\kappa} _{AVV}/\kappa _{\rm SM}) \cdot \tan \alpha =0,\pm 5;  
\tilde{\kappa} _{HVV} =0)$ are shown in Figure~\ref{fig:oo_g4}.
The contributions of all backgrounds considered in this analysis are also included.  By construction the $TO_2$ observables 
are sensitive to the modulus of the $\tilde{\kappa} _{HVV}/\kappa _{\rm SM}$ and $(\tilde{\kappa} _{AVV}/\kappa _{\rm SM})\cdot\tan {\alpha}$ ratios: 
their distributions change with the strength of the respective coupling.  These observables are
insensitive to the relative sign of $\tilde{\kappa} _{HVV}$ and $\tilde{\kappa} _{AVV}$ with respect to $\kappa_{\rm SM}$. 
The sign sensitivity comes from the $TO_1$ observables, which are based on the interference terms: 
their distributions feature pronounced sign-dependent asymmetries.  It was also found that the observables 
$TO_1(\tilde{\kappa}_{HVV})$ and $TO_2(\tilde{\kappa}_{HVV})$ are linearly correlated.
To maximise the population of analysis histograms with  currently available Monte Carlo event samples, it is 
desirable to reduce this correlation. This is achieved by considering the modified observables 
$TO_1(\tilde{\kappa}_{HVV}) + TO_2(\tilde{\kappa}_{HVV})$ and  $TO_1(\tilde{\kappa}_{HVV}) - TO_2(\tilde{\kappa}_{HVV})$
in the current analysis.
  
The analysis is performed in several steps. First, multi-dimensional histograms of observables are created 
in $81$ bins of $\KtildeH$ and $\KtildeA$ for all fits. The predicted shapes of the observables for  the signal are produced 
by reweighting the base Monte Carlo sample described in Section~\ref{sec:MCsamples}. 
The corresponding weights are derived using the analytical calculation of the \hZZ\ matrix elements at leading order in perturbative QCD.
The weights are calculated and applied at the Monte Carlo generator level. The observables used in the analysis are evaluated
after detector simulation, accounting for the detector acceptance, resolution and reconstruction efficiency.  The distributions 
of observables for backgrounds are estimated using Monte Carlo (for the irreducible background) 
and data-driven techniques (for the reducible backgrounds) described in Section~\ref{sec:fixed_hypo} and Refs.~\cite{MassPaper,Aad:2014eva}.

The distributions  of observables are three-dimensional: $TO_{1}(\tilde{\kappa}_{AVV}$, $\alpha), TO_{2}(\tilde{\kappa}_{AVV}, \alpha)$,
 ${\rm BDT(}ZZ\rm{)}$ and $TO_1(\tilde{\kappa}_{HVV}) \allowbreak + TO_2(\tilde{\kappa}_{HVV})$, $TO_1(\tilde{\kappa}_{HVV}) - TO_2(\tilde{\kappa}_{HVV})$,  
  ${\rm BDT(}ZZ\rm{)}$ respectively.  To obtain a reliable description for bins with an insufficient number of  Monte Carlo events, the Kernel 
 Density Estimation~\cite{Cranmer:2000du} smoothing procedure is applied to signal and background multi-dimensional histograms. 
In the smoothing procedure the smearing is done separately in four bins of ${\rm BDT(}ZZ\rm{)}$, preserving the original normalisation.

The final pdfs used in the fits are obtained by applying linear histogram
interpolation between the multi-dimensional bins of $\KtildeH$ and $\KtildeA$.
The individual likelihood functions per centre-of-mass energy ($\sqrt{s}$) and final state (FS) are:
 \begin{equation}
 \mathcal{L}\left (\bar{\Omega}\middle | \frac{\tilde{\kappa}_{HVV}}{\kappa _{\rm SM}}, \frac{\tilde{\kappa}_{AVV}}{\kappa _{\rm SM}}\tan{\alpha},\bar{\theta} \right ) =
\prod\limits_{i} P\left [ \bar{ \Omega}_{i} \middle | s_{i} \left (\frac{\tilde{\kappa}_{HVV}}{\kappa _{\rm SM}}, \frac{\tilde{\kappa}_{AVV}}{\kappa _{\rm SM}}\tan{\alpha} , \bar{\theta} \right) + b_{i}(\bar{\theta}) \right ],
 \label{eq:meobs_l} \end{equation} 
where $P$ is the probability density function for the data vector $ \bar{\Omega}$, given the signal model $s$ and
background model $b$. The index $i$ runs over all the bins of multi-dimensional histograms of
observables and $\bar{\theta}$ represents the vector of nuisance parameters
corresponding to systematic uncertainties. 
Fits to data are performed by minimising the negative log-likelihood function with respect
to the ratios of the couplings: 
\begin{equation} L\left
(\bar{\Omega}\middle | \frac{\tilde{\kappa}_{HVV}}{\kappa _{\rm SM}}, \frac{\tilde{\kappa}_{AVV}}{\kappa _{\rm SM}}\tan{\alpha}, \bar{\theta} \right )
=-2 \ln \prod\limits_{\sqrt{s}} \prod\limits_{\rm FS} \mathcal{L}\left
(\bar{\Omega} \middle | \frac{\tilde{\kappa}_{HVV}}{\kappa _{\rm SM}}, \frac{\tilde{\kappa}_{AVV}}{\kappa _{\rm SM}}\tan{\alpha}, \bar{\theta} \right
).   \label{eq:meobs_LN}
\end{equation}

The test statistic $q' = -2\ln(\lambda) $ is defined as the profiled value of $L$ of Eq.~(\ref{eq:meobs_LN}).
To ensure the correctness of the statistical treatment and the absence of significant biases, a series of tests were performed before
applying the fit to the data. Asimov datasets~\cite{asimov,asimovErratum} created from independently generated Monte Carlo samples with 
$\tilde{\kappa}_{HVV} /{\kappa _{\rm SM}}$ and $(\tilde{\kappa} _{AVV}/\kappa _{\rm SM}) \cdot \tan {\alpha}$  
equal to $0, \pm 2, \pm 4, \pm 6, \pm 8 $ and $ \pm 10$ were injected into the analysis procedure. The tests were repeated for
samples corresponding to 1 and 100 times the LHC Run-I integrated luminosity.  
In all cases the fitted values of coupling constants were found to be in agreement with the injected values within statistical uncertainties.

The results of the matrix-element-observable fit were validated and cross-checked using a nine-dimensional matrix-element method (9D fit).
The method implements a multivariate per-event extended likelihood that is
sensitive to both the $\KtildeH$ and $\KtildeA$ mixing parameters and is based on nine experimental 
observables. 
The probability model is constructed with separate components for signal, the SM $ZZ^*$ background and the reducible
background. The background components are assumed to be independent
of the Higgs boson tensor structure, so all of the sensitivity to
mixing parameters comes from the signal component. Each component
depends on nine experimental observables: $\mfl$, $\ptfl$, $\etafl$, $\cth$, $\cthone$, $\cthtwo$, $\Phi$, $\mone$ and $\mtwo$ 
(described in Section~\ref{sec:fixed_zz}). 

The main sources of systematic uncertainty 
for the tensor structure measurements are the same as discussed in Section~\ref{sec:fixed_hypo} since they are based 
on the same four-lepton variables. 
Several additional sources of uncertainty, specific to each of the methods, are also taken into account.
For the matrix-element-observable fit, the uncertainty related to the Kernel Density Estimation smoothing 
procedure  applied to signal and background multi-dimensional histograms is considered. To estimate the 
influence of this uncertainty on the final result, a procedure similar to the one described in 
Section~\ref{sec:fixed_hypo} is employed. 
The impact of the different sources of systematic uncertainty on the final results is evaluated 
by comparing the BSM exclusion limits obtained with a specific systematic uncertainty included or excluded in 
the fit, while excluding all other systematic uncertainties. 
A similar conclusion holds in the fixed hypothesis test:
the systematic uncertainties have a
very limited impact on the final result. The most important uncertainties 
are related to the estimates of the reducible backgrounds. The relative
impact of these uncertainties on the final $95\%$~CL exclusion limit on BSM couplings was 
found to be around $\pm 1\%$. The second most important group of  sources of systematic 
uncertainty is related to the theoretical uncertainties on the production cross section of the $ZZ^{*}$ background process. 
Their relative impact on the final result is found to be less than $\pm 1\%$. 
The precision of the tensor structure analysis is thus dominated by the statistical errors.

In this paper, only results based on the matrix-element-observable approach are reported. 
The 9D approach was used as a cross-check and produced results compatible with the matrix-element approach. 

\subsection{Individual and combined results} 
\label{sec:tensor_comb}

The results of the tensor structure analyses performed in the \hWW\  channel are reported 
in Ref.~\cite{spincp_ww_paper} and, for completeness, they are also summarised in Table~\ref{tab:res_tens_ww}.

\begin{table}[htbp!]
\centering
 \begin{tabular}{lccc}
 \hline\hline
Coupling ratio &  Best-fit value & \multicolumn{2}{c}{$95\%$ CL Exclusion Regions} \\
  \hWW\        & Observed        & Expected &Observed\\
 \hline
  $\KtildeH$  &$-1.3$            &$[-1.2, -0.7]$ &$(-\infty, -2.2] \bigcup [-1, -0.85]  \bigcup [0.4, \infty)$ \\
  $\KtildeA$  &$-0.2$            & {\rm n.a. }            &$(-\infty, -6]   \bigcup [5, \infty)$ \\

   \hline\hline
 \end{tabular}
\caption{\label{tab:res_tens_ww} Fitted values of $\KtildeH$ and $\KtildeA$ and  $95\%$~CL excluded regions obtained in \hWW\ analysis.
The expected values are estimated for the signal strength measured in data and assuming best-fit values for all other nuisance parameters. 
Only data collected at  $\sqrt{s}=8$~\TeV\ are used. The symbol "n.a." denotes the absence of $95\%$~CL sensitivity.}
\end{table}

\begin{figure*}[htbp!]
 \centering
\begin{minipage}[h]{0.45\linewidth}
\subfloat[~]{\includegraphics[width=1.0\linewidth]{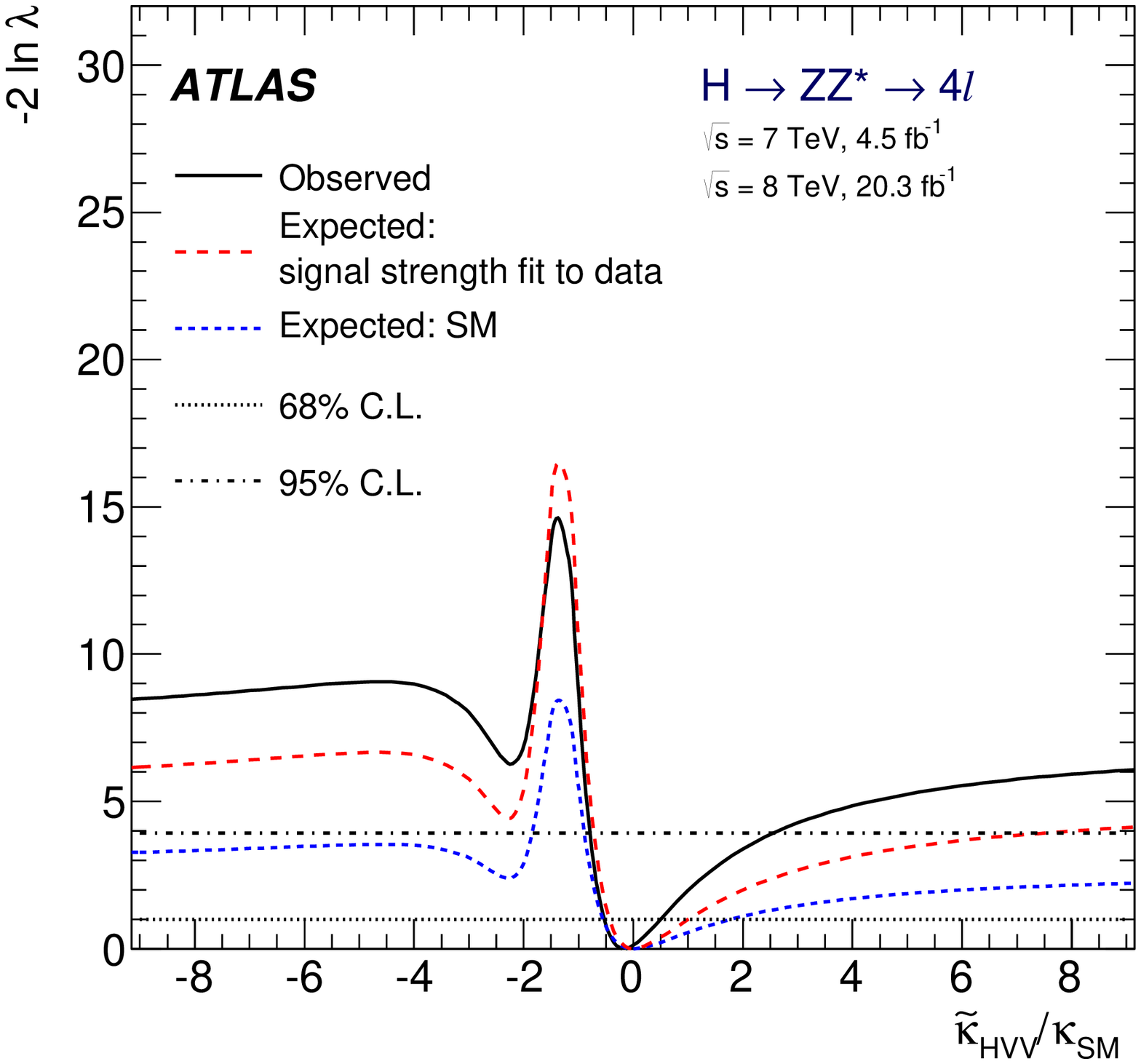}}  
\end{minipage}
\hfill
\begin{minipage}[h]{0.45\linewidth}
\subfloat[~]{\includegraphics[width=1.0\linewidth]{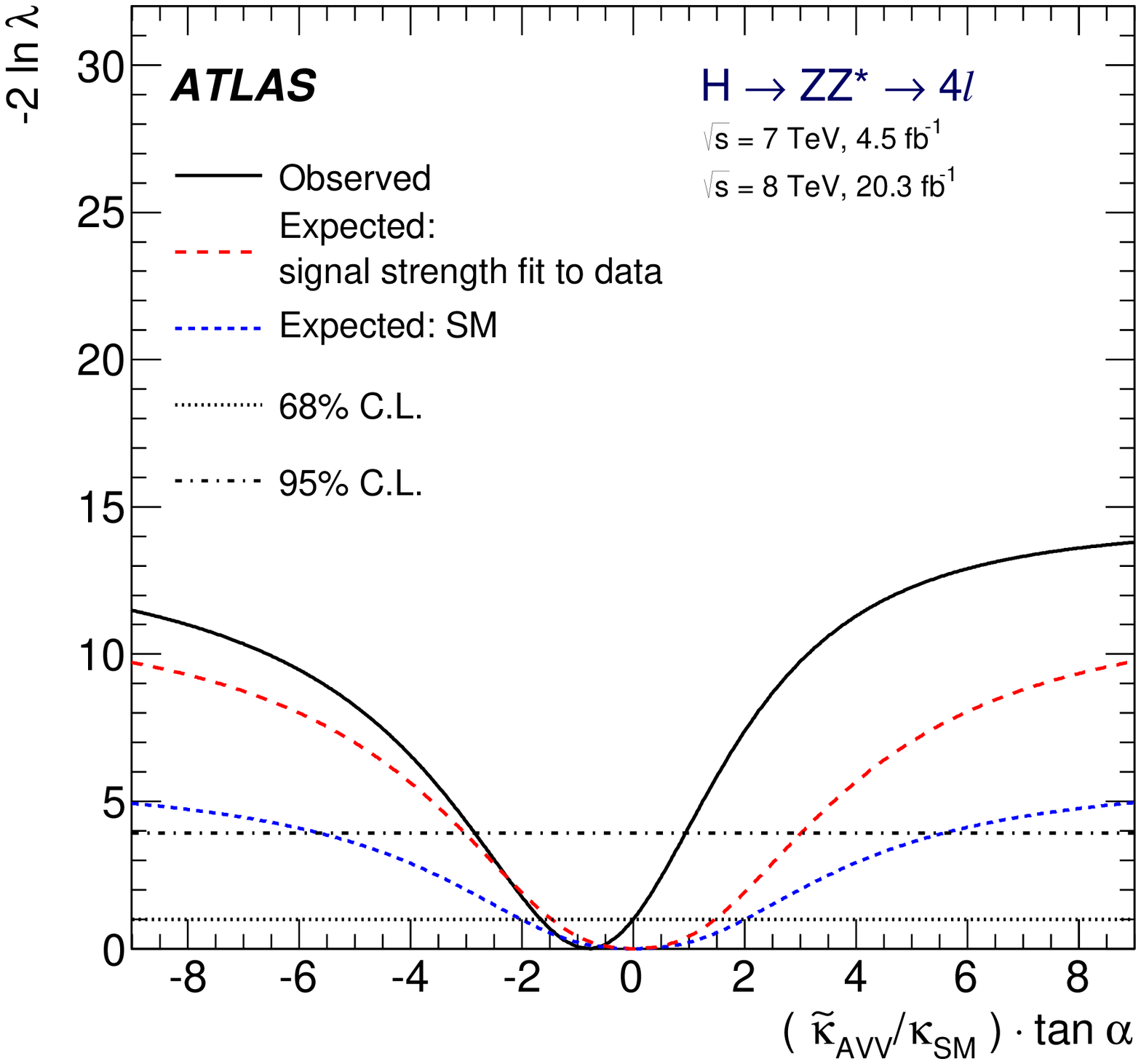}}  
\end{minipage}
   \caption{ \label{fig:res_cond}  Expected and observed distributions of the test statistic 
for fits of  (a) $\KtildeH$ and  (b) $\KtildeA$ for the \hZZ\ analysis.
The expected curves are calculated assuming the SM $J^P=0 ^+$ signal and  produced with the SM 
signal strength $\mu =1$ and with the signal strengths fitted to data.  
The horizontal dotted black lines represent the levels of $-2 \ln \lambda$ above which the values of coupling ratios under 
study are excluded above $68\%$ and $95\%$~CL, respectively.} 
\end{figure*}

The distributions of the test statistic for fits of $\KtildeH$ and $\KtildeA$ 
measured in the \hZZ\ analysis are shown in Figure~\ref{fig:res_cond}.
The expected curves are calculated assuming the SM $J^P=0 ^+$ signal, both with the SM 
signal strength, $\mu =1$, and with the signal strength fitted to data, $\hat{\mu}$. 
The fitted values of  $\KtildeH$ and $\KtildeA$, together with the intervals where these couplings are excluded at above the  $95\%$~CL, 
are reported in Table~\ref{tab:res_cond}. The fitted values agree with the SM predictions within uncertainties. 
\begin{table}[htbp!]
\centering
 \begin{tabular}{lccc}
 \hline\hline
Coupling ratio &  Best-fit value & \multicolumn{2}{c}{$95\%$ CL Exclusion Regions} \\
  \hZZ\        & Observed        & Expected &Observed\\
 \hline
  $\KtildeH$   & $-0.2$ & $(-\infty, -0.75]\bigcup [6.95, \infty)$ & $(-\infty, -0.75] \bigcup  [2.45, \infty)$\\
   $\KtildeA$  & $-0.8$ & $(-\infty, -2.95]\bigcup [2.95, \infty)$  &$(-\infty, -2.85] \bigcup [0.95, \infty) $\\
   \hline\hline
 \end{tabular}
\caption{\label{tab:res_cond} Expected and observed best-fit values of $\KtildeH$ and $\KtildeA$ and  $95\%$~CL excluded regions obtained in the \hZZ\ analysis.
The expected values are estimated for the signal strength measured in data and assuming best-fit values for all other nuisance parameters. 
The data for $\sqrt{s}=7$~\TeV\ and $\sqrt{s}=8$~\TeV\ are combined.}
\end{table}

\begin{figure}[htbp]
  \centering
 \subfloat[~]{\includegraphics[width=0.49\textwidth]{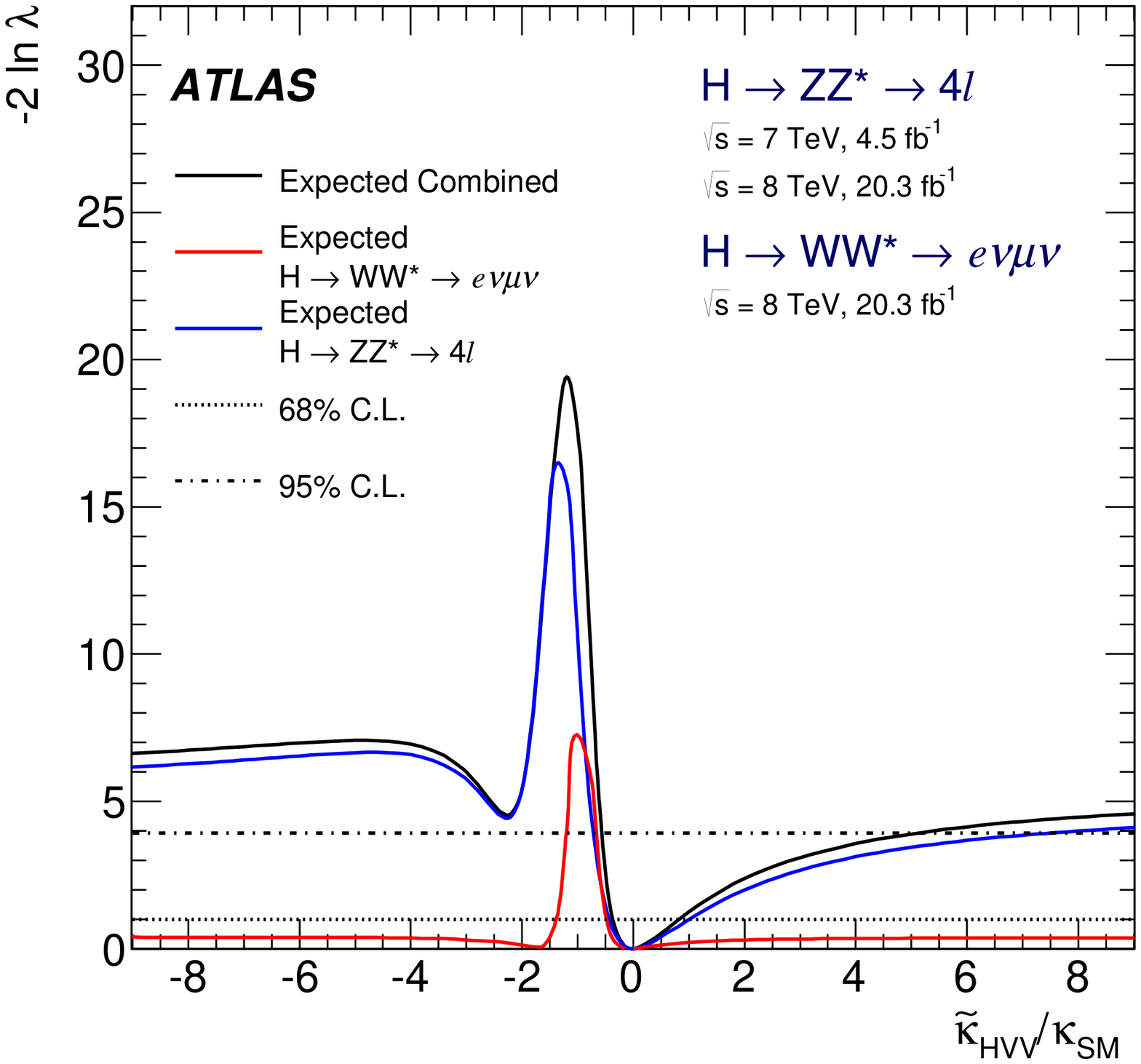}}
  \subfloat[~]{\includegraphics[width=0.49\textwidth]{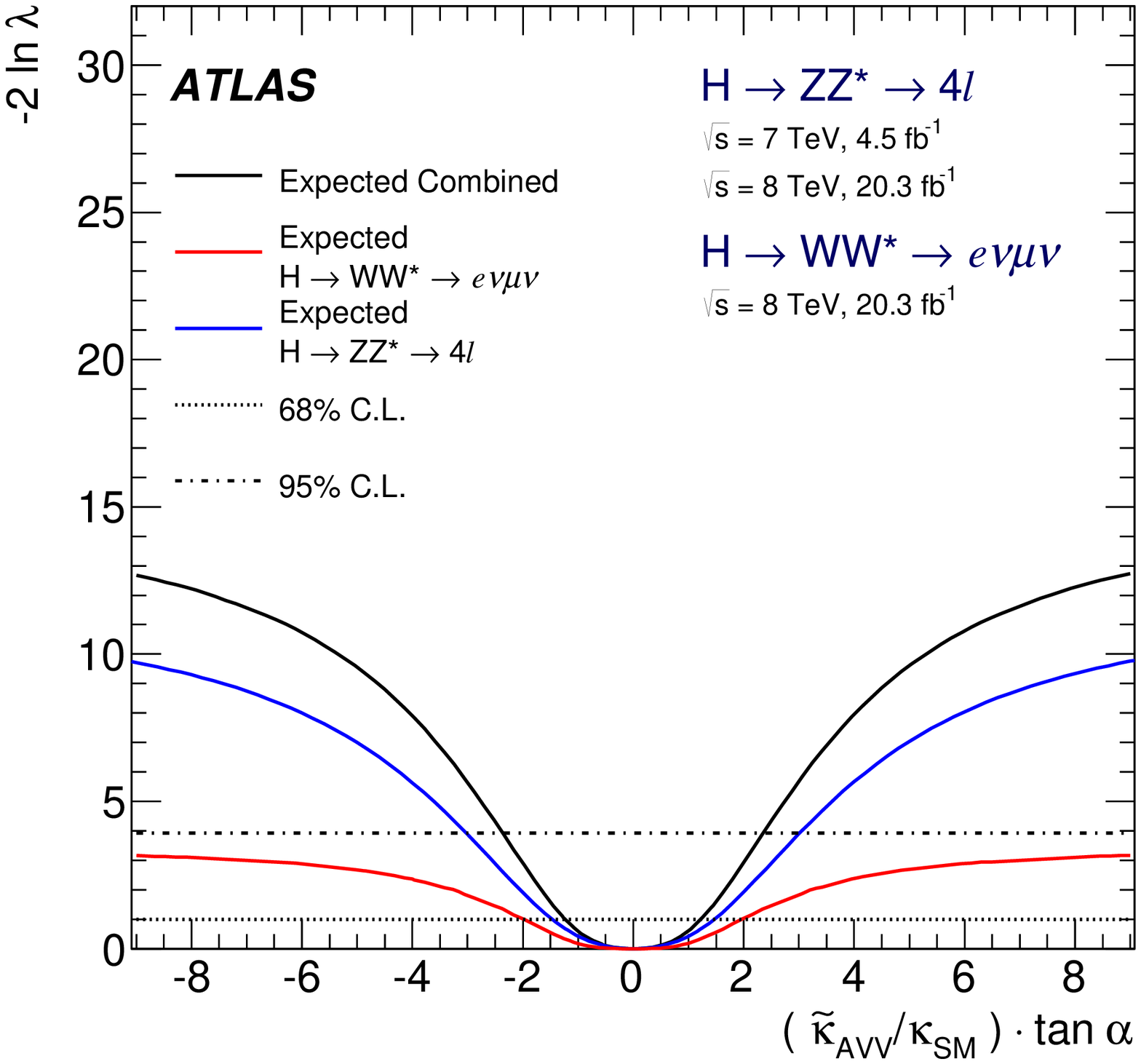}}\\
  \caption{Expected distributions of the test statistic for the combination of \hWW\ and \hZZ\ analyses 
   as a function of BSM coupling ratios (a) \KtildeH\ and (b) \KtildeA . 
   The expected values are estimated for the signal strengths measured in data and assuming best-fit values for all other nuisance parameters.  
  The 68\% and 95\% CL exclusion regions  are indicated as lying above the corresponding horizontal lines.  
    The individual distributions for \hWW\ and \hZZ\ channels  are shown.\label{fig:tensor_comb_exp}}
\end{figure}

\begin{figure}[htbp!]
  \centering
 \subfloat[~]{\includegraphics[width=0.49\textwidth]{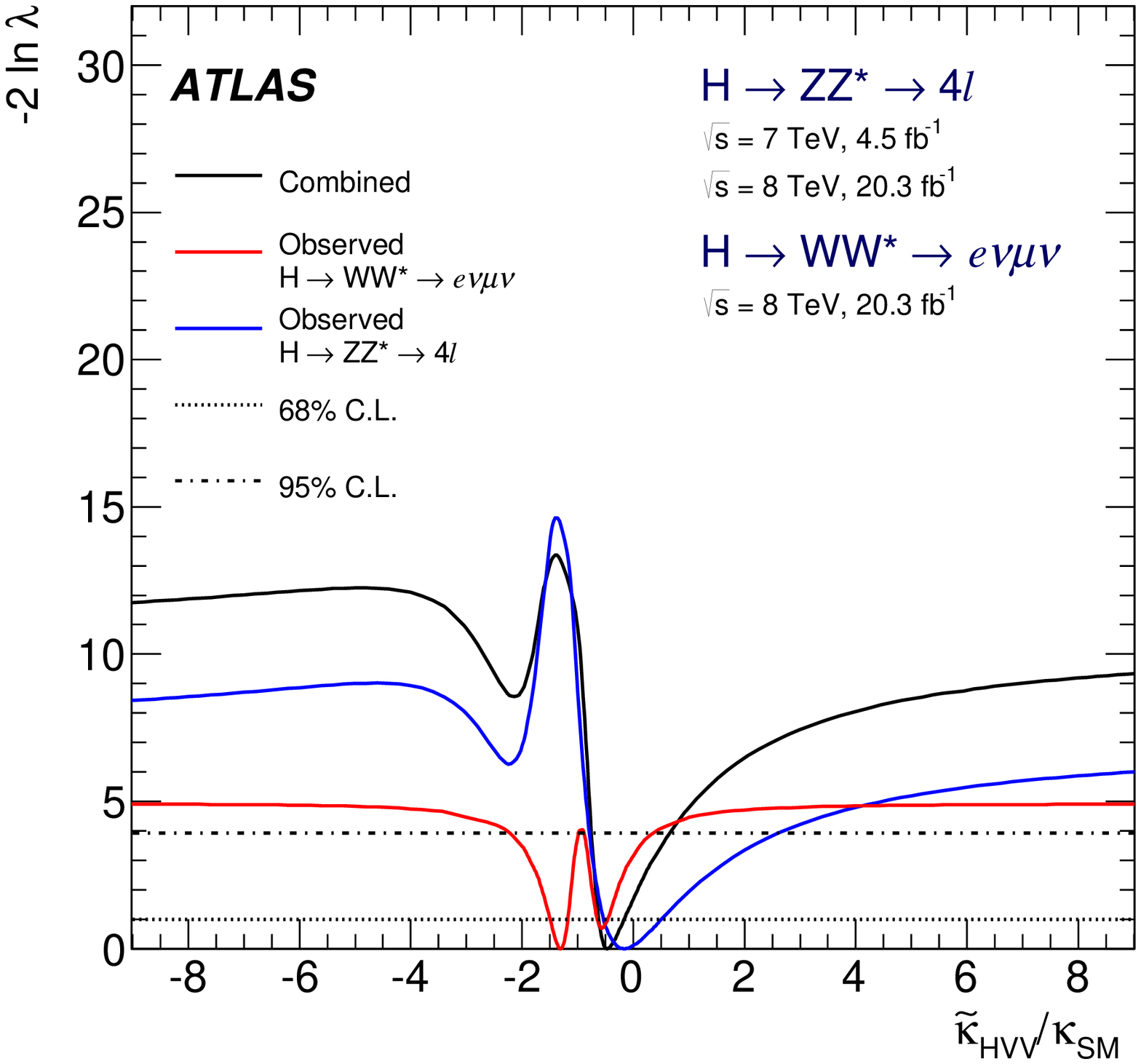}}
   \subfloat[~]{\includegraphics[width=0.49\textwidth]{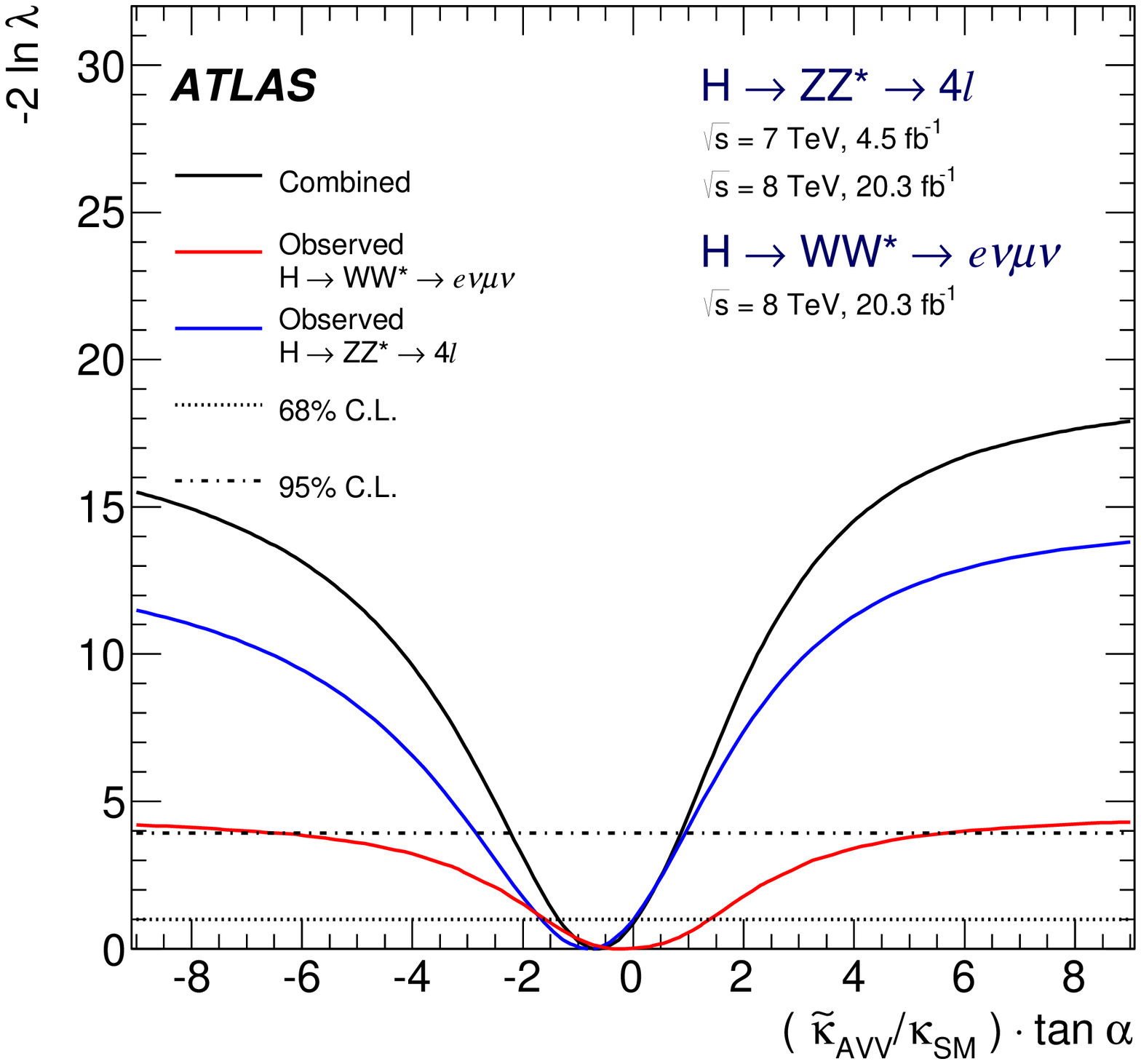}}\\
  \subfloat[~]{\includegraphics[width=0.49\textwidth]{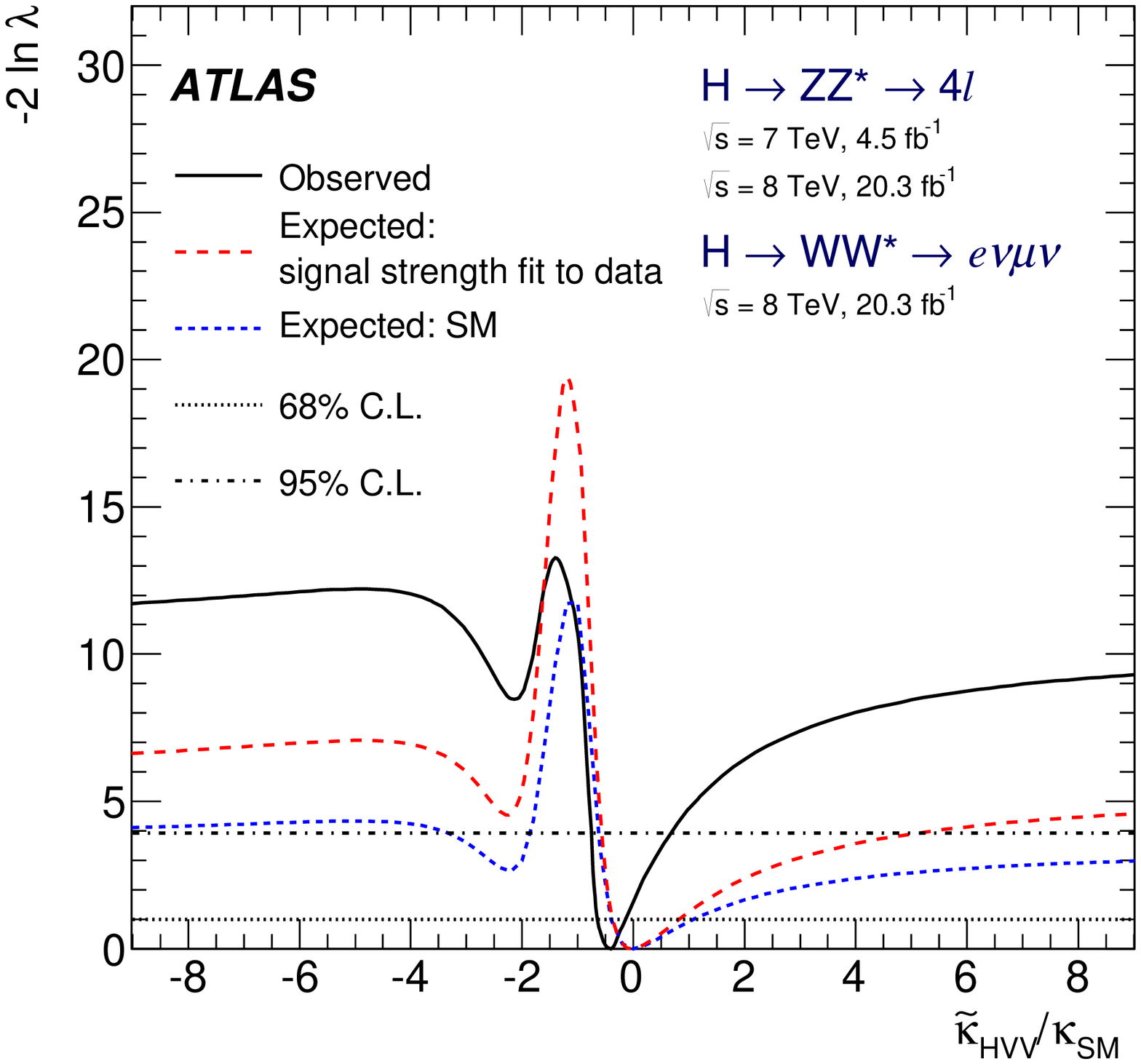}}
   \subfloat[~]{\includegraphics[width=0.49\textwidth]{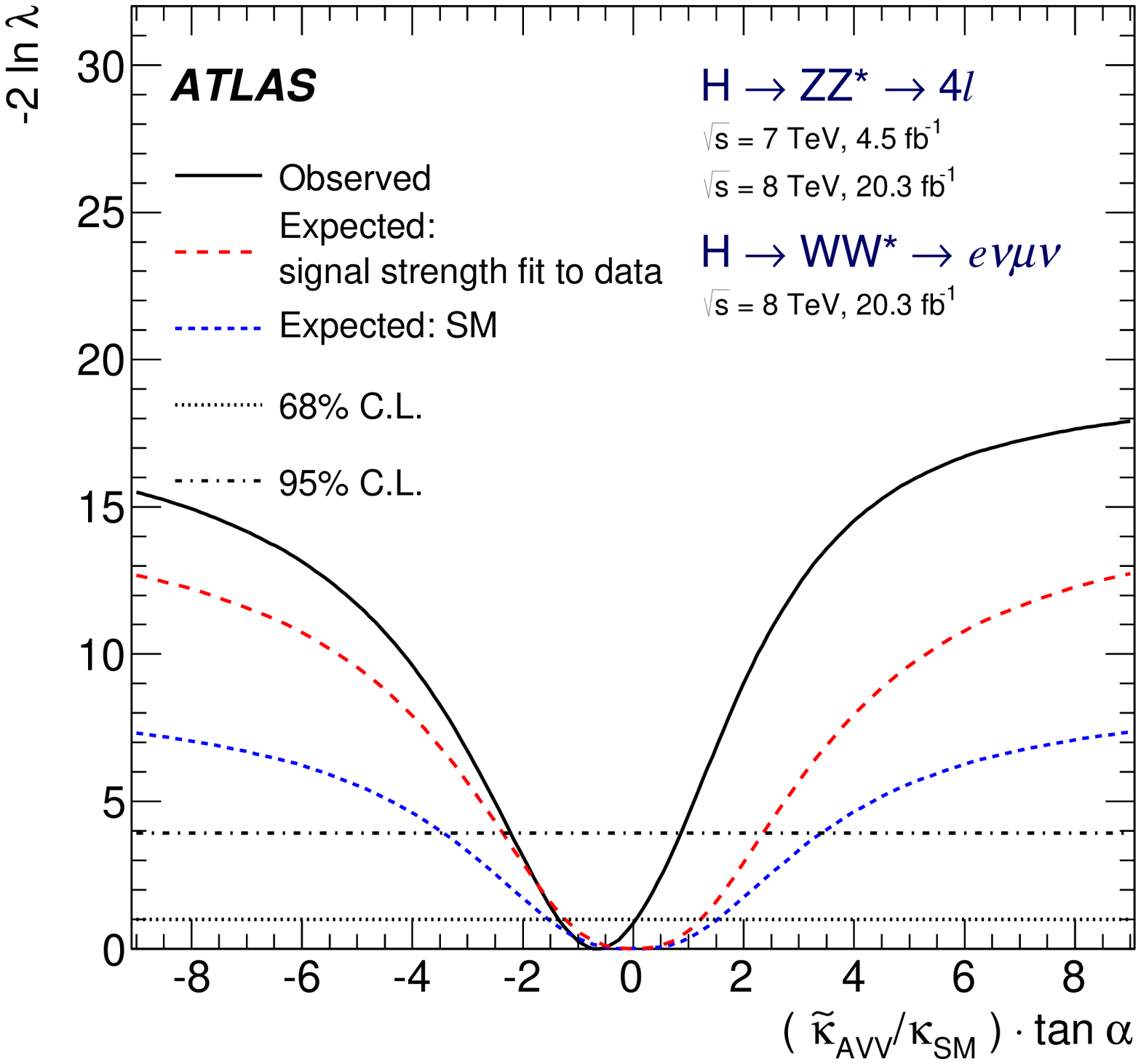}}\\
  \caption{Expected and observed distributions of the test statistic for \hWW\ and \hZZ\ analyses 
  and their combinations. 
  The distributions are shown as a function of the BSM coupling ratios \KtildeH\ and \KtildeA . 
  The 68\% and 95\% CL exclusion regions  are indicated as lying above the corresponding horizontal lines.  
  (a) and (b): individual  \hWW\ , \hZZ\  and combined observed distributions. 
  (c) and (d): expected and observed combined distributions. The expected distributions are presented 
  for the SM signal strength  $\mu =1$ and for the signal strengths obtained from the fit to data. \label{fig:tensor_comb_obs}}
\end{figure}

The measurements from the \hWW\ and \hZZ\ channels are combined under the assumption that the 
BSM  ratios of couplings \KtildeH\ and \KtildeA\  are the same for the $W$ and $Z$ vector bosons. 
A common test statistic is obtained by combining the profiled likelihoods of the individual channels. 
The expected distributions of the likelihoods, for the signal strength values obtained from the fits to the data ($\mu = \hat{\mu}$), 
are presented in Figure~\ref{fig:tensor_comb_exp}.
The observed distributions of profiled likelihoods for the combination of \hWW\ and \hZZ\ measurements
 are presented in Figure~\ref{fig:tensor_comb_obs}.
The asymmetric shape of the expected and observed limits in the \KtildeH\ results is mainly due to the 
interference between the BSM and the SM contributions that gives maximum deviation from the SM
predictions for negative relative values of the BSM couplings. 
Here the signal normalisations are treated as independent nuisance parameters of the different 
decay channels and the different centre-of-mass energies. The other nuisance parameters related to the 
experimental and theoretical uncertainties are treated as correlated when appropriate. 
The resulting $95\%$~CL exclusion regions for the combinations of \hWW\ and \hZZ\ channels 
are listed in Table.~\ref{tab:mixing_exclusion}.
\begin{table}[htbp!]
\centering
\begin{tabular} {lccc}
\hline
\hline
 Coupling ratio &  Best-fit value & \multicolumn{2}{c}{$95\%$ CL Exclusion Regions} \\
 Combined       &  Observed       &Expected&Observed\\
\hline
 $\KtildeH$     & $-0.48$  &   $(-\infty , -0.55]\bigcup [ 4.80, \infty) $& $(-\infty , -0.73]\bigcup [ 0.63, \infty)$\\                          
 $\KtildeA$     &  $-0.68$   &  $(-\infty , -2.33]\bigcup [ 2.30, \infty) $& $(-\infty , -2.18]\bigcup [ 0.83, \infty)$\\
\hline
\hline
\end{tabular}
\caption{Expected and observed best-fit values of  (a) $\KtildeH$ and  (b) $\KtildeA$ and  $95\%$~CL excluded regions obtained in
the combination of \hZZ\  and \hWW\ analyses. The expected values are estimated for the signal strengths measured in data 
and assuming best-fit values for all other nuisance parameters. The signal strengths are treated independently per decay 
channel and per collision energy. \label{tab:mixing_exclusion} }
\end{table}

\FloatBarrier

\section{Conclusion}
\label{sec:conclusion}
Studies of the spin and parity of the observed Higgs boson in the \hZZ , \hWW\ and  \hgg\ decay processes are presented. 
The investigations are based on  $4.5\;{\rm fb}^{-1}$ and $20.3\;{\rm fb}^{-1}$ of $pp$ collision data collected by the ATLAS experiment  at the LHC 
at $\sqrt{s}=7$~TeV and $\sqrt{s}=8$~TeV, respectively. 
The SM Higgs boson hypothesis, corresponding to the quantum numbers $J^{P}=0^{+}$, 
is tested against several alternative spin and parity models.
The models considered include non-SM spin-0 and spin-2 models with universal and non-universal couplings to 
quarks and gluons. The combination of the three decay processes allows the exclusion
of all considered non-SM spin hypotheses at a more than 99.9\% CL in favour of the SM spin-0 hypothesis.

The tensor structure of the $HVV$ interaction in the spin-0 hypothesis is also investigated using the \hZZ\ and \hWW\ decays. 
Only one BSM tensor coupling is investigated at a time, while the other one is set to zero.
The observed distributions of the variables sensitive the ratios of the BSM to SM tensor couplings, \KtildeH\ and \KtildeA, 
are compatible with  the SM predictions.

Values of the BSM tensor couplings outside of the intervals $-0.75< \KtildeH < 2.45$  and $-2.85< \KtildeA < 0.95$ are excluded at the 95\% CL 
for the \hZZ\ process.  For the \hWW\ process the ranges $-2.2< \KtildeH < -1.0$ 
and $-0.85< \KtildeH < 0.4$  and $-6.0< \KtildeA < 5.0$ are excluded at the 95\% CL.   

The results from the \hWW\ and \hZZ\ decay channels are combined under the 
assumption that the \KtildeH\ and \KtildeA\ couplings have the same values 
for the $HWW$ and $HZZ$ processes.
As a result of this combination, the regions outside of $-0.73< \KtildeH < 0.63$  and $-2.18< \KtildeA < 0.83$ intervals 
are excluded at the 95\% CL. 
The corresponding expected not-excluded intervals at the 95\% CL, assuming the SM Higgs boson 
hypothesis and  the signal strength values measured in data, are $-0.55< \KtildeH < 4.80$  and $-2.33< \KtildeA < 2.30$.

\section*{Acknowledgements}

We thank CERN for the very successful operation of the LHC, as well as the
support staff from our institutions without whom ATLAS could not be
operated efficiently.

We acknowledge the support of ANPCyT, Argentina; YerPhI, Armenia; ARC,
Australia; BMWFW and FWF, Austria; ANAS, Azerbaijan; SSTC, Belarus; CNPq and FAPESP,
Brazil; NSERC, NRC and CFI, Canada; CERN; CONICYT, Chile; CAS, MOST and NSFC,
China; COLCIENCIAS, Colombia; MSMT CR, MPO CR and VSC CR, Czech Republic;
DNRF, DNSRC and Lundbeck Foundation, Denmark; EPLANET, ERC and NSRF, European Union;
IN2P3-CNRS, CEA-DSM/IRFU, France; GNSF, Georgia; BMBF, DFG, HGF, MPG and AvH
Foundation, Germany; GSRT and NSRF, Greece; RGC, Hong Kong SAR, China; ISF, MINERVA, GIF, I-CORE and Benoziyo Center, Israel; INFN, Italy; MEXT and JSPS, Japan; CNRST, Morocco; FOM and NWO, Netherlands; BRF and RCN, Norway; MNiSW and NCN, Poland; GRICES and FCT, Portugal; MNE/IFA, Romania; MES of Russia and NRC KI, Russian Federation; JINR; MSTD,
Serbia; MSSR, Slovakia; ARRS and MIZ\v{S}, Slovenia; DST/NRF, South Africa;
MINECO, Spain; SRC and Wallenberg Foundation, Sweden; SER, SNSF and Cantons of
Bern and Geneva, Switzerland; NSC, Taiwan; TAEK, Turkey; STFC, the Royal
Society and Leverhulme Trust, United Kingdom; DOE and NSF, United States of
America.

The crucial computing support from all WLCG partners is acknowledged
gratefully, in particular from CERN and the ATLAS Tier-1 facilities at
TRIUMF (Canada), NDGF (Denmark, Norway, Sweden), CC-IN2P3 (France),
KIT/GridKA (Germany), INFN-CNAF (Italy), NL-T1 (Netherlands), PIC (Spain),
ASGC (Taiwan), RAL (UK) and BNL (USA) and in the Tier-2 facilities
worldwide.

\clearpage
\appendix
\part*{Appendix A}
\addcontentsline{toc}{part}{Appendix A}
\label{app:conv}
To compare the exclusion limits obtained in this analysis to other existing studies, the final results of this analysis are also expressed in 
terms of effective cross-section fractions $(f_{g2},\phi_{g2})$ and $(f_{g4},\phi_{g4})$.
The definitions proposed in Section~11.4.2 of Ref.~\cite{YR3} and Section II of Ref.~\cite{Anderson:2013afp} are used:
\begin{equation}
f_{gi}=\frac{|g_i|^2 \sigma _i}{|g_1|^2 \sigma _1 + |g_2|^2 \sigma _2 + |g_4|^2 \sigma _4},\; \phi _i =\arg{\left ( \frac{g_i}{g_1} \right )}.
\label{eq:fg}
\end{equation}
Here the symbols  $g_1$,  $g_2$ and $g_4$ denote the SM, BSM CP-even and BSM CP-odd tensor
couplings of the $HVV$ scattering amplitude, respectively. The numeric coefficients $\sigma _1$, $\sigma _2$ and $\sigma _4$ are effective cross 
sections of the $HVV$ interactions  calculated when only the $g_1$-, $g_2$- or $g_4$-related terms are present in the amplitude, respectively, such that  $g_i=1,g_{i \neq j} =0$. 

When, in addition to the SM term, only one CP-even or CP-odd  BSM contribution is present, the 
conversion between the parameterisation used in this analysis and the $(f_{gi},\phi _{gi})$ parameterisation  is given by Eq.~(\ref{eq:fg})
rewritten in the following way:
\begin{equation}
 f_{g_i}=\frac{r_{i1}^2}{1+r_{i1}^2}; \;\;\;(i=2,4),
 \label{eq:f_to_g}
 \end{equation}
 where $r_{41}$ and $r_{21}$ are chosen such that:
 \begin{equation}
r_{21}^2 = \frac{\sigma _{HVV}}{\sigma _{\rm SM}} \left ( \frac{ \tilde {k}_{HVV}}{k_{\rm SM}}  \right )^2, \;\;{\rm and}\;\;  r_{41}^2 =   \frac{\sigma _{AVV}}{\sigma _{\rm SM}} \left ( \frac{ \tilde {k}_{AVV}}{k_{\rm SM}}  \right )^2 \tan ^2{\alpha}.
\end{equation}
 The numeric coefficients $\sigma _{\rm SM}$, $\sigma _{HVV}$ and $\sigma _{AVV}$ are effective cross sections of the $HVV$ interaction  calculated when
 only each of the $\kappa_{\rm SM}$-, $\kappa_{HVV}$- and $\kappa_{AVV}$-related terms is present in the Lagrangian.

For consistency with previous measurements reported  in Ref.~\cite{CMS_Spin}, the expected and observed  results of the current analysis 
of the \hWW\ and \hZZ\ channels and for their combination  are expressed in terms of $f_{gi}$ and $\phi _{gi}$ parameters 
for the  \hZZ\ decay, $(f_{g2}^{ZZ} ,\phi _{g2} ^{ZZ})$ and $(f_{g4}^{ZZ},\phi _{g4}^{ZZ})$. These parameters are  
denoted  hereafter by $(f_{g2},\phi _{g2})$ and $(f_{g4}, \phi _{g4})$.  The corresponding results are presented in Tables~\ref{tab:g4zz_exp} and \ref{tab:g4zz_obs}.  
\begin{table}
\centering
\begin{tabular}{ccc}
\hline\hline
\multicolumn{3}{c}{Expected $95\%$~CL limits}\\
\hline
\multicolumn{3}{c}{\hWW\ }\\
\hline
{\rm n.a.} \;\;\;\;\; for  $\phi_{g2} = 0$ &and& $f_{g2}<0.15;\; f_{g2}>0.33$ for $\phi_{g2} = \pi $\\
{\rm n.a.} \;\;\;\;\; for $\phi_{g4} = 0 $ &and& {\rm n.a.} \;\;\;\;\; for $\phi_{g4} = \pi $\\
\hline
\multicolumn{3}{c}{\hZZ\ }\\
\hline
$f_{g2}<0.94$  for  $\phi_{g2} = 0$ &and& $f_{g2}<0.16$ for  $\phi_{g2} = \pi $\\
$f_{g4}<0.56$ for $\phi_{g4} = 0 $ &and&$f_{g4}<0.56$ for $\phi_{g4} = \pi $\\
\hline
\multicolumn{3}{c}{Combination of \hZZ\  and \hWW\ }\\
\hline
$f_{g2}<0.89$ for $\phi_{g2} = 0 $&and&$f_{g2}<0.096$ for $\phi_{g2} = \pi $\\
$f_{g4}<0.43$ for $\phi_{g4} = 0 $&and&$f_{g4}<0.44$ for $\phi_{g4} = \pi $\\
\hline\hline
\end{tabular}
\caption{\label{tab:g4zz_exp} Expected limits on $(f_{g2},\phi _{g2})$ and $(f_{g4}, \phi _{g4})$ parameters defined in Ref.~\cite{YR3}
obtained in the analyses of the \hWW\ and \hZZ\ channels and for their combination. The symbol "n.a." denotes the absence of $95\%$~CL sensitivity. } 
\end{table}
\begin{table}
\centering
\begin{tabular}{ccc}
\hline\hline
\multicolumn{3}{c}{Observed $95\%$~CL limits}\\
\hline
\multicolumn{3}{c}{\hWW\ }\\
\hline
$f_{g2}<0.053$ for $\phi_{g2} = 0 $&and& $f_{g2}<0.20;\; 0.26<f_{g2}<0.63$ for $\phi_{g2} = \pi $\\
$f_{g4}<0.78$ for $\phi_{g4} = 0 $ &and& $f_{g4}<0.84$ for $\phi_{g4} = \pi $\\
\hline
\multicolumn{3}{c}{\hZZ\ }\\
\hline
$f_{g2}<0.68$  for  $\phi_{g2} = 0$ &and& $f_{g2}<0.16$ for  $\phi_{g2} = \pi $\\
$f_{g4}<0.11$ for $\phi_{g4} = 0 $ &and&$f_{g4}<0.54$ for $\phi_{g4} = \pi $\\
\hline
\multicolumn{3}{c}{Combination of \hZZ\  and \hWW\ }\\
\hline
$f_{g2}<0.12$ for $\phi_{g2} = 0 $&and&$f_{g2}<0.16$ for $\phi_{g2} = \pi $\\
$f_{g4}<0.090$ for $\phi_{g4} = 0 $&and&$f_{g4}<0.41$ for $\phi_{g4} = \pi $\\
\hline\hline
\end{tabular}
\caption{\label{tab:g4zz_obs}  Observed imits on $(f_{g2},\phi _{g2})$ and $(f_{g4}, \phi _{g4})$ parameters defined in Ref.~\cite{YR3}
obtained in the analyses of the \hWW\ and \hZZ\ channels and for their combination.} 
\end{table}
To obtain these results, the effective cross sections  $\sigma _{\rm SM}$, $\sigma _{HVV}$ and $\sigma _{AVV}$  of the $HZZ$  interaction are calculated using the \mgaMC\ Monte Carlo generator
\cite{MG5} at leading order. The ratios of cross sections used in the calculation are: $\sigma _{HVV}/\sigma _{\rm SM} = 0.349$ and $\sigma _{AVV}/\sigma _{\rm SM} = 0.143 $, respectively.

\clearpage

\printbibliography

 \newpage 
\begin{flushleft}
{\Large The ATLAS Collaboration}

\bigskip

G.~Aad$^{\rm 85}$,
B.~Abbott$^{\rm 113}$,
J.~Abdallah$^{\rm 151}$,
O.~Abdinov$^{\rm 11}$,
R.~Aben$^{\rm 107}$,
M.~Abolins$^{\rm 90}$,
O.S.~AbouZeid$^{\rm 158}$,
H.~Abramowicz$^{\rm 153}$,
H.~Abreu$^{\rm 152}$,
R.~Abreu$^{\rm 116}$
% ...
\bigskip
\\
% ...
$^{11}$ Institute of Physics, Azerbaijan Academy of Sciences, Baku, Azerbaijan\\
% ...
$^{85}$ CPPM, Aix-Marseille Universit{\'e} and CNRS/IN2P3, Marseille, France\\
% ...
$^{90}$ Department of Physics and Astronomy, Michigan State University, East Lansing MI, United States of America\\
% ...
$^{107}$ Nikhef National Institute for Subatomic Physics and University of Amsterdam, Amsterdam, Netherlands\\
% ...
$^{113}$ Homer L. Dodge Department of Physics and Astronomy, University of Oklahoma, Norman OK, United States of America\\
% ...
$^{116}$ Center for High Energy Physics, University of Oregon, Eugene OR, United States of America\\
% ...
$^{151}$ Institute of Physics, Academia Sinica, Taipei, Taiwan\\
$^{152}$ Department of Physics, Technion: Israel Institute of Technology, Haifa, Israel\\
$^{153}$ Raymond and Beverly Sackler School of Physics and Astronomy, Tel Aviv University, Tel Aviv, Israel\\
% ...
$^{158}$ Department of Physics, University of Toronto, Toronto ON, Canada\\
% ...
\end{flushleft}


\end{document}